\documentclass[12pt]{article}%
\usepackage[nosort]{cite}
\usepackage{graphicx}
\usepackage{float}
\usepackage{subfig}
\usepackage{scalefnt}
\usepackage{wrapfig}
\usepackage{fancybox}
\usepackage{multicol}
\usepackage{multirow}
\usepackage{amsfonts}
\usepackage{amssymb}
\usepackage{amsmath}
\usepackage{heck}
\usepackage[all]{xy}
\xyoption{arc}
\usepackage{setspace}
\usepackage{verbatim}
\usepackage{color}
\usepackage{epsfig}
\usepackage[margin=1in]{geometry}
\usepackage{titletoc}
\usepackage{lscape}
\usepackage{tikz}
\usepackage[debug,pageanchor=false]{hyperref}
\providecommand{\U}[1]{\protect\rule{.1in}{.1in}}
\numberwithin{equation}{section}
\def\Z{\mathbb{Z}}

\renewcommand{\a}{\alpha}
\renewcommand{\b}{\beta}
\renewcommand{\l}{\lambda}

\newcommand{\be}{\begin{equation}}
\newcommand{\ee}{\end{equation}}

\def\t{\tau}
\def\l{\lambda}

\newcommand{\IR}{\mathbb{R}}
\newcommand{\IC}{\mathbb{C}}
\newcommand{\IH}{\mathbb{H}}
\newcommand{\IZ}{\mathbb{Z}}
\newcommand{\IN}{\mathbb{N}}

\def\={\;  = \;}
\def\+{\, + \,}
\def\inn{\,\in\,}
\newcommand{\CN}{\mathcal{N}}
\newcommand{\CH}{\mathcal{H}}
\newcommand{\CA}{\mathcal{A}}

\newcommand{\vth}{\vartheta}
\newcommand{\D}{\Delta}
\newcommand{\wt}{\widetilde}

\definecolor{link}{rgb}{0,0,0}
\hypersetup{colorlinks=true,linkcolor=link,citecolor=link,urlcolor=link,linktocpage}
\begin{document}

\date{September 2015}

\title{F-Theory, Spinning Black Holes\\
and Multi-string Branches}

\institution{HARVARD}{\centerline{${}^{1}$Jefferson Physical Laboratory, Harvard University, Cambridge, MA 02138, USA}}

\institution{HARVARDMATH}{\centerline{${}^{2}$Department of Mathematics, Harvard University, Cambridge, MA 02138, USA}}

\institution{KINGS}{\centerline{${}^{3}$Department of Mathematics, King's College London, The Strand, London WC2R 2LS, U.K.}}

\institution{UTRECHT}{\centerline{${}^{4}$Institute for Theoretical Physics, Utrecht University, 3508 TD Utrecht, The Netherlands}}

\authors{Babak Haghighat\worksat{\HARVARD, \HARVARDMATH}\footnote{e-mail: {\tt babak@physics.harvard.edu}},
Sameer Murthy\worksat{\KINGS}\footnote{e-mail: {\tt  sameer.murthy@kcl.ac.uk}},
Cumrun Vafa\worksat{\HARVARD}\footnote{e-mail: {\tt  vafa@physics.harvard.edu}},
Stefan Vandoren\worksat{\UTRECHT}\footnote{e-mail: {\tt  S.J.G.Vandoren@uu.nl}}}

\abstract{We study 5d supersymmetric black holes which descend from strings of generic $\mathcal{N}=(1,0)$ supergravity in 6d.  These strings have an F-theory realization in 6d as D3 branes wrapping smooth genus $g$ curves in the base of elliptic 3-folds.  They enjoy $(0,4)$ worldsheet supersymmetry with an extra $SU(2)_L$ current algebra at level $g$ realized on the left-movers. 
When the smooth curves degenerate they lead to multi-string branches and
we find that the microscopic worldsheet theory flows in the IR to disconnected 2d CFTs having different central charges.
The single string sector is the one with maximal central charge, which when wrapped on a circle, leads to a 5d spinning BPS black hole whose horizon volume agrees with the leading entropy prediction from the Cardy formula. However, we find new phenomena
 where this branch meets other branches of the CFT. These include multi-string configurations which have no bound states in 6 dimensions but are bound through KK momenta when wrapping a circle, as well as loci where the curves degenerate to spheres. These loci lead to black hole configurations which can have total angular momentum relative to a Taub-Nut center satisfying $J^2 > M^3$ and whose number of states, though exponentially large, grows much slower than those of the large spinning black hole.}

\maketitle

\tableofcontents

\enlargethispage{\baselineskip}

\setcounter{tocdepth}{2}
\section{Introduction \label{sec:INTRO}}
A lot has been learned about BPS black holes in 5 and 4 dimensions in the past twenty years.  In particular
the entropy of BPS black holes which preserve 4 of the supersymmetries have been matched to the expectations
based on the Bekenstein-Hawking entropy formula.  This has been achieved for $d=5$ supergravity theories
with ${\cal N}=4$ and ${\cal N}=2$ supersymmetry (coming in particular from type II strings on $T^5$ or $K3\times S^1$)
\cite{Strominger:1996sh}. Similarly for ${\cal N}=1$ supersymmetry, in the context of F-theory on $M\times S^1$ where $M$ is an elliptic CY 3-fold, this has also been
shown to give rise to the expected macroscopic entropy of the BPS black holes \cite{Vafa:1997gr}.  Finally, also for ${\cal N}=2$ supergravities in $d=4$ realized as type II strings on a CY 3-fold $X$, viewed as M-theory on $X\times S^1$, the entropy has been matched to the microscopic theory \cite{Maldacena:1997de}.  
These results all rely heavily on the realization of the black holes as BPS strings wrapped on a circle in one higher dimension.
In the case of $d=5$ the strings arise by wrapping $D3$ branes on a 2-cycle $C$ in the base of $M$ 
and in the case of $d=4$
they arise\footnote{This has led to a misconception
that black holes always arise from strings in one higher dimension.  That this cannot
be the case is clear if for example one considers M-theory on the quintic 3-fold.  In this case there
is only one charge direction that the $M2$ brane can wrap, so the dual 5d black hole cannot
arise from oscillator modes of a wrapped string in one higher dimension which would require at least two charges.}  by wrapping $M5$ branes on a 4-cycle in $X$.  The two pictures are related by duality between
 F-theory on $M\times S^1$  with M-theory on $X=M$, where D3 branes wrapped on a surface go over to either $M2$ branes if
they wrap the circle, or $M5$ branes wrapped on a 4-cycle (elliptic fibration over the surface) if they do not wrap the circle\footnote{However the configurations needed to lead to macroscopic black holes in 4d require wrapping on a different type of
4-cycles than those that arise from strings in 6d.}.
Spinning BPS black holes exist in 5d.  Moreover, the macroscopic solution for a spinning black hole has been found and matched to the microscopic
count of the BPS strings in 6d, for the cases of ${\cal N}=2,4$ supersymmetries in 5d \cite{Breckenridge:1996is} but
not for the minimal ${\cal N}=1$ case.  One aim of this paper is to remedy this gap and extend
the computation for 5d black holes in \cite{Vafa:1997gr} to the spinning case.  The main ingredient
in this direction is the identification of an $SU(2)_L$ current algebra on the CFT side
and computing its level.  We show how this arises from $D3$ branes wrapped around a genus $g$ curve
on the base of F-theory, leading to an $SU(2)_L$ current algebra of level $k_L=g$.

However, we encounter a number of surprises:  we find that there are, in addition to single
string configurations, multi-string configurations in 6d
which do not bind in 6d but are bound through KK momenta once they wrap
a circle.  This is based on recent discoveries in the context
of strings of 6d $(1,0)$ SCFT's \cite{Haghighat:2013gba,Haghighat:2013tka,Haghighat:2014pva,Kim:2014dza,Haghighat:2014vxa,Hohenegger:2015cba,Gadde:2015tra} as well as related computations of topological strings for compact elliptic 3-folds \cite{Huang:2015sta}. The multi-string configurations can be viewed as coming from 2d CFT's which do not have normalizable ground states.  In gauge theory terminology, they arise as we go from one Higgs branch to another and are localized
where the Higgs branches meet.
In particular we find an intricate structure of such branches of CFT's which lead to distinct theories in the IR with a diverse range of central charges.  
 A similar phenomenon, where the two branches
were Coulomb and Higgs branches, has already been pointed out in \cite{Witten:1997yu}.  The case we are seeing
here is similar in that the phases get disconnected, but it involves transitions between Higgs branches.
Unlike the case of $AdS_3\times S^3\times K3 $ and $AdS_3\times S^3\times T^4$
for which the Coulomb branch generically play no crucial role \cite{Seiberg:1999xz}, in the case of $AdS_3\times S^3\times B$
where $B$ is the base of F-theory compactifications, we expect that the singularities in the Higgs moduli spaces, which
signal the existence of other branches, are not removable.
The effective central charge for the CFT's in various branches are lower than the one which corresponds
to the single centered black hole.    Generically if we distribute
the charges of a BPS black hole to various centers, the entropy goes down, as the entropy grows
as $Q^{3/2}$.  In general such multi-centered black holes would not be bound.   In our case, however, we find that the KK momenta (which
from the 5d perspective can be viewed as another charge for the black hole) binds
these multi-black holes into one object. 

\begin{figure}[here!]
  \centering
	\includegraphics[width=\textwidth]{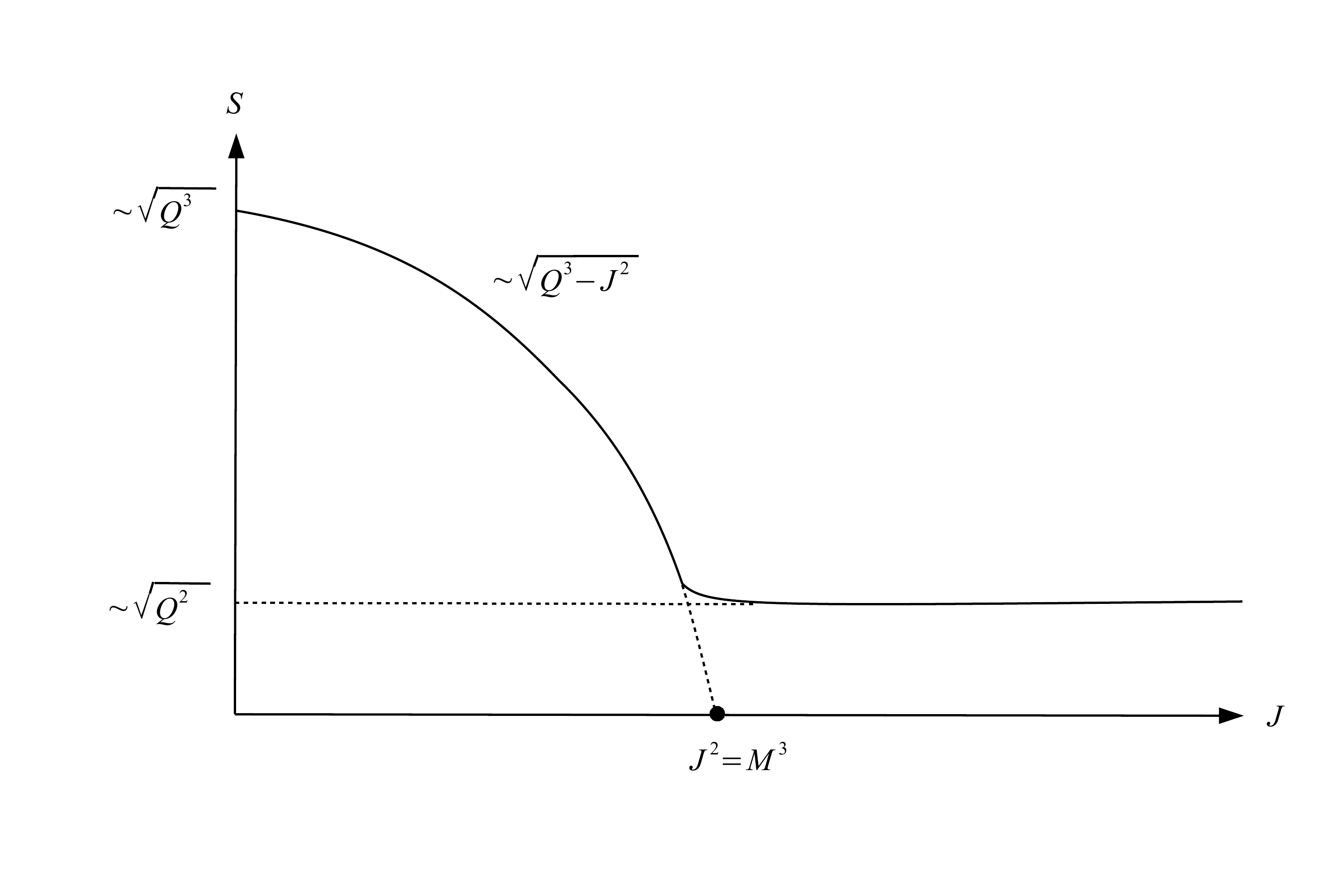}
  \caption{The entropy of a spinning black hole corresponding to large charge $Q$
  with $SU(2)_L$ spin $J$ is schematically plotted here for fixed charge $Q$ as a function of $J$.
  We see that the states $J^2 \gg Q^3$ have an entropy which grows linearly with $\sqrt{Q^2}$ independent of $J$. These large values of $J$ are due to large orbital angular momentum. For black holes arising from strings in 6d, $\sqrt{Q^3-J^2}\sim \sqrt{C^2n-J^2}$ where $C$ denotes the string charge and $n$ is the momentum around the circle and
  the large angular momentum states scale as $\sqrt{Q^2}\sim \sqrt{Cn}$.
  }
  \label{fig:ccb}
\end{figure}

Another surprise we encounter is that a certain subset of configurations involving KK momenta bound to strings leads to black hole states which can have total angular momentum satisfying $J^2>M^3$, but whose number of states scales much slower than those of the large spinning black hole. Naively, this seems to indicate a violation of the cosmic censorship bound (CCB)\footnote{For an account on CCB violating states in the context of black hole entropy see \cite{moulting}.}, but there is no contradiction if such states have large orbital angular momentum relative to a Taub-Nut center\footnote{The Taub-NUT geometry is introduced here as a regulator so that we have an origin with respect to which we can measure orbital angular momentum. The results, in particular the topological string partition function, do not depend on the radius of the Taub-NUT geometry and we will take this radius to be infinite in order to arrive at $\mathbb{R}^4$.}.  Within our setup the total angular momentum can be interpreted as momentum along the Taub-NUT circle and we observe states with arbitrarily large angular momentum for fixed black hole charges, i.e. $J$ is not bounded from above by the charge. Also, we find that these states do contribute to the index, contrary to the states observed in \cite{moulting}. In particular if we consider the black hole entropy for a fixed charge class $C$ and a fixed momentum $n$ around the 6d circle, as a function of (BPS protected) $SU(2)_L$ spin $J$ we find that the entropy goes as shown in Figure \ref{fig:ccb}.
Namely for $J^2 < M^3$ the entropy is dominated by the large spinning black hole satisfying $S\sim \sqrt{C^2n-J^2}$ but for $J$ violating this bound the entropy does not vanish and
to leading order is independent of $J$ and grows as $S\sim \sqrt{Cn}$, no matter how large $J$ is. In the first phase $J$ denotes the internal angular momentum or the internal spin of the black hole whereas in the second phase $J$ is the total angular momentum including the orbital one. States of the second phase arise when the curve in the base of F-theory picks up momenta which are localized at the intersection of the curve with D7 branes.  These intersection points lead to $C\cdot [D7]=C\cdot 12c_1(B)$ complex fermions whose momenta around the circle lead to states violating the cosmic censorship bound.
This leads to a degeneracy of such states going like $exp( 2\pi \sqrt{2c_1(B)\cdot C\cdot n})$ which scales as $exp(a' Q)$, and is subdominant to the macroscopic black hole whose entropy scales as $Q^{3/2}$. In the dual M-theory, these correspond to higher genera curves which degenerate to a sphere by attaching to tori at loci where the tori have degenerated. More generally, we argue that for M-theory compactified on any Calabi-Yau threefold, when a genus $g$ M2 brane
degenerates to a sphere, it leads to black hole states with large orbital angular momenta violating the bound $M^3 > J^2$. We will henceforth sometimes denote these states as CCB violating states although this is not true in the strictest sense of the word as explained above.
These have a simple interpretation:  They arise as the usual angular momentum states of individual BPS particles comprising the black hole ensemble.  In fact the puzzle is if there are always orbital angular momentum states for any value of $J_{orb}$ why the entropy is not a constant function of angular momentum taking always the maximal possible value $Q^{3/2}$ corresponding to $J_{spin}=0$ and $J_{tot}=J_{orb} $? That is the total angular momentum for fixed charge $Q$ is fully comprised of orbital components and the internal spin is zero to yield maximum entropy.
The answer to this turns out to be very interesting and relates to the fact that the index of the black hole states is all that is protected in various phases and the particles which comprise the bulk of the black hole have an index contribution that cancels the orbital angular momentum contributions except for a smaller subset whose growth is subdominant to the usual black hole entropy as was mentioned above.\footnote{This employs and elucidates a feature that was observed in \cite{Vafa:1997gr}
where it was noted that the entropy index of the black hole is subdominant for genus 0 curves.}  We point out that the same states also exist for type IIB compactification on $K3\times S^1$ where they are dual to certain Dabholkar-Harvey states \cite{Dabholkar:1989jt} 
of heterotic strings on $T^5$. 

  Similarly in 4d, the strings studied in \cite{Maldacena:1997de}
correspond to single string configurations which account for the macroscopic
entropy of single-center 4d BPS black holes.  However, the branches involving multi-M5 branes lead to multi-strings
wrapped around the circle which lead to bound states. 
The possibility that wrapped M5 branes should lead
to many 2d CFT's was raised
as an option to resolve the entropy enigma \cite{Denef:2007vg} for 4d black holes.   Here we see that these degenerate configurations, even
though generally lead to lower entropy states, nevertheless do make up the dominant contribution to the black hole entropy in certain phases.

The organization of this paper is as follows:  In Section 2 we discuss the basic setup and preview the main results of the paper. In Section 3 we develop some technical aspects
of 6d strings which arise from wrapping D3 branes around holomorphic 2-cycles in the base of F-theory geometry. Moreover, we show the existence of an $SU(2)_L$ current algebra and we demonstrate the existence of the multi-string phases. Section 4 relates these phases to properties of meromorphic Jacobi forms, and we compare the F-theory setup with the more traditional K3 compactifications.   We also use the relation between the  topological string and spinning black holes \cite{Gopakumar:1998ii,Gopakumar:1998jq,Katz:1999xq} to draw some general lessons about the black hole entropy and in particular the growth of states for large angular momentum at infinity.
In Section 5 we discuss the example of elliptic 3-folds over a ${\mathbb P}^2$ base. Macroscopic aspects of the spinning black holes are discussed in Section 6 
and finally in Section 7, we end with some concluding thoughts.  Some technical details are postponed to the Appendix.

\section{Basic Setup}
\label{sec:basics}

We are interested in this paper in theories with 8 supercharges, in $d=6,5$ and $4$
dimensions with ${\cal N}= (1,0),1,2$ supersymmetries respectively.  In particular we would be interested in realizing
these theories as F-theory, M-theory and IIA strings on elliptic Calabi-Yau 3-folds respectively.

Let us first consider F-theory on an elliptic CY 3-fold, consisting of a complex 2 dimensional
base $B$ and the elliptic fiber varying over it, which captures the coupling constant of type IIB theory up to $SL(2,\mathbb{Z})$ duality.
Non-compact versions of this theory have been recently studied \cite{Heckman:2013pva,DelZotto:2014hpa,Heckman:2015bfa} leading to a conjectured classification of all $(1,0)$ SCFT's in 6 dimensions.  These theories have strings which arise from $D3$ branes wrapped over 2-cycles of $B$.
Moreover the structure of the BPS strings in these theories have been investigated in \cite{Haghighat:2013gba,Haghighat:2013tka,Haghighat:2014pva,Kim:2014dza,Haghighat:2014vxa,Gadde:2015tra}.  These strings carry a $(0,4)$ SCFT
on their worldsheet.  But there is more structure:  Such strings in six dimensions have four transverse directions.  Therefore the $SO(4)$ rotational
symmetry should act on the worldsheet.  Let us denote this global $SO(4)$ symmetry as
$SO(4)=SU(2)_L\times SU(2)_R$.  Some of this symmetry will be realized as acting
on the position of the center of mass and the associated oscillators, and the rest of it
will lead to a current algebra acting on the internal degrees of freedom.  As we will discuss in detail
in the next section this internal $SU(2)_L$ gives rise to a left-moving $SU(2)$ current algebra
on the worldsheet, and $SU(2)_R$ gives rise to a right-moving $SU(2)$ current algebra, which
can be viewed as part of the ${\cal N}=4$ right-moving supersymmetry algebra on the worldsheet\footnote{
In the non-compact case, there is in addition an extra $SU(2)'$  R-symmetry which commutes with the
supersymmetry currents on the worldsheet and is a left-moving symmetry.  This will be absent for the
compact case which is the main focus of this paper.}.

The totality of such strings correspond to all the possible ways we can wrap a curve $C \subset B$.
As in \cite{Vafa:1997gr} we will be interested in the situation where $C$ can be deformed inside $B$, unlike the SCFT case
where $C$ is rigid.   Apart from the center of mass degree of freedom,
we expect to have in the IR a conformal theory with a discrete spectrum.  Moreover, we find that the $SU(2)_L$ global symmetry
of the transverse $\mathbb{R}^4$ rotation is realized, apart from its action on the center of mass of the strings, as a left-moving current
algebra whose level is $k_L=g(C)$ where $g(C)$ is the genus of $C$.

Next we consider compactifying this theory on a circle of radius $R$.  By the duality between
F-theory and M-theory, this is equivalent to M-theory on an elliptic 3-fold where the area of the elliptic 
fiber is given by $1/R$.  The strings of 6d will now have two options:  Either they wrap the circle
or they don't.  If they wrap it, they can also carry momentum along the circle $p=n/R$.  In 5d they correspond to BPS particles.  In the M-theory description of the 5d theory, they will correspond to M2 branes wrapping the class $C$ in the base
and glued to $n$ tori fibers.  If they do not wrap the circle, they will still be a string in 5d and in the M-theory setup they correspond to M5 branes wrapping the four cycle $\widehat C$, which is the total space of the elliptic fibration over $C$.  

On the other hand there is a relation between the partition function of BPS states and the topological string
\cite{Gopakumar:1998ii,Gopakumar:1998jq,Dijkgraaf:2006um}.  Alternatively, the partition function of the topological string can be interpreted as computing
the partition function of spinning BPS black holes in 5 dimensions \cite{Katz:1999xq}. More precisely,
we consider compactifying the theory on one more circle, where as we go around the circle we rotate the 2-planes of $\mathbb{R}^4$
by angles $(\lambda,-\lambda)$, which can be interpreted as the partition function of the Cartan of $SU(2)_L$.
On the other hand, for elliptic threefolds, this same partition function can be viewed from the perspective of the
6d BPS strings as was shown in \cite{Haghighat:2013gba}.  Let $Z_C(\tau, \lambda )$ denote the elliptic
genus \cite{Schellekens:1986xh,Witten:1987cg} of the $(0,4)$ strings arising from $D3$ branes wrapping a class $C$ of the base, where
$\tau$ denotes the modulus of the torus and $\lambda$ denotes turning on Wilson lines in the Cartan
of $SU(2)_L$:
$$Z_C(\tau,\lambda)={\rm Tr} (-1)^{F} q^{L_0} {\overline q}^{\overline L_0}\ e^{2\pi i J_3^L\cdot \lambda }\ ,$$
where the trivial curve class $C$ gives $Z_C=1$. This leads to the formula \cite{Haghighat:2013gba,Haghighat:2014vxa}
$$Z_{top}(\tau,\vec{t},\lambda)=Z^{5d}_{BPS}=Z_{BH}=Z_0(\tau, \lambda)\sum_C Z_C(\tau, \lambda) e^{-\vec{t}\cdot C}\ ,$$
where $Z_0$ is the contribution from the massless multiplet in 6d compactified on a circle\footnote{
We are suppressing the classical cubic term and the one loop linear term.}:
$$Z_0(\tau, \lambda)=\prod_{m>0} (1- e^{2\pi i m \lambda})^{-m \chi/2}\prod_{n>0,m>0} (1-q^n e^{2\pi i m \lambda})^{-m (\chi-2\chi_B)}\prod_{n>0,m>0} (1-q^n e^{2\pi i (m \pm 1)\lambda})^{-m\chi_B}\ ,$$
where $\chi$ is the Euler characteristic of the CY, $X$: $\chi =2h^{1,1}(X)-2h^{2,1}(X)$ and $\chi_B$ is the Euler number of the base $B$ of the CY. 
This can be obtained from finding the contribution of massless 6d modes (hypers and vectors contribute to the second term and tensor
contributes to the third term), when combined with their KK modes.
Moreover $(\tau,\vec{t})$ denote the K\"ahler classes of the elliptic fiber and the base $B$.
Here $Z_{BH}$ denotes the 2nd quantized partition function for BPS spinning black holes where
$\lambda$ is dual to the Cartan of $SU(2)_L$ of the BPS black hole.

The fact that $Z_C(\tau, \lambda)$ is the elliptic genus of a worldsheet implies that it has to have nice
modular properties, and they transform as Jacobi forms.  Indeed we have 
$$Z_C(-1/\tau,\lambda/\tau)=e^{2\pi i k \lambda^2/\tau} Z_C(\tau, \lambda)\ ,$$
where $k=g(C)-1$ is the effective `level' of $SU(2)_L$ or more precisely its anomaly coefficient (the $g(C)$ comes from a current algebra, while
$-1$ comes from the contribution of the center of mass of the string).
As was shown in \cite{Haghighat:2013gba} this modular property of the topological string partition function reexpressed
in terms of the elliptic genus of 6d strings explains the observations of \cite{Hosono:1999qc,Alim:2012ss,Klemm:2012sx} about holomorphic anomaly of the topological string partition function for elliptic Calabi-Yau threefolds.
Notice that, as written, the sum over $C$ leads to an object which does
not have a well-defined modular property because different $C$'s have different levels.\footnote{As discussed in \cite{Haghighat:2013gba} to restore modularity we have to rescale by a non-holomorphic piece in $\tau$:
$$Z_C \rightarrow {\hat Z}_C =exp(\pi k \lambda^2/\tau_2)Z_C\ .$$
It is easy to check that under modular transformation this cancels the prefactor of the effective level
and leads to a modular invariant object.  This is because $\lambda^2\rightarrow \lambda^2/\tau^2$ and
$\tau_2\rightarrow \tau_2/|\tau|^2$.  Note also that this prefactor disappears in the asymmetric limit where
we fix $\tau$ but send ${\overline \tau}\rightarrow \infty$, which is as expected for the holomorphic anomaly.
In this paper we will not worry about this distinction and ignore this non-holomorphic prefactor, except when
we talk about modular properties of the full partition function $Z$.}

 As the curve $C$ deforms inside $B$ we may get singular curves, i.e. it 
may degenerate to $n_C$ curve components $\Sigma_i$ with multiplicity $d_i$. Of course the class of $[C]$
must be equal to
$$[C]=\sum_{i=1}^{n_C} d_i [\Sigma_i]\ .$$
In such a case we get $\sum_{i=1}^{n_C} d_i$ distinct strings which lead to a $2d$ theory with gauge group on the worldsheet:
$$\prod_{i=1}^{n_C} U(d_i)\ ,$$
where the maximal abelian gauge group is $U(1)^{n_{CM}}$ with $n_{CM}=\sum_{i=1}^{n_{C}} d_i$ the center of mass modes.
Here there is a further substructure:  Namely for the $d_i$ strings wrapping $[\Sigma_i]$
we have a $U(d_i)$ gauge symmetry with adjoint fields whose vevs lead to Higgs branches  leading to $d_i$ separable strings.
For a single center of mass, as we noted before there is a contribution of $-1$ to the index of
the modular form.  But for the $U(d_i)$, as we will show, there is a more negative contribution:
$$k_i^{CM}=-d_i(d_i+1)(2d_i+1)/6\ ,$$
leading to a total contribution of the generalized center of mass to the index as
$$k^{CM}=\sum_i k_i^{CM}\ .$$
Of course the net index cannot change and is still $k=g-1$ and so there is a compensating positive contribution to $k$ from the numerator in $Z_C$.
Taking into account the effect of center of mass, which is the only source of potential divergence in the partition
function of the 2d theory, and estimating the order of the pole of the modular form, allows us to restrict the form
of the contribution of various CFT branches to the black hole entropy.  In the computation of the total elliptic genus for the
full CFT, we receive contributions from all branches.  These include the single string branch as well as the other branches.
At first sight one may think that to match to the black hole entropy the only contribution we need to take into account
is that of the single string branch, as the other ones would have multi-centers and are thus not bound.  However, as already
noted we find evidence that where the single string branch meets the other branches there are additional contributions
to the elliptic genus which correspond to bound states of such black holes.  Indeed we can subtract from $Z_d$ the
contribution from disconnected strings (which is part of $Z_d$).  But even after such a subtraction, it turns out
there are contributions to bound states which cannot be viewed as wrapping a single string $d$ times around
the circle.  Moreover it can be seen that without the KK momenta, these bound states disappear.
So the KK mode binds them.

\begin{figure}[here!]
  \centering
	\includegraphics[width=\textwidth]{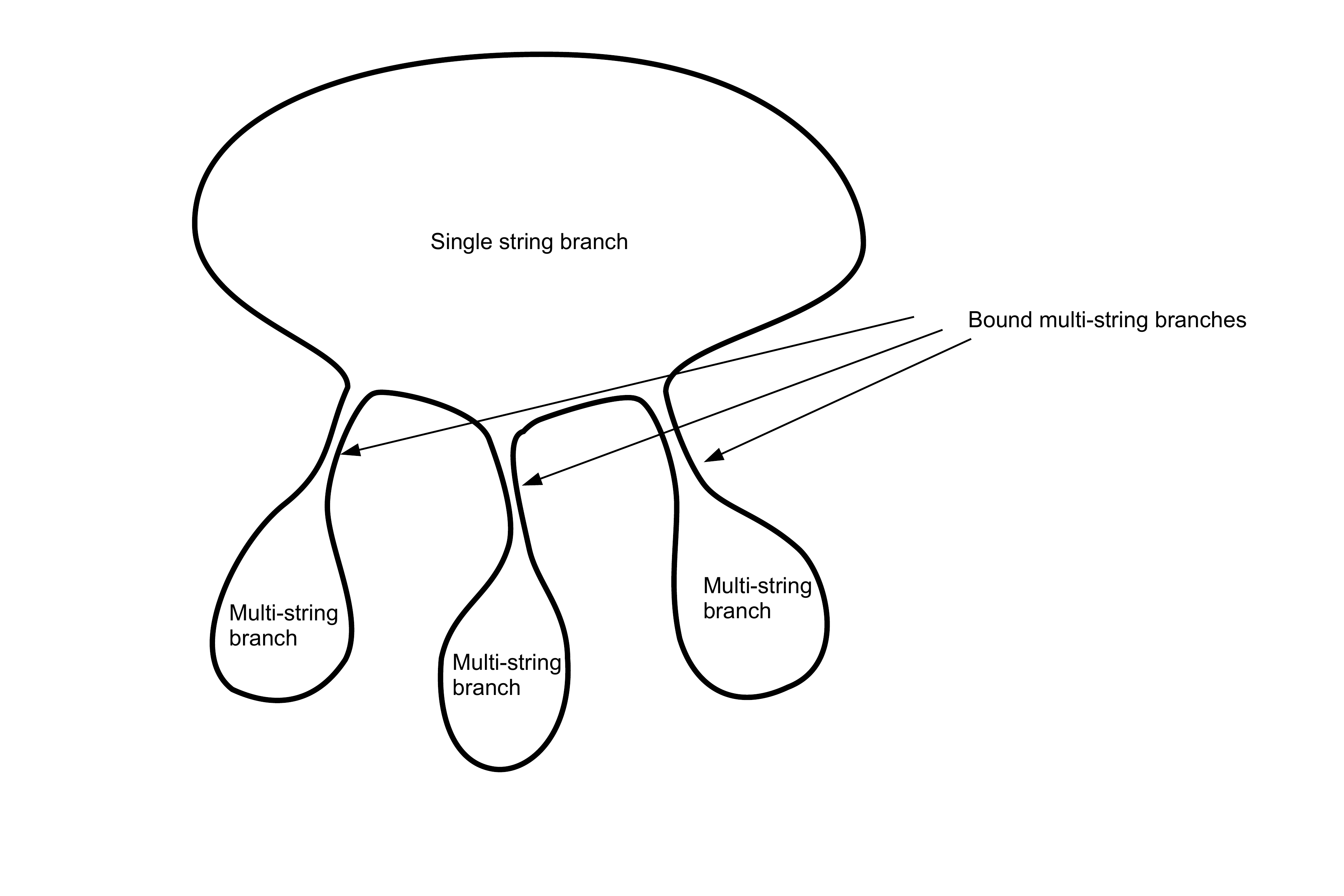}
  \caption{The single string branch is connected to multi-string branches in the UV. In the IR they lead to disconnected CFT's. However, there are additional contributions to the elliptic genus of the full CFT where the branches meet which correspond to multi-string states which are bound by KK-momenta. The effective central charge of these bound strings is expected to be the same as that of the multi-string branches.}
  \label{fig:branches}
\end{figure}

 In other words, the
2d theory ${\cal W}$ decomposes to a number of disconnected CFT's ${\cal W}_\alpha$,
where each $\alpha$ corresponds to a decomposition  $[C]=\sum_{i=1}^{n_{C}} d_i [\Sigma_i]$.  We have
$${\cal W}=\oplus_\alpha {\cal W}_\alpha\ .$$
All of these CFT's have a continuous spectrum, corresponding to generalized center of mass degrees of freedom
of the string.  The one which has only 4 center of mass degrees of freedom corresponds
to a single string.    We will denote this with ${\cal W}_{max}$ which corresponds to the case where $C$ consists
of a single component, and thus with gauge group $U(1)$.   This branch will give us the maximal central charge $c_{max}(C)$.   It may sound surprising that
a given 2d QFT flows to disconnected CFT's.  This was already argued for in \cite{Witten:1997yu} for certain 2d QFT's based
on the fact that Higgs and Coulomb branch of the corresponding theories have different central charges.  This situation also
occurs for type IIB theory compactified on $K3$ or $T^4$ (where the elliptic fibration of F-theory is constant) \cite{Seiberg:1999xz}.  In those cases the Coulomb branch can be avoided by turning on suitable  B-fields.  This
however is not possible in our case, and where ${\cal W}_{max}$ meets the other Higgs branches, there are additional contributions to the elliptic genus which shows that they
cannot disappear by deformations.    Therefore in these cases we expect that the corresponding Higgs branches 
correspond to sigma models which have non-removable singularities.  Their contributions to the elliptic genus arises from configurations which in the CFT language
arises from states of various CFT's localized near the singularities.  These are precisely where the effective central charge can be lower than the bulk.  We find evidence,
through the computation of elliptic genus, that the effective central charge associated with these states are given by central charges of the multi-string branches:
\begin{equation}
c^{eff}=\sum_i d_i c_{max}(\Sigma_i) < c_{max}(C)\ ,
\end{equation}
where $[C]=\sum d_i[\Sigma_i]$.
Figure \ref{fig:branches} illustrates this multi-branch structure.

An interesting subset of these branch degenerations correspond to when $C$ degenerates to a sphere.
We will see later that these are the configurations which lead to violations of the cosmic censorship bound $M^3>J^2$.

Having set these preliminaries, we are now ready for a more detailed discussion in the next section.

\section{Strings from D3-branes \label{sec:D3}}

In this section we study the effective two-dimensional theory arising by wrapping D3-branes on curves inside the base $B$ of an elliptic Calabi-Yau threefold compactification in F-theory. We will begin with deriving the field-content of the two-dimensional theory of a single D3-brane wrapping a genus $g>0$ curve $C$ inside $B$.  We will first assume that $C$ is non-degenerate.
This is somewhat complicated, because the S-duality acts on the worldvolume of the D3 brane, as its coupling
constant is varying and jumping by $SL(2,{\bf Z})$ transformations.  Nevertheless as we shall see some aspects of the
resulting theory can be deduced easily in this way, including the amount of supersymmetry and the existence of
an $SU(2)_L$ current algebra, including its level.  But for some other aspects of the theory it is more convenient
to use its duality with M5 brane wrapping the 4-cycle ${\widehat C}$ (the elliptic fibration over $C$), which was studied in \cite{Vafa:1997gr}, extending
the cases studied in \cite{Maldacena:1997de} to non-ample divisors.

Let us start with the D3 brane description and consider the following brane configuration in type IIB string theory:
\begin{center}
\begin{tabular}{c|cccccccccc}
& \multicolumn{2}{c}{$\mathbb{R}_{||}$} & \multicolumn{4}{c}{$\mathbb{R}^4_\perp$}&\multicolumn{4}{c}{$B$}\\
&  $ X^0 $& $ X^1 $& $ X^2 $& $ X^3 $& $ X^4 $& $ X^5 $& $ X^6 $& $ X^7 $& $ X^8 $ & $X^{9}$\\
\hline
$D3$ & $\times$ & $\times$ & $-$ &$-$ &$-$ & $-$ & $\times$ & $\times$ & $-$ & $-$\\
\end{tabular}
\end{center}
The R-symmetry of the D3-brane worldvolume theory is:
\begin{equation}
	SO(6) \equiv SU(4)_R \rightarrow SO(4)_R \times U(1)_R\ ,
\end{equation}
where $SO(4)_R \equiv SU(2)_L \times SU(2)_R$ arises from rotations in the $\mathbb{R}^4_{\perp}$ plane transverse to the D3 brane and $U(1)_R$ corresponds to rotations in the $X^8$, $X^9$ plane. Furthermore, we identify the Lorentz symmetry of the two-dimensional worldvolume $\mathbb{R}^2_{||}$ of the strings as $U(1)_{||}$ and the canonical line bundle of $C_g$ gives rise to a $U(1)_{C}$-symmetry rotating in the plane $X^6$, $X^7$. Next, note that for D3-branes in F-theory there is a further $U(1)$, which we denote by $U(1)_D$, studied in \cite{Martucci:2014ema}. This is due to the fact that the four-manifold $B$ has non-trivial first Chern class and hence one has to vary the axio-dilaton field over $B$ in order to preserve supersymmetry. This leads to a non-trivial connection 
\begin{equation}
	\mathcal{A} = \frac{1}{2 \textrm{Im}\tau} {\rm d}\textrm{Re}\tau\ ,
\end{equation}
defining for non-constant $\tau$ a $U(1)_D$ line bundle $L_D$. In fact we have $c_1(L_D) = c_1(B)$. Fields which transform with $U(1)_D$-charge $q_D$ can be regarded as sections of $L^{q_D}_D$.

We next want to decompose the supercharges and field content on the D3-brane worldvolume with respect to the full group
\begin{equation}
	G = SU(2)_L \times SU(2)_R \times U(1)_{||} \times U(1)_{C} \times U(1)_R \times U(1)_D\ .
\end{equation}
The resulting theory will be twisted \cite{Martucci:2014ema} in order to preserve supersymmetry. Initially, before the twisting, the field content of a D3-brane consists of a gauge field $A_{\mu}$, six scalars $\varphi^i$, and eight fermions $\Psi^I_A$ and $\tilde \Psi_{I \dot{A}}$, where $A=1,2$ ($\dot{A} = \dot{1},\dot{2}$) denote the left-(right-)moving Weyl indices, and $I=1,\ldots,4$ transform in the $\mathbf{4}$ or $\bar{\mathbf{4}}$ of the internal R-symmetry group $SU(4)_R$. These fields transform as follows under the combined (Euclidean) group $SO(4)\times SU(4)_R = SU(2) \times SU(2) \times SU(4)_R$:
\begin{equation}
	A_{\mu} \in (\mathbf{2},\mathbf{2},1) \quad \varphi^i \in (\mathbf{1},\mathbf{1},\mathbf{6}_v) \quad \Psi^I_A \in (\mathbf{2},\mathbf{1},\mathbf{4}) \quad \tilde \Psi_{I \dot A} \in (\mathbf{1},\mathbf{2},\bar{\mathbf{4}})~,
\end{equation}
and the sixteen supercharges transform as 
\begin{equation}
	Q_{A I} \in (\mathbf{2},\mathbf{1},\bar{\mathbf{4}}) \quad \tilde Q^I_{\dot A} \in (\mathbf{1},\mathbf{2}\ ,\mathbf{4})\ .
\end{equation}
Furthermore, as explained in \cite{Martucci:2014ema}, the supercharges $Q_{A I}$ and $\tilde Q^I_{\dot{A}}$ have $q_D$-charges $+\frac{1}{2}$ and $-\frac{1}{2}$ and the pair ($\Psi^I_A, \tilde \Psi_{I \dot{A}}$) transforms with $U(1)_D$-charges ($+\frac{1}{2},-\frac{1}{2}$). 

Next, let us look how the supercharges transform under the group $G$ above:
\begin{eqnarray}
	Q_{A I} & \longrightarrow & (2,1,+,+,+,+) \oplus (1,2,+,+,-,+) \oplus (2,1,-,-,+,+) \oplus (1,2,-,-,-,+) \nonumber \\
	\tilde Q^I_{\dot{A}} & \longrightarrow & (2,1,+,-,-,-) \oplus (1,2,+,-,+,-) \oplus (2,1,-,+,-,-) \oplus (1,2,-,+,+,-) \nonumber \\
\end{eqnarray}
Now we define a twisted theory by modifying the generators as follows:
\begin{equation}
	T'_{C} = T_{C} + T_R, \quad T'_D = T_D + T_{R}\ .
\end{equation}
Then we find that the surviving supercharges which are neutral under $U(1)'_{C}$ and $U(1)'_D$ are $(1,2,+,+,-,+)$ and $(1,2,+,-,+,-)$. We thus see that the effective two-dimensional theory on $\mathbb{R}^2_{||}$ has $(0,4)$ supersymmetry and that the supercharges transform as doublets of $SU(2)_R$. To summarize, we find that the supercharges transform under the modified group 
\begin{equation}
	G' = SU(2)_L \times SU(2)_R \times U(1)_{||} \times U(1)'_{C} \times U(1)'_D
\end{equation}
as
\begin{equation}
	(1,2)_{+,0,0}, \quad (1,2)_{+,0,0}~.
\end{equation}
Analogously, we find for the scalars $\varphi$ and the fermions $\Psi$:
\begin{eqnarray}
	\varphi^i & \longrightarrow & \{ \varphi^{\alpha \dot{\beta}} \in (2,2)_{0,0,0}\} \oplus \{ \sigma \in (1,1)_{0,1,1}\} \oplus \{ \bar{\sigma} \in (1,1)_{0,-1,-1}\} \nonumber \\
	\Psi^I_A & \longrightarrow & (1,2)_{+,1,1} \oplus (2,1)_{+,0,0} \oplus (1,2)_{-,0,1} \oplus (2,1)_{-,-1,0} \nonumber \\
	\tilde \Psi_{I \dot{A}} & \longrightarrow & (1,2)_{+,-1,-1} \oplus (2,1)_{+,0,0} \oplus (1,2)_{-,0,-1} \oplus (2,1)_{-,1,0} ~.
\end{eqnarray}
In a dimensional reduction of this theory to $\mathbb{R}^2_{||}$ the components of the gauge field parallel to $C$ transform like the scalars $\sigma$ and $\bar{\sigma}$ and thus pair up with these. We next want to study such a dimensional reduction. This is done by counting zero-modes of the relevant fields on the Riemann surface $C$ which amounts to counting zero-sections of the corresponding line bundles. Fields with charge vector $(q_{C},q_D)$ transform as sections of line bundles $K_{C}^{q_{C}} \otimes L_D^{q_D}$. Let us first focus on the charge sector $(q_{C},q_D)=(1,1)$. Then, using Riemann-Roch, we get 
\begin{equation}
	\chi(C,K_C \otimes L_D) = \int_C c_1(K_C) + c_1(L_D) - \frac{c_1(K_C)}{2} = g - 1 + c_1(B) \cdot C,
\end{equation}
where $g$ is the genus of $C$ and we have used $c_1(L_D) = c_1(B)$. Now $\chi = h^0 - h^1 $ but in our case we can use Kodaira vanishing since $L_D$ is an ample line bundle and we get
\begin{equation}
	h^1(C, K_C \otimes L_D) = 0\ .
\end{equation}
Therefore, we get
\begin{equation}
	h^0(C, K_C \otimes L_D) = g - 1 + c_1(B) \cdot C \qquad and \qquad h^0(C,L_D^{-1})=0\ .
\end{equation}
We can also immediately deduce
\begin{equation}
	h^0(C,K_C^{-1} \otimes L_D^{-1}) = 0\ , 
\end{equation}
as negative powers of ample line bundles have no non-zero sections.  Similarly we get
\begin{equation}
	h^0(C,K_C) = g \quad and \qquad h^0(C,K_C^0)=1\ .
\end{equation}
Using these we obtain the following table:\footnote{Note that in computing the charge
for the right-movers on $C$ we have to flip the signs in the index computations above.}
\begin{center}
\begin{tabular}{|c|c|c|}
\hline
~ & bosons & fermions \\
\hline
\multirow{2}{*}{L} & $1 \times (2,2)$ &  \multirow{2}{*}{$2g \times (2,1)$} \\
                            & $ (4g-4+4 c_1(B)\cdot C) \times (1,1)$                    & \\
\hline 
\multirow{2}{*}{R}  & $1 \times (2,2)$ & $2 \times (2,1)$ \\
                            & $ (4g-4+4 c_1(B)\cdot C) \times (1,1)$                         & $ (2g-2+2c_1(B)\cdot C) \times (1,2)$ \\
\hline
\end{tabular}
\end{center}
Note that the left-moving fermions include $2g$ copies of $SU(2)_L$ doublets, where the
$2g$ copies come from the $2g$ 1-cycles on $C$.  We will see later  how these fermions lead to the
$SU(2)_L$ current algebra of level $g$.  Note that the right-moving bosons and fermions come in equal
numbers, which is a reflection of the $(0,4)$ supersymmetry on the worldsheet.  In particular note that
these 4-supercharges transform as $2 \times (1,2)$ under $SU(2)_L\times SU(2)_R$.

 Counting the contributions to the left and right central charges of the resulting 2d theory for all fields in the table we find:
\begin{equation}
	c_L^{\textrm{D3}} = 6 g + 4 c_1(B) \cdot C, \quad c_R^{\textrm{D3}} = 6g  + 6 c_1(B) \cdot C.
\end{equation}
This count does not include the contribution of left-chiral bosons localized at points where D7-branes pierce the Riemann surface $C$.  For each $D7$ brane intersecting the $D3$ brane we expect to receive a left-moving contribution.  However,
these are not all independent modes.  To get the full count of these degrees of freedom
it is convenient to consider compactifying the theory on a cirlce and not wrap the string
around the circle.  Then we have a dual M-theory on the same elliptic 3-fold where  the same
string arises from wrapping M5 brane on $\widehat C$ which is the total space of the elliptic fibration over
$C$. The counting of these degrees of freedom has been carried out in \cite{Vafa:1997gr} (which is slightly
different from the case studied in \cite{Maldacena:1997de} because the divisor in our case is not ample) and it is found there that
\begin{equation} \label{eq:cL}
	c_L = 6g + 12 c_1(B) \cdot C,
\end{equation}
which is now the total left-moving central charge.  Let us review this result.  Notice that going down to 5 dimensions the center of mass
of the string, which is not wrapping the $S^1$, is ${\mathbb R}^3$.  From the perspective of the 6d string whose center of mass is ${\mathbb R}^4$
the extra ${\mathbb R}$ is now realized as a point on the compactified $S^1$ and so ${\mathbb R}^4\rightarrow {\mathbb R}^3\times S^1$.  The $S^1$
will be realized, in the M5 brane setup as one self-dual and one component of an anti-self-dual $h^{1,1}(\widehat{C})$ coming from the base and fiber classes.
Apart from the center of mass degree of freedom, the left-moving bosonic modes can be subdivided into non-compact and compact ones. The first class comes from deformations of $\widehat{C}$ inside the Calabi-Yau manifold giving rise to $h^{2,0}(\widehat{C})$ modes. The fields in the second class come from anti-self dual two-forms and are thus equal to $h^{1,1}(\widehat{C}) - 2$. Left-moving fermions come from odd cohomologies of $\widehat{C}$, giving a total of $4 h^{1,0}(\widehat{C})$ real fermions. These numbers were computed in \cite{Vafa:1997gr} in terms of the cohomology class of the curve $C$ and the first Chern class of the base $B$, namely $c_1(B)$. They are given by
\begin{eqnarray}
	h^{2,0}(\widehat{C}) & = & \frac{1}{2}(C \cdot C + c_1(B)\cdot C) \nonumber \\
	h^{1,0}(\widehat{C}) & = & \frac{1}{2}(C \cdot C - c_1(B)\cdot C) + 1 \nonumber \\
	h^{1,1}(\widehat{C}) & = & C \cdot C + 9 c_1(B) \cdot C+ 2\ .
\end{eqnarray}
One can easily check that the above numbers give rise to the left-moving central charge
\begin{equation}
	c_L = 3 C\cdot C + 9 c_1(B)\cdot C + 6\ ,
\end{equation}
 which is equal to (\ref{eq:cL}) upon using the identity $g(C)= \frac{1}{2}(C\cdot C - c_1(B)\cdot C) + 1$. Similarly, one can check that there are $2h^{2,0}+1$ right-moving compact bosons coming from self-dual two-forms and $2h^{2,0} + 3$ right-moving non-compact bosons coming from deformation degrees of freedom and center of mass motion. Using supersymmetry on the right-moving side we see that the right-moving central charge becomes
\begin{equation}
	c_R = 3 C \cdot C + 3 c_1(B) \cdot C + 6\ ,
\end{equation}
which is equal to $6g + 6 c_1(B) \cdot C$.

\subsection{The left-moving current algebra}

In this subsection we construct the left-moving currents arising from the $2g$ left-moving fermions, each being a doublet under $SU(2)_L$.
Recall that the $2g$ fermions transform as $(2,1)$ of $(SU(2)_L,SU(2)_R)$, so they can be labeled as $\psi^\alpha_A$
where $\alpha=1,2$ labels the fundamental of $SU(2)_L$ and $A=1,...,2g$ labels the 1-cycles on $C$.  However as the curve $C$ moves
inside $B$ the cycles undergo monodromy, so each individual $\psi^\alpha_A$ is not well defined.  However we can combine
the fermions to invariant combinations which are independent of the markings of the 1-cycles in $C$.  Let $\epsilon^{AB}$ denote
the skew-symmetric pairing of the 1-cycles on $C$, and consider
$$J_L^{\alpha\beta}=\sum_{A,B} \psi^\alpha_A \psi^\beta_B \epsilon^{AB}\ .$$
Note that since $\psi^\alpha_A$ are fermionic and $\epsilon^{AB}$ is anti-symmetric, $J_L^{\alpha \beta}$ is symmetric in $\alpha \leftrightarrow \beta$,
i.e. it transforms as the adjoint of $SU(2)_L$.  This is the $SU(2)_L$ current we were after.  Each pair of conjugate $A,B$-cycles
leads to an $SU(2)$ with level $k=1$.  Since the above is a diagonal sum of $g$ such pairs, it immediately follows that $J^{\alpha\beta}_L$
has level $k_L=g$.

This result will become important later not only for the single string branch but also for the multi-string branches because the level of the current algebra comes from anomaly matching and is thus invariant under RG-flow.  Note that there is also a contribution from the center of mass.  However, $SU(2)_L$ acting on the center
of mass, does not lead to a current algebra, as is familiar from how rotation symmetries are realized
in string theory.  Even though there is no current algebra for the action of $SU(2)_L$ on the center of mass
modes, nevertheless it acts, as far as the anomaly of the $SU(2)_L$ current is concerned, as if it has $k=-1$.
We sometimes loosely write $k_L^{CM}=-1$ bearing in mind that there is no current algebra for the center of mass
and we are only including its contribution to the anomaly of the $SU(2)_L$ global symmetry.
The total contribution to the $SU(2)_L$ anomaly is
$$k_L=k_L^{ferm.}+k_L^{CM}=g-1\ .$$
Note that due to the fact that center of mass carries $SU(2)_L$ charge, the relation $L_0\geq Q^2/4k_L$, where $Q$ is the
$J_3$ charge,
is no longer true.  For example if $\alpha_{-1}$ is an oscillator from the center of mass, then it carries charge $+1$ and
if we consider $\alpha_{-1}^n$ it carries $L_0=n$, $Q=n$ which will violate this relation for large enough $n$.
These oscillator modes translate, on the macroscopic side, to black hole states which violate the $M^3\geq J^2$ bound.

To summarize, we have found that if we have a non-degenerate $C$ we get a $(0,4)$ supersymmetric
theory with 
$$(c_L,c_R) = (6g + 12 c_1(B) \cdot C, 6g + 6 c_1(B) \cdot C)\ ,$$
and 
$$(k_L,k_R)=(g-1,g-1+c_1(B)\cdot C)\ .$$
Notice the relation
$$c_L-c_R=6 c_1(B)\cdot C\ ,$$
which will be connected to the bulk gravitational anomaly in the bulk in Section 7.

\subsection{Multi-string phases}

Up to now we have focused on the effective two-dimensional theory of a single D3-brane wrapping a genus $g$ curve $C$. We can parametrize the class of $C$ in terms of cohomology classes $[C_i]\in H_2(B,\mathbb{Z})$
\begin{equation}
	[C] = \sum_{i=1}^{h_2(B)} Q^i[C_i]\ , 
\end{equation}
with integral coefficients $Q_i$. Then the genus of the curve $C$ is computed by
\begin{equation}
	g(C) = {1\over 2}(C\cdot C-c_1(B)\cdot C)+1 
	   = {1 \over 2}(Q^iQ^j \eta_{ij}- Q^ic_i) + 1\ ,
\end{equation}
where $\eta_{ij} = C_i \cdot C_j$ is the intersection form on $B$ and $c_i = c_1(B) \cdot C_i$. We see that the genus is quadratic in the charges $Q^i$ and hence also the left-moving central charge $c_L$. 

However, this is if the curve $C$ is non-degenerate.  In fact $C$ can degenerate to multiple copies of lower genus
curves. Consider one such degeneration, given by
$$C=\sum_{i=1}^{n_C} d_i \Sigma^i\ ,$$
where $n_C$ denotes the number of such curves\footnote{To avoid too many different indices, we use the same index for the number of curve components and for the elements in $H_2(B,\mathbb{Z})$. It should be clear from the context what is meant.}.  Such a splitting we sometimes
denote by $\alpha =(d,\Sigma)$.  If this happens since we have multiple $D3$ branes,
we will have a non-abelian gauge symmetry on the $D3$ branes given by
$$G=\prod_{i}^{n_C} U(d_i)\ ,$$
with additional matter fields. In particular the transverse bosons are now in two copies of the adjoint
representation of $G$ and one can use them to go to a new Higgs branch ${\cal W}_\alpha$, and separate the strings to a total of
$$n_{CM}=\sum_i^{n_C} d_i\ ,$$
copies. We now want to study the central charges these different phases will give rise to. Naively one may think that these
branches do not contribute to normalizable BPS states when we consider the 2d CFT on a circle, and that they
simply give rise to tensor products of CFT's each of which has its own center of mass mode. This turns
out to be false as was discovered in the context of strings of the 6d SCFT's \cite{Haghighat:2013gba} and also even
for the compact case \cite{Huang:2015sta}.   The results found there can be interpreted as the statement that
even though in the IR we may flow to disconnected CFT's, there are nevertheless bound states which reside
at the singular loci where these multi-string branches meet the single-string branch.
Thus we expect the same here. The multi-string branch structure translates at the level of the elliptic genus to a splitting such that for each sub-sector $\alpha$ the elliptic genus is given by a meromorphic Jacobi-from $Z^{\alpha}$ which is itself given by the quotient of two weak Jacobi-forms $N_{\alpha}$ and $Z^{\alpha}_0$. Thus we have 
$$Z^\alpha = N_{\alpha}/Z^{\alpha}_0.$$
The total elliptic genus is then given as a sum of these sub-sectors
\begin{equation}
	Z_C = \sum_{\alpha} Z^{\alpha}.
\end{equation}
We will analyze the growth of the coefficients of $Z_C$ and use this to estimate the central charges of the sub-sectors $\mathcal{W}_{max}$ and $\mathcal{W}_{min}$ which come from the multi-string branches. We refer to a more thorough discussion of meromorphic Jacobi forms to Appendix A.

\subsubsection{Evaluation of $Z^{\alpha}_0$}
There are two sources of singularity in the elliptic genus $Z^{\alpha}$.  One arises because
of the contribution of additional center of mass modes, and the other arises by the contribution of the
vacuum energy, which contributes a pole $q^{r}$ for some $r<0$.  
Finding $r$ is relatively easy:  If we go down on a $T^2$ to 4d and wrap the string only on one circle and consider the dual type IIA theory picture, involving wrapped
$D4$ branes on $\widehat C$ (the elliptic fibration over $C$), the most singular term in the elliptic genus simply measures the
$D0$ brane charge induced on the $D4$ brane charge as in \cite{Vafa:1995zh,Green:1996dd}.  In this case
we find that
$$r={p_1({\widehat C})\over 48},$$
where $p_1(\widehat{C})$ is the Pontryagin class of $\widehat{C}$. For $\widehat C$ elliptic one can show that $p_1(\widehat C)=-2\chi ({\widehat C})=-24c_1(B)\cdot C$, which leads to
$$r=-{1\over 2} c_1(B)\cdot C$$
(this can also be anticipated from the Gotsche formula for the partition function of symmetric products of $\widehat C$).
If we consider 
$$Z^{\alpha}\cdot \eta(\tau)^{12c_1(B)\cdot C},$$
the resulting partition function will have no poles as $q\rightarrow 0$.

Now we come to estimating the structure of the singular contribution due to the center of mass modes. If we ignore
the $SU(2)_L$ parameter $\lambda$, up to the divergent volume factors, this should be simply
$$1/\eta(\tau)^{4d}\ ,$$
for $4\cdot d$ transverse bosons.  Of course these come from part of $12c_1(B)\cdot C$ discussed above
in computing the singular piece of $Z^{\alpha}$.
Now we wish to restore the $\lambda$ dependence. First
let us consider the case of a single string, so $d=1$.  In this case we have just the usual center of mass of a single string and noting
that the transverse direction splits to two copies of the spin 1/2 representation of $SU(2)_L$, this leads to the contribution (with~$y=e^{2\pi i \l}$):
$$Z_1^{CM}={1\over y q^{1/6}\prod_{n=1}^{\infty} (1-q^n y)^2(1-q^{n-1} y^{-1})^2} \; .$$
%
To get rid of the negative power of $q$ which is already accounted for
in $r$ by the $\eta^{-12c_1(B)\cdot C}$ term, we consider
$${\hat Z}_1^{CM}={\prod_{n=1}^{\infty} (1-q^n )^4\over y \prod_{n=1}^{\infty} (1-q^n y)^2(1-q^{n-1} y^{-1})^2} \; , $$
which accounts for the $y$-dependence of the singularity for the single string case. 
The function~${\hat Z}_1^{CM}$ can be written in terms of well-known~$\vartheta_{1}$ and~$\eta$ functions
as (with~$q=e^{2\pi i \tau}$, $z=e^{2\pi i \l}$):
\be\label{Z1CM}
{\hat Z}_1^{CM}(\t,\l)={\eta(\tau) ^6\over \vartheta_1(\tau, \lambda)^2} ={1\over \varphi_{-2,1}(\t,\l)} \; .
\ee
As indicated in this equation, ${\hat Z}_1^{CM}$ is the inverse of the weak Jacobi form $\varphi_{-2,1}$ 
of weight $w=-2$ and index $k=1$. 
Some properties of this function are discussed in the appendix along with other relevant facts about Jacobi forms .

As is clear from the above representations, the function~${\hat Z}_1^{CM}(\t,\l)$ is \emph{meromorphic} 
in the~$\l$ variable, with poles at~$\l=\IZ\tau+\IZ$. 
The key physical distinction between holomorphic Jacobi forms and meromorphic Jacobi forms is that the 
states captured by the holomorphic ones satisfy  $L_0>J_L^2/4k$ (up to a
shift of $L_0$ by $c/24$) but those of the meromorphic ones only satisfy $L_0>0$ (again up to a shift by $c/24$).
This is clearly necessary for conformal theories which include rotations acting on non-compact generalized center of mass degrees of freedom.
Now consider the contribution to the center of mass where we have, say $d$, identical strings.  In this case we may think that we simply get $(Z_1^{CM})^d$.  But this does not take into account that we need to symmetrize the identical strings.
In particular if we denote the transverse position of a string in a transverse complex plane by $z_i$, the invariant
combinations we get are $\sum_{i=1}^d z_i^s$  for $s=1,...,d$.  This leads to a different $SU(2)_L$ contribution of the
form
$${\hat Z}_d^{CM}=\prod_{s=1}^d \hat Z_1^{CM}(\tau, s\lambda) = \prod_{s=1}^d {1\over \varphi_{-2,1}(\tau, s\lambda)} \, .$$
So taking into account all the distinct strings we get
$${\hat Z}^\alpha_{CM}= \prod_{i=1}^{n_C}\prod_{s_i=1}^{d_i} {1\over \varphi_{-2,1}(\tau, s_i\lambda)}\ .$$
Putting these two ingredients together we find that the singular contribution to $Z_{\alpha}$ is given by
$$Z_0^\alpha ={1\over \eta(\tau)^{12c_1(B)\cdot C}\prod_{i=1}^{n_C}\prod_{s_i=1}^{d_i} \varphi_{-2,1}(\tau, s_i\lambda)}$$
\noindent
We then define $N_\alpha =Z^\alpha/Z_0^\alpha$ which gives a non-singular weak Jacobi form.  As we have argued above for each decomposition $\alpha$ of $C$ to curves we expect
the elliptic genus to be expressible as
\be \label{Zalpha}
Z^\alpha(\t,\l)={N_\alpha(\tau,\lambda)\over \eta(\tau)^{12c_1(B)\cdot C}\prod_{i=1}^{n_C}\prod_{s_i=1}^{d_i} \varphi_{-2,1}(\tau, s_i\lambda)} \; ,
\ee
where $N_\alpha$ is a weak Jacobi form. 
We now determine the weight and index of $N_\alpha$.  Since $Z^\alpha$ has
modular weight zero, we learn that the modular weight is $6c_1(B)\cdot C-2n_{CM}$ (where $n_{CM}=\sum_i d_i$).  Moreover the
index of the denominator can be computed using the fact that the index of $\varphi_{-2,1}(\tau,s_i\lambda)$ is $s_i^2$, so 
$$k_\text{denom}\=\sum_{i=1}^{n_c}\sum_{s_i=1}^{d_i}s_i^2\={1\over 6}\sum_{i=1}^{n_c}d_i(d_i+1)(2d_i+1)\ .$$
Since $k_L=g-1$, as we have argued before, and this should not change in various branches,
it implies that the index of $N_\alpha$ is given by
$$k_N\=g-1+{1\over 6}\sum_{i=1}^{n_c}d_i(d_i+1)(2d_i+1)\ .$$
Thus we see that the weight and index are given by
\be \label{wkNum}
(w_N,k_N)=(6c_1(B)\cdot C-2n_{CM}, g-1+{1\over 6}\sum_{i=1}^{n_c}d_i(d_i+1)(2d_i+1))\ .
\ee
We note, for later use, that the weight is positive and the index is non-negative.

On the other hand the full $Z^{\alpha}$ is a meromorphic Jacobi form and has weight and index
\begin{equation}
	(w_Z,k_L) = ( 0,g-1)\ .
\end{equation}
Here we have discussed the contribution to the elliptic genus for each branch $\alpha$ corresponding to
a particular degeneration of $C$.  The total elliptic genus for a $D3$ brane wrapping $C$ is given by
$$Z_C(\tau, \lambda)=\sum_\alpha Z^\alpha (\tau, \lambda)\ .$$
Sometimes it is convenient to combine these contributions to one object, by combining
the denominators together. 

For example, for the case of $B={\mathbb P}^2$ when we consider
a degree $d$ curve we can write $Z_d$ as
\be \label{defZd}
Z_d={N_d(\tau,\lambda)\over \eta^{36d} \prod_{s=1}^d \varphi_{-2,1}(\tau,s\lambda)}\ ,
\ee
where $N_d(\tau,\lambda)$ is a weak Jacobi form of weight and index 
$$(w,k)=(16d,g-1+{1\over 6}d(d+1)(2d+1))\ ,$$
where $g={1\over 2}(d^2-3d)+1$.

The bottom line is that the generating functions of BPS degeneracies~\eqref{defZd} are merormorphic
Jacobi forms. The gravitational physics associated to such meromorphic Jacobi forms is somewhat different 
from that of classical, holomorphic, Jacobi forms like the ones considered in~\cite{Strominger:1996sh}. 
We turn to their analysis in the following section.

\section{Black holes from spinning strings} \label{sec:BHCen}

Our main goal in this section is to explain the various aspects of the five-dimensional 
gravitational physics associated with the microscopic BPS partition functions of the F-theory compactifications 
that we found in~Section \ref{sec:D3}. We first start by recalling the connection between the topological
string partition function and the black hole degeneracy index.  In that subsection we also explain special
features for the case where Calabi-Yau is elliptic.  
In the next subsection we review the simpler case of~$K3$ compactifications as a warm-up and as a benchmark,
and go on to our F-theory compactifications in the third subsection. 
While most of the features of the microscopic 
partition function in the $K3 \times S^{1}$ compactifications are explained by a large spinning 
BMPV black hole that obeys the usual $M^{3}>J^{2}$ bound, there are configurations
which violate this bound which grow as $exp(a Q)$, as long as the KK momenta are 0.  
We find that the F-theory partition function contains, in addition configurations having arbitrary high values of spin for any value~$n$ of KK momentum, 
with a degeneracy of states growing as $exp(a\sqrt{n Q})$ for large~$n$.

\subsection{Topological string and spinning black hole degeneracy index \label{toplspinBH}}
In this subsection we review the general connection between topological string degeneracies and spinning black holes
\cite{Gopakumar:1998ii,Gopakumar:1998jq,Katz:1999xq}.   We will start with the general setting, and do not
restrict to elliptic CY.  BPS particles in 5d carry $SO(4)$ rotation quantum
numbers of the form
$$(j_L,j_R)\otimes I_1\ ,$$
where
$$I_1=2 (0,0)\oplus (1/2,0)\ ,$$
and denotes the quantization of the four fermion zero modes corresponding to the broken supercharges. $(j_L,j_R)$ denotes the $SU(2)_L\times SU(2)_R$ quantum number of the BPS state.  A BPS state
has a charge $Q$.  Let $N^{Q}_{(j_L,j_R)}$ denote the degeneracy of such BPS states.
The index is obtained by considering a $(-1)^F$ insertion and in particular summing over the $SU(2)_R$ quantum numbers.
As such, it is convenient to consider the indexed degeneracy 
$$N^Q_{j_L}=\sum_{j_R} (-1)^{2j_R} (2j_R+1) \ N^Q_{(j_L,j_R)}\ .$$
If we consider an isolated genus $g$ curve in CY, it gives rise to the $j_L$ representation given by \cite{Gopakumar:1998ii,Gopakumar:1998jq}
$$S_g=[({1\over 2})\oplus 2(0)]^{\otimes g}$$
In general we can use this as a basis and expand any $N^Q_{j_L}$ in this basis:
$$\sum_{j_L} N^Q_{j_L}[j_L]=\sum_g N^Q_g S_g$$
where we interpret $N^Q_g$ as the net number of states of charge $Q$ of genus $g$.
Another convenient thing to do is to consider individual $J^3_L$ quantum numbers.  Let 
$$N^Q_m=\sum_{j_L\geq |m|,\, j_L-m\in {\bf Z}} (-1)^{2j_L} N^Q_{j_L}.$$ 
Note that the contribution of internal spin of $S_g$ to this index is
$$sTr_{S^g} y^{2J^3_L}=[-(y^{-1/2}-y^{1/2})^{2}]^g=[2 \sin ({\pi \lambda})]^{2g}$$
where we identify $y=exp(2\pi i \lambda)$ where $\lambda$ is the topological string coupling constant.
 Then the partition function of the topological string is given by
$$Z^{top}(t,\lambda)=\prod_{n=1}^{\infty}\prod_{Q}\prod_{m} (1-y^{n-1+2m} {\rm exp}(-t \cdot Q))^{nN^Q_m}\ .$$
Here $t$ denotes the K\"ahler class of the CY.   The extra product over $n$ in the above expression is physically meaningful \cite{Dijkgraaf:2006um}:
It reflects the BPS content of spacetime angular momenta for individual BPS particles.  For each orbital left spin $n-1$ there are only $n$ contributions
(coming from the count of $SO(4)$ spins adding up to $n$ which preserve the BPS condition and can be
viewed in complex coordinates of the 4 dimensional space as $z_1^{m_1} z_2^{m_2}$ and lead to left spin: $(m_1+m_2)=n-1$ ).  
This is sometimes ignored
in counting of spinning black holes, as one only considers the leading behaviour of entropy.  However, it turns
out that these are precisely the modes which are responsible for violations of cosmic censorship bound.  Indeed
even for a single BPS state, the orbital spin can be arbitrarily large, independent of its mass, e.g. by taking $\vec r$ large in the angular momentum $\vec L = \vec r \times \vec p$.  
These orbital angular momenta modes, which are BPS lead to violations of the bound $M^3>J^2$.  Even though one may think this leads to a macroscopically
observable violation of the cosmic censorship bound, because there are as many of them as the number
of BPS black holes, we will show momentarily that they are always subleading
contributions to the index due to cancellations in the computation of the index.

The expression for topological string partition function can be written in a more illuminating way as
$$Z=sdet(1-y^{2J^3_L}{\rm exp}(-M))\ ,$$
where $sdet$ denote the super-determinant (taking into account fermions versus bosons) over all BPS states and $M=t \cdot Q$,
including the internal and orbital contribution to $J^3_L$ states.  From this form, it is clear that this is a second quantized
partition function and to get to individual black holes, we need to consider the free energy $F$
\be \label{defF}
F=-{\rm log}Z=- {\rm str}\big[{\rm log}(1-y^{2J^3_L}\ {\rm exp}(-M))\big]=\sum_{k>0}{1\over k}{\rm str}\big[\ y^{2kJ^3_L}\ {\rm exp}(-kM)\big]\ .
\ee
From this one can extract a more natural object for black hole entropy index which is simply the $k=1$ term
above, ${\tilde F}={\rm str}\big[\ y^{2J^3_L}\ {\rm exp}(-M)\big]$.  The distinction between $F$ and ${\tilde F}$
becomes relevant in subleading terms, as the leading term for a fixed charge comes from the $k=1$ term.
In the context of an elliptic Calabi-Yau, one of the $t$'s is the K\"ahler structure of the elliptic fiber 
given by $\tau$.

It is well known, using the $S_g$ basis, that $F$ can be expanded  as \cite{Gopakumar:1998ii,Gopakumar:1998jq}:
$$F=\sum_{Q,k,g\geq 0}^{\infty} N^Q_g\  {1\over k}{\rm exp}(-ktQ) \cdot [2\sin (k\pi \lambda)]^{2g-2}\ ,$$
where the $[2\sin (k\pi \lambda)]^{2g}$ factor comes from the trace over $S_g$ and the $[2\sin (k\pi \lambda)]^{-2}$
comes from the orbital contributions:
$$\sum_{n>0} ny^{kn}={y^k\over (1-y^k)^2}={-1\over [2\sin (k\pi \lambda)]^2}.$$
Note that $F$ has a second order pole in $\lambda$ coming from $N^Q_{g=0}$ BPS states.  
As just discussed, the divergence in $\lambda$ can be traced to the orbital angular momenta of individual BPS states. 
 In particular
this divergence is a signature that arbitrarily large angular momenta contribute for a fixed $Q$, which leads
to a maximal violation of cosmic censorship bound coming from arbitrarily large $J$ values.  Note that the states contributing
to $N^Q_{g>1}$ {\it do not} contribute to this pole.  The reason is that, even though these
states also have arbitrary angular momentum states, their contribution to the {\it index} cancels between
adjacent spins for large angular momenta. This was already noted in \cite{Vafa:1997gr}  and amounts to a cancellation between states with internal and orbital angular momentum. To illustrate this phenomenon consider the following spin-content and index contribution of two BPS states respectively:
\begin{equation} \label{eq:microexa}
	\left[\left(\frac{1}{2},0\right)\oplus \left(0,\frac{1}{2}\right)\right] \rightarrow \frac{y + y^{-1}}{(y^{\frac{1}{2}} - y^{-\frac{1}{2}})^2} -\frac{2}{(y^{\frac{1}{2}} - y^{-\frac{1}{2}})^2} =\frac{ (y^{\frac{1}{2}}-y^{\frac{1}{2}})^2}{(y^{\frac{1}{2}} - y^{-\frac{1}{2}})^2} = 1,
\end{equation}
where we have used that the contribution of the orbital angular momenta to the index is of the form
\begin{equation} \label{eq:microexb}
	\frac{y}{(1-y)^2} = \frac{1}{(y^{\frac{1}{2}} - y^{-\frac{1}{2}})^2}.
\end{equation}
The two contributions each give rise to an infinite set of states to the entropy but we see that there is a perfect cancellation and the divergence at $y=1$ or $\lambda = 0$ vanishes in the index. The only surviving divergences thus come from the $g=0$ sector of the topological string. It is well known that $N^Q_{g=0}$ grows as $\sim {\rm exp}(a Q)$ for large $Q$ so this growth is mild compared to the growth of BPS black holes which scales as $exp(a Q^{3/2})$. We will discuss the macroscopic implications of this in section 6.

For the case of elliptic CY, one of the $t$'s can be replaced by $\tau$. Moreover, as we have discussed before,
the topological string partition function becomes a meromorphic Jacobi form for a fixed charge sector $Q$.
Moreover, as we discuss in section 4.2 and 4.3, we find explicitly that all contributions with arbitrarily large angular momentum come from the polar term and that there are no additional contributions to it.
In particular they arise from genus 0 holomorphic curves in the base $B$ attached to degenerate elliptic
curves.  Let $C$ denote the class of the curve in the base.  The number
of points over $C$ where the elliptic fiber degenerates is $12c_1(B)\cdot C$.
 If we have $C$ attached to $n$ elliptic curves, since we can attach $n_i$ curves over the
$i$-th degenerate elliptic curve, we can deduce that the combinatorics of such choices goes as the $q^n$ coefficient of
$1/\eta(q)^{12c_1\cdot C}$ (with further subleading contributions coming from the submoduli of $C$ in $B$ which fixes the genus to be 0).
Indeed we find this confirmed in the example of elliptic curves over ${\mathbb P}^2$ later in the paper.

In the case of CY being $K3\times T^2$, we can define a modified version of the topological string \cite{Katz:1999xq}
(taking into account extra supersymmetry in this case) which counts the BPS states of M2 branes
wrapped on a genus $g$ curve in $K3$ and bound to $n$ M2 branes in the elliptic fiber.  This is given by
the elliptic genus of $Sym^g(K3)\times \mathbb{R}^4$.  Note that in this case the elliptic fiber is constant
and does not degenerate, and so we cannot form a genus 0 curve by attaching any number of elliptic curves.
This translates to the statement that with KK momenta included there cannot be any states with arbitrarily large angular momenta. Note that at zero momentum there still can be states with arbitrarily large angular momentum coming from curves of genus $g$  in $K3$ which degenerate to spheres that are not bound to $T^2$ and whose growth go as the Euler characteristic of $Sym^g(K^3)$.  This can be viewed as a special case of the above construction if we view this as an F-theory model with base
$B=T^2\times P^1$ where we have exchanged the $T^2$ of the F-theory with elliptic fiber description of $K3$
over $P^1$.  The genus 0 curves arise from attaching the genus 0 curve wrapped around $P^1$ to any of the 24 degenerate elliptic fibers
over the $P^1$ leading to 
the generating function 
$1/\eta^{24}$.

In the next section we review the $K3$ case from the viewpoint of meromorphic Jacobi form, which
confirms the above picture, and at the same time leads to the natural setting to discuss the case of elliptic CY 3-folds.

\subsection{K3 compactifications \label{sec:K3}}

We begin with the five-dimensional theory with 16 supersymmetries obtained by compactifying Type II string theory 
on~$K3 \times S^{1}$. Consider the D1-D5 brane system wrapping the $K3$ with charge~$Q_{1} Q_{5} \equiv g$, or 
equivalently, D3 branes wrapping a curve of genus $g$ in $K3$---thus giving us a string in the effective six-dimensional theory. 
This string, wrapped on the~$S^{1}$, with~$n$ units of momentum around it, and with left-moving spacetime spin~$J_{L}=r$, is a $\frac14$-BPS configuration.
The worldvolume theory of such strings has $\mathcal{N}=(4,4)$ superconformal symmetry, 
and~$n$ and~$r$ are the eigenvalues of the left-moving~$L_{0}$ 
and~$J^{3}_{0}$, respectively. 

\begin{figure}[here!]
  \centering
	\includegraphics[width=\textwidth]{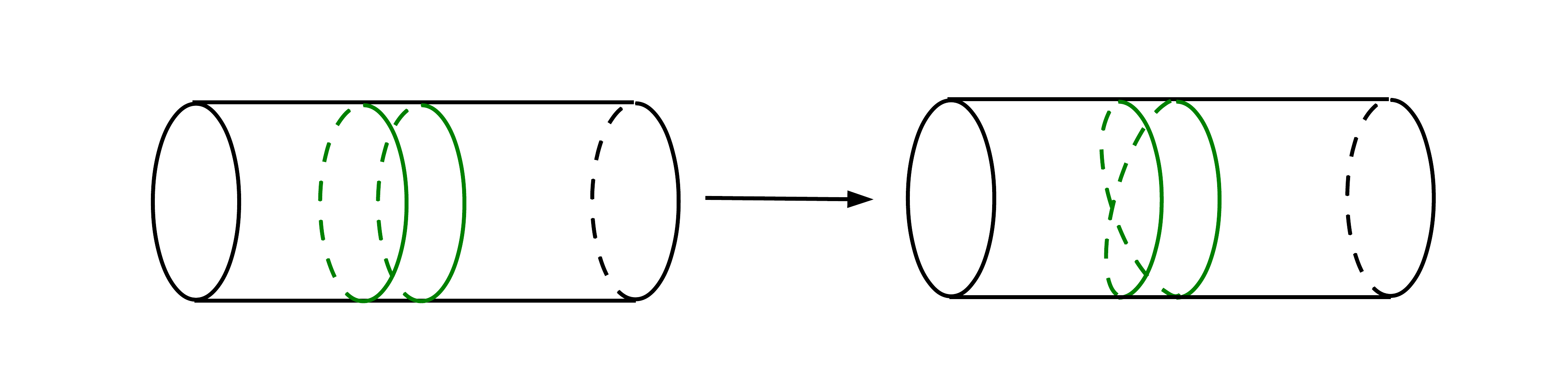}
  \caption{D3-branes wrapping curves in K3 give rise to strings in 6d. Upon wrapping a circle in a compactification to five dimensions several strings can join to form a multi-wound string.}
  \label{fig:multiwound}
\end{figure}

The partition function of this system factors into two pieces---one associated with the internal dynamics of 
the effective string that is captured by the symmetric product of the~$K3$ SCFT,  
and the other associated with the center of mass modes of the string and including all its 
oscillators. 
The latter piece is simply
the~$\IR^{4}$ SCFT whose elliptic genus is the 
function~$1/\varphi_{-2,1}(\t,\l)=\eta(\t)^{6}/\vth_{1}(\t,\l)^{2}$ that we already encountered in Equation~\eqref{Z1CM}. 
For the former piece, it is convenient to consider a grand canonical ensemble allowing for arbitrary charge~$g$,
whose partition function is~\cite{Dijkgraaf:1996xw}:
\be\label{defZ5d}  
\widehat Z(\sigma,\t,\l) := \sum_{g=0}^\infty \chi(\textrm{Sym}^{g}(K3); \t,\l)\,e^{2 \pi i g \sigma} \, 
\ee
where $\chi$ here denotes the Elliptic genus.
The states of this system can be divided into sectors 
made up of multi-wound strings, depicted pictorially in Figure \ref{fig:multiwound}, and multi-particle 
states thereof. 
The grand canonical partition function of the~$\frac14$ BPS states can thus be written as\footnote{There can also be additional 
discrete charge invariants that the brane configuration carries. The modular structure of the 
general  BPS partition function for arbitrary charge configurations~\cite{Banerjee:2008pu, Dabholkar:2008zy}  
is very close to the simplest case and we shall only discuss that here.} the product of two factors:
\be\label{Z5d}  
Z^{5d}(\sigma,\t,\l) \= \frac{\widehat Z(\sigma,\t,\l)}{\varphi_{-2,1}(\t,\l)} \, ,
\ee
where~$(\sigma, \t,\l)$ are the chemical potentials for~$(g,n,r)$.

The  fixed-$g$ partition function~$\psi_{g-1}(\t,\l)$, obtained by 
expanding the partition function as
\be\label{reciproigusa} 
Z^{5d}(\sigma,\t,\l) \= \sum_{g=0}^{\infty} \psi_{g-1} (\t,\l) \, e^{2 \pi i g \sigma}  \, ,
\ee
is a Jacobi form of weight~$w=2$ and index~$k=g-1$. 
The BPS index for given charges~$(n,r)$ is the coefficient~$c_{g-1}(n,r)$ in the Fourier expansion 
(with~$q=e^{2 \pi i \tau}$ and~$y=e^{2\pi i \l}$):
\be \label{Jac}
\psi_{g-1}(\t,\l)  \= \sum_{n,r} c_{g-1}(n,r) \, q^n \, y^r \, .
\ee 
Since the supersymmetric index does not change under continuous 
changes of the moduli, one expects that 
there is a \emph{macroscopic} interpretation of the numbers~$c_{k}(n,r)$ as the total indexed degeneracies of all 
gravitational configurations that exist for a given value of charges~$(g,n,r)$.

An important question that immediately arises is: which gravitational configurations exist for a given set of 
charges? Let us focus on a fixed~$k=g-1$, and consider states with~$n>0$. 
When $\Delta \equiv 4kn-r^{2}>0$ 
the gravitational ensemble is dominated\footnote{Classical gravity only applies, of course, for~$\Delta \gg 0$, but 
we could allow for a conservative treatment of quantum gravity where there is a black hole for every value of 
positive~$\Delta$ \cite{Sen:2010mz, Bringmann:2012zr}. } 
by a single black hole with degeneracy~$c_{g}(n,r)$.  
On the other hand, when $\Delta$ is negative there is no smooth black hole horizon in accordance with the cosmic censorship bound. 
States with negative~$\Delta$ can be generated as follows. 
The four bosonic fields of the stringy center of mass carry orbital angular momentum,
and their zero modes can be excited so as to make~$r$ large at no cost in energy. 
The energy~$n$ can be put into all the oscillator modes of the~$\IR^{4}$ SCFT (with no net angular momentum), 
which have the usual exponential growth of states governed by~$c=6$. 
A similar argument would say that the growth of states of the microscopic partition function~$\psi_{g-1}$ would 
be governed by~$c_L=6g+6$, the two terms being associated to the two SCFTs appearing in~\eqref{Z5d}. 

This, however, is not the case. As was explained in~\cite{Dabholkar:2010rm}, the function~$\psi_{g-1}$
which is really a supersymmetric \emph{index} has a growth of states governed by~$c=6k=6g-6$. The states 
of the~$\IR^{4}$ SCFT with~$c=6$ cancel an equal contribution from the internal degrees of freedom of 
the~$K3$ symmetric product SCFT. 
These two sets of cancelling modes both live in the region \emph{exterior} to the black hole horizon, so that 
in fact the growth of the microscopic index-degeneracy agrees precisely with the index-degeneracy of the 
the gravitational modes \emph{inside} the horizon of the black hole~\cite{Castro:2008ys, Dabholkar:2010rm}. 
In black holes with at least four supercharges, one can further clarify this from the gravitational theory 
using the near horizon~$AdS_{2}$. Considerations of the $AdS_2$ ensemble suggest that states that contribute to the black hole index-degeneracy actually have the same value of fermion number, and therefore this suggests that by the time we get to strong coupling, the unpaired states of the microcanonical ensemble with different fermion numbers have paired up~\cite{Sen:2009vz, Dabholkar:2010rm}\footnote{This index  in the gravitational theory may not always be easy to compute, as 
one needs to know which modes live inside or outside the horizon~\cite{Banerjee:2009uk,Jatkar:2009yd}.}.
We shall perform a similar analysis for the F-theory compactifications in  Section \ref{macroscopics}. 

\vspace{0.2cm}

The above conclusion can be understood directly and at a detailed level in the microscopic theory by using the 
analysis of meromorphic Jacobi forms~\cite{Dabholkar:2012nd}. 
The partition function~$\psi_{k}(\tau,\l)$ contains states whose spin is bounded above by the mass,
as well as states having arbitrarily large spin for a given mass. The microstates of the black hole (with~$\D>0$) 
are associated with the former type, while the latter states (with~$\D <0$) violate the cosmic censorship bound infinitely. 
The main decomposition theorem of~\cite{Dabholkar:2012nd}
allows us to separate these two types of states in an elegant manner into the \emph{finite part} and \emph{polar part}
of the meromorphic Jacobi form, as we now describe using the function~$\psi_{k}$ as a prototype example.

We recall that the two defining transformation properties of a Jacobi form~$\varphi_{w,k}$ of weight~$w$ and 
index~$k$ \footnote{Our conventions 
differ from the mathematics literature, in which~$(k,m)$ is usually used instead of~$(w,k)$. Our notation
follows the physics literature in which $k$ usually refers to level, and coincides with the index of the Jacobi form.} 
are the modular transformations
\begin{equation} \label{modular1}
	\varphi_{w,k}\left(\frac{a\tau+b}{c\tau+d},\frac{\lambda}{c\tau+d}\right)=(c\tau+d)^w e^{\frac{2\pi i k c \lambda^2}{c\tau+d}} \varphi_{w,k}(\tau,\lambda) \quad \forall \left(\begin{array}{cc}a & b\\c & d\end{array}\right) \in SL(2,\mathbb{Z}) \, , 
\end{equation}
and the elliptic transformations
 \be\label{elliptic1}  \varphi(\t, \l+a\tau+b)\= e^{-2\pi i k(a^2 \t + 2 a \l)} \, \varphi(\t, \l)
  \qquad \forall \quad a,\,b \in \Z \, . \ee
The latter transformation is simply a statement 
of the fact that the complex parameter~$\l$ actually naturally lives on the torus~$\IC/(\Z\tau+\IZ)$. 
The main idea of the decomposition theorem is to first build a function~$\psi_{k}^{\rm P}(\t,\l)$, 
called the \emph{polar part} that has the same poles and residues as well as the same elliptic transformation properties 
as~$\psi_{k}(\tau,\l)$. 
The basic building blocks of such functions are called \emph{Appell-Lerch sums}. In our case here, the polar part is:
  \be \label{simfun} \psi_{k}^{\rm P}(\t,\l) \= D(\tau) \; \CA_{2,k}(\t,\l) \, ,  \ee
where the Appell-Lerch sum~$\CA_{2,k}$ is given by:
  \be \label{defBP} \CA_{2,k}(\t,\l)\= \sum_{s\in\Z} \, \frac{q^{ks^2 +s} \, y^{2ks+1}}{(1 -q^s y)^2} \, , \ee
and the function~$D(\tau)$ is the the Laurent coefficient of~$\psi_{k}(\t,\l)$ at~$\l=0$. 
In the present case, we see from 
Equations~\eqref{Z5d}, \eqref{defZ5d} that~$D(\t)$ is simply the constant~$p_{24}(k+1)$, the number of partitions 
of~$k+1$ into integers of 24 different colors.

The Appell-Lerch sum has the Fourier expansion 
\be \label{FexpAL}
\CA_{2,k}(\t,\l) \=  \biggl(\sum_{s \ge 0}  \sideset{}{^{*}} \sum_{\ell \ge 0} \, - \, \sum_{s < 0}  \sideset{}{^{*}} 
\sum_{\ell \le 0} \biggr) \, \ell \, q^{ks^2 + \ell s} \, y^{2ks+\ell} \, , 
\ee
where the asterisk on the summation sign means that one should count the term $\ell=0$ with multiplicity $1/2$.
This formula manifestly shows that each term in the Appell-Lerch sum has~$4kn-r^{2}=-\ell^{2}$
for some~$\ell$, so that it is always non-positive. 
Further, we see that its Fourier coefficients grow extremely slowly (linearly in~$\ell$). 
In particular, we see that the growth of states in the polar part is given, up to polynomial pre-factors that we will ignore 
from now, by~$p_{24}(k+1) \sim e^{4 \pi\sqrt{k}}$ for large~$k$, as consistent with the discussion in Section \ref{toplspinBH} 
(recall that $k\sim Q^{2}$). 
We also see that, for fixed~$k$, there is no exponential growth of states as a function of~$n$, as consistent with the macroscopic 
index vs degeneracy analysis above. 

Having separated the states living outside the horizon, we are left with the finite part, which contains the 
black hole degeneracies\footnote{The focus of the physics 
in~\cite{Dabholkar:2012nd} was four-dimensional $\frac14$-BPS black holes wherein there are discontinuous 
jumps at special values of moduli space. In that case, the black hole degeneracies, as defined by the attractor contour, 
agreed precisely with the degeneracies of the finite part~$\psi^\text{F}$. 
Further, the polar part had an additional~$1/\eta(\t)^{24}$ multiplying it, and it had an interpretation as bound states 
of two~$\frac12$-BPS small black holes. 
Combined with the argument~\cite{Dabholkar:2009dq} that the only gravitational configurations that contribute 
to theories with~$\CN = 4$ supersymmetry in four-dimensions are either single-centered $\frac14$-BPS black 
holes or two-centered bound states of $\frac12$-BPS black holes, this furnished a complete physics 
understanding in the~$\CN=4$ theory.}.
One thus has the \emph{additive} decomposition of the partition function~$\psi_{k}(\t,\l)$:
\be \label{decomppsi} 
\psi_{k}(\tau,\l) \; = \;  \psi_{k}^\text{F}(\tau,\l) \, + \, \psi^\text{P}_{k}(\tau,\l) \, .
\ee
These two pieces are not modular by themselves---one has separated out a part of the spectrum of the 
original theory and therefore it is not surprising that one breaks the original modular symmetry. 
The non-trivial fact is that they have \emph{mock} modular properties---which means 
that one can \emph{complete} these two functions by adding a~$\overline{\tau}$-dependent function 
(which depends on another function called the \emph{shadow})
to the first piece (and subtracting it from the second):
\be \label{decomppsi2} 
\psi_{k}(\tau,\l) \; = \;  \widehat{\psi^\text{F}}_{k}(\tau,\l) \, + \, \widehat{\psi^\text{P}}_{k}(\tau,\l) \, ,
\ee
so that the completed functions~$\widehat{\psi^\text{F}}_{k}$ and~$\widehat{\psi^\text{P}}_{k}$
are modular (but not holomorphic). 
This phenomenon is similar to the holomorphic anomaly and has been linked to 
the non-compactness of the target space of the SCFT~\cite{Dabholkar:2012nd, Murthy:2013mya}. 

The most important part of the decomposition theorem is that finite piece~$\psi^\text{F}_{k}$ is a \emph{mixed mock Jacobi form} 
with shadow~$D(\t) \, \sum_{\ell \in \IZ/2k\IZ} \overline{\vartheta_{k,\ell}(\t,0)} \, \vartheta_{k,\ell}(\t,z)$.
We explain the meaning of this statement in some detail in the appendix, and here we note that it has the following important 
consequences. Firstly (as is obvious from its construction), $\psi^\text{F}_{k}$ is holomorphic in~$\l$ and obeys the elliptic 
transformation property~\eqref{elliptic1}. This implies that its Fourier coefficients defined 
by~$\psi_{k}^\text{F}(\t,\l)=\sum_{n,r} c^\text{F}_{k}(n,r) \, q^{n} \, y^{r}$ are only non-zero for states that obey 
the bound\footnote{The function~$\psi^\text{F}_{k}$ also contain a finite number of states with~$4kn-r^{2}<0$, 
as for any weak Jacobi form that is holomorphic in~$z$---these are also often called polar states in the literature (as they correspond 
to poles in~$q$). In contrast, the polar states discussed in this section correspond to poles in~$z$, and for each given mass, 
there are infinitely many such states that violate the cosmic censorship bound.}~$4kn-r^{2}\ge -\mu^{2}$ 
(with $\mu \equiv r$ mod~$2k$), which is simply the statement that~$L_{0}>J_{L}^{2}/4k$, up to a shift of the zero-point energy. 
Secondly, the correction term that we add to~$\psi^{F}$ is small---in the sense that, as far as the asymptotic growth of 
the Fourier coefficients are concerned,~$\psi^{F}$ behaves like a regular Jacobi form! Therefore the degeneracy of black hole states grow 
as~$c^\text{F}_{k}(n,r) \sim \exp \bigl(\pi \sqrt{4kn-r^{2}} \bigr)$ as~$\sqrt{4kn-r^{2}} \to \infty$.

We note that the polar states in the K3 compactification are~$\frac12$-BPS states dual to Dabholkar-Harvey states 
having orbital angular momentum. Thus our decomposition can also be viewed as separating out objects with enhanced 
supersymmetry\footnote{Such a separation was noted in~\cite{Banerjee:2009uk, Dabholkar:2010rm}, in which the states
were called~$Z^\text{extra}$.}. 
In F-theory, on the other hand, we only have $\frac12$-BPS black holes, and the separation into black hole states 
and CCB violating states takes on a more physical meaning independent of supersymmetry.

\subsection{F-theory compactifications}
\label{sec:Ftheory}

We now turn to our F-theory compactifications of Section \ref{sec:D3}, which preserve~8 supersymmetries. 
As in the K3 case, the presence of meromorphic Jacobi forms provides us a powerful handle on the analysis. 
It allows to extract the growth of black hole degeneracies easily, and also refine the topological string 
estimate for states with very large orbital angular momentum presented in Section \ref{toplspinBH}. 

The corresponding strings come from D3 branes wrapping a curve $C$ inside the base $B$ of an elliptic Calabi-Yau 
and have $\mathcal{N}=(0,4)$ worldvolume supersymmetry.
In Section \ref{sec:D3} we presented the elliptic genus of these strings:
\be
Z_{C}(\t,\l) \={N_C(\tau,\lambda)\over \eta(\tau)^{12c_1(B)\cdot C}\prod_{s=1}^{d} \varphi_{-2,1}(\tau, s\lambda)} \, , 
\ee
where to simplify the discussion here we are assuming that there is only one class $C$ with degree $d$.  This
is a meromorphic Jacobi form of weight~$0$ and index~$k=g-1$, with~$g= \frac{1}{2}(C\cdot C - c_1(B)\cdot C) + 1$. 
In contrast to the K3 case, the function~$Z_{C}$ has poles of order~$2d$ at~$\l=0$, as well as lower order poles at
torsion points~$\alpha \tau + \beta$, for some $\alpha, \beta \in \mathbb{Q}$. This makes the analysis slightly more complicated,
but as we shall see, the main conclusions are similar. 

We begin by decomposing the function~$Z_{C}(\t,\l)$ into finite and polar parts as before:
\be 
	Z_{C}(\t,\l) \= Z_{C}^\text{F}(\t,\l) + Z_{C}^\text{P}(\t,\l) \, . 
\ee
The finite part~$Z_{C}^\text{F}(\t,\l)$ is essentially a mixed mock Jacobi form\footnote{The higher order poles implies that 
these Jacobi forms are defined over the ring of quasi modular forms instead of modular forms, 
see \S 9.6 of~\cite{Dabholkar:2012nd} for details.}, 
\be 
	Z_{C}^\text{F}(\t,\l) \= \sum_{n,\ell} \, c^\text{F}(n,\ell) \, q^n \, y^{\ell} \, , 
\ee
to which we can add an explicit correction term to get its modular completion~$\widehat{Z_{C}}(\t,\l)$ that transforms as a 
holomorphic Jacobi form, but suffers a non-holomorphic anomaly. The microstates of spinning black holes are contained 
in the function~$Z_{C}$, and its Fourier coefficients have the growth
\begin{equation}\label{BHentropy}
	c^\text{F}(n,\ell) \sim \exp\bigl({2\pi \sqrt{ (k + 2 c_1(B) \cdot C)(n- \ell^{2}/4k)}} \bigr)\, , \quad \textrm{as} \quad n- \ell^{2}/4k \gg 0,
\end{equation}
We thus find a simple entropy formula for the spinning black hole, that is controlled by a combination of~$6k$ from the level of the 
current algebra and the~$12 c_1(B) \cdot C$ free bosons.

Now we come to the polar part $Z_{C}^\text{P}(\t,\l)$, which is a sum over 
all the poles of the function~$Z_{C}$ (with multiple poles counted separately):
\be \label{ZCPolar} 
Z_{C}^{\rm P}(\t,\l) \= \sum_{p\in \text{Poles}(Z_{C})} D_{p}(\tau) \; \CA^{p}_{k}(\t,\l) \, . 
\ee
The Appell-Lerch sums~$\CA^{p}_{k}$ are again simple functions that have the same poles and have the 
same elliptic transformation properties as the meromorphic Jacobi forms (see~\cite{BringFol, Olivetto} for a detailed analysis
of the higher pole cases).
As before, these functions have the two important properties mentioned after~\eqref{simfun}---they capture states with 
arbitrarily high spin for any~$L_{0}$, and that their Fourier coefficients have extremely small growth. 
All the growth of the polar degeneracies thus comes from the Laurent coefficients~$D_{p}(\t)$. Recall that in the K3 
case, the only Laurent coefficient was at~$\l=0$, and was a constant. Here in the F-theory situation, the coefficients 
are modular forms that have their own growth. 

We now discuss the pole at~$\l=0$ in some detail, and postpone a discussion of the other poles to Section \ref{ObsBnd}. 
Near~$\l=0$, the function~$Z_{C}$ has the expansion:
\be
Z_{C}(\t,\l) = {1 \over \eta(\tau)^{12c_1(B)\cdot C}} \, \frac{1}{(2\pi i \l)^{2d}} \,
\bigl(a_{0}(\tau) + a_{2}(\t)\, (2 \pi i \l)^{2} + a_{4}(\t)\, (2 \pi i \l)^{4} + \cdots \bigr) 
\ee
The various Laurent coefficients~$a_{2i}$ are found by computing the Taylor expansion of the functions~$N_{C}(\t,\l)$ 
and~$\prod_{s=1}^{d} \varphi_{-2,1}(\tau, s\lambda)$ at~$\l=0$, and taking a ratio. We know that~$N_{C}$ is a weak Jacobi form of
positive weight and even non-negative index~implying that it has an expansion of the 
form~$N_{C}(\t,\l) = \sum_{n \ge 0} \, (2\pi i \l)^{2n} \, b_{2n}(\tau)$, where~$b_{2n}(\t)$ are quasi-modular forms~\cite{Eichler:1985ja}. 
The denominator function~$\prod_{s=1}^{d} \varphi_{-2,1}(\tau, s\lambda)$ is known explicitly. In particular, it has an 
expansion of the form~$(2\pi i \l)^{2d} \, \sum_{n \ge 0} \, (2\pi i \l)^{2n} \, c_{2n}(\tau)$, where~$c_{2n}$ are quasi-modular forms  
of weight~$2n$ (so that~$c_{0}$ is a constant). 
Putting these facts together we deduce that the Laurent coefficients~$a_{2n}$ are quasi-modular forms, and 
consequently\footnote{This can be seen explicitly by the fact that the ring of 
quasi-modular forms is generated by the functions~$E_{2}(\t)$, $E_{4}(\t)$, $E_{6}(\t)$.} 
do not have any exponential growth of states. Therefore the growth of any of the Laurent coefficients of~$Z_{C}$ at~$\l=0$ 
is governed by the central charge~$12c_1(B)\cdot C$. 

It is important to note that the elliptic genera $Z_C(\tau,\lambda)$ do not count degeneracies only of single particles,
but also include degeneracies of multi-particle states. In order to count only single-particle states, one has to consider the 
\textit{free energy}~$F_{C}(\tau,\l)$ given by the logarithm of the sum of topological string partition functions:
\begin{equation}
	\log\Bigl(\sum_C Z_C(\tau,\lambda) \, e^{-\vec{t} \, \cdot \, C}\Bigr) \= \sum_C F_C(\tau,\lambda) \, e^{-\vec{t} \, \cdot \, C}.
\end{equation}
On inverting this, we find an expression for the free energy~$F_{C}$ as a linear combination of products~$\prod_{i} \, Z_{C_{i}}$, 
with~$C=\sum_{i} C_{i}$. The free energy~$F_{C}(\t,\l)$ only has a second order pole at~$\l=0$, but it is no longer a Jacobi form 
as it is formed by adding Jacobi forms of different indices. We can, however, still deduce the growth of degeneracies. 
For large charges, the index of~$Z_{C}$ grows quadratically in the charges~$C$,
and therefore the leading growth of states of~$F_{C}$ with large charges for a fixed spin is the same as that of~$Z_{C}$. 
Thus the black hole entropy formula~\eqref{BHentropy} still holds.

For the polar states at~$\l=0$, we note that the growth of the Laurent coefficient of~$Z_{C}$ is linear in~$C$, and therefore the 
products~$\prod_{i} \, Z_{C_{i}}$ in~$F_{C}$ all have the same growth. We thus obtain that, generically, the growth 
of polar states of~$F_{C}$ is controlled by the central charge~$12c_1(B)\cdot C$. 
In principle there could be cancellations which reduce the growth, in which case this is only an upper bound. 
On the other hand, we saw from general considerations from the topological string in Section \ref{toplspinBH} 
that the central charge must be linear in charges.  The analysis in this section is consistent with this expectation.

We end this section by summarizing the main degeneracy formulas:
\begin{itemize}
\item \emph{Spinning black hole}: The Fourier coefficients of the 
finite part of the index~$Z_{C}^\text{F}$ (i.e. the index of spinning BPS states with $SU(2)_L$ spin $r$ and
charges $n,C$) have the asymptotic form (for~$k>0$):
\be \label{cFinite}
c^\text{F}(n,r) \sim \exp\Bigl(2 \pi \sqrt{\frac{c^\text{index}}{6}\bigl(n- r^{2}/4k \bigr)} \Bigr) \, , \qquad \text{as} \quad n -r^{2}/4k \to \infty \, , 
\ee
where 
\be \label{clpos}
c^\text{index} \=6k+12c_1(B)\cdot C \,  
\ee
and $k=g-1={1\over 2} (C\cdot C-c_1(B)\cdot C)$.
These coefficients~$c^\text{F}(n,r)$ are the degeneracies of the large spinning black hole.  
\item \emph{States with $J^2 > M^3$}: The index of the states with $r^2/4k>n$ for
large $n$ are to leading order independent of the spin $r$ and have the form:
\be
	c^\text{bound}(n,r) \sim  \exp\Bigl(2 \pi \sqrt{2c_1(B)\cdot C \, n} \Bigr) \, ,  \qquad \text{as} \quad n \to \infty \, .
\ee
\end{itemize}

\section{An Illustrative Example:  Black holes for M-theory on elliptic ${\mathbb P}^2$}  \label{sec:P2}
In this section we give an illustrative example of the main ideas discussed in previous sections. 
We consider M-theory on elliptic threefold fibered over ${\mathbb P}^2$ and consider the BPS states consisting of M2 branes wrapped
around 2-cycles.  As is well known, the topological string captures the BPS content of spinning black holes in 5d \cite{Gopakumar:1998ii,Gopakumar:1998jq,Katz:1999xq}.
So the question of microscopics accounting of the entropy amounts to whether one can compute topological string amplitudes.
Unfortunately a full expression is not available yet, but a conjectural structure of the result
for this case was suggested recently \cite{Huang:2015sta} for the $M2$ branes wrapped around an arbitrary class  but computed only for low degree (up to degree 5) in the base of ${\mathbb P}^2$.   These results were in turn motivated
from the associated string description in 6d \cite{Haghighat:2013gba,Haghighat:2014pva} which arises from the duality with F-theory
and is in agreement with the general structure we have found in the previous section.

\begin{figure}[here!]
  \centering
	\includegraphics[width=0.5\textwidth]{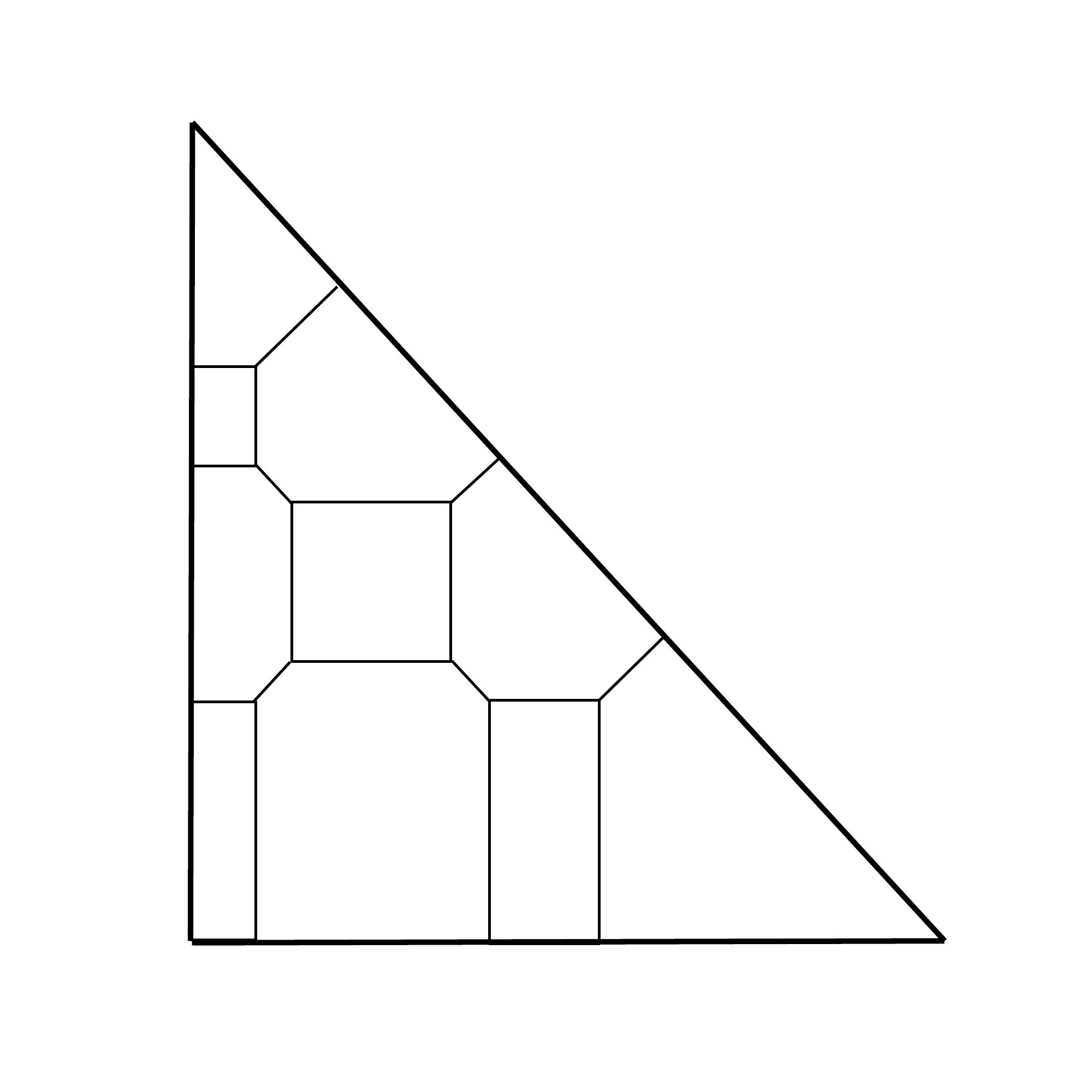}
  \caption{The toric skeleton of a degree $d=3$ hypersurface in $\mathbb{P}^2$. The number of lines ending on each edge indicates the degree of the curve in such a description. One can see that $d\geq 3$ curves necessarily form a genus $g \geq 1$ Riemann surface (in this case $g=1$ as one can see from the single hole in the middle of the graph).}
  \label{fig:toric3}
\end{figure}

Elliptic 3-fold over ${\mathbb P}^2$ has two Kahler classes:  The elliptic fiber, and the base, whose basic
integral generators we denote by $E$ and $H$.  We will
denote the class of the $M2$ brane as 
$$[M2]=n[E]+d[H]$$
From the viewpoint of the dual F-theory, this state arises by considering the base ${\mathbb P}^2$ and compactifying
on $S^1$ and wrapping
a D3 brane wrapped around $S^1\times C$ where $[C]=d[H]$ and we consider momentum $n$ along
the circle.  The generic $C$ in that class has a genus
$$g(C)={1\over 2}(d^2-3d) +1$$
In particular for $d=1,2$ we have $g=0$ and for $d\geq 3$  we get $g\geq 1$.
These curves can be visualized using the toric realization of ${\mathbb P}^2$, see Figure \ref{fig:toric3}.
As discussed in the previous section the general structure expected for the partition function in the sector with $[C]=d[H]$
leads to the elliptic genus
$$Z_{d}={N_d(\tau,\lambda)\over \eta(\tau)^{36d}\prod_{s=1}^{d} \varphi_{-2,1}(\tau,s\lambda)}$$
where $N_d(\tau,\lambda)$ is a weak Jacobi form of index $g(C)-1+d(d+1)(2d+1)/6$
and weight $16 d$. Motivated
from M-strings, this
structure was in fact conjectured for this case in \cite{Huang:2015sta}.   Moreover for $d< 6$ the actual $N_d$ was
determined.  

%
%
%

We can now flesh out the analysis of Section \ref{sec:Ftheory} in these explicit examples.
The functions~$Z_{d}(\t,z)$ have index~$k_\text{mero}=\frac12 \, d(d-3)$, 
and $12c_1(B)\cdot C = 36 d$ in this case, so that the Laurent coefficient at~$z=0$ is~$1/\eta(\t)^{36d}$. 
Our analysis in~Section \ref{sec:Ftheory} says that for large positive values of~$4kn-r^{2}$, the degeneracies of Fourier coefficients 
of~$Z_{d}$, which we interpret as the degeneracies of a single-centered black hole grows with a central charge
\be \label{clpos}
c^{index} \= 6 k_{mero} + 36 d = 3d^{2}+27d \, , \quad d \ge 4 \, . 
\ee
The special cases~$d=1,2$ with $k_\text{mero}<0$ and~$d=3$ with $k_\text{mero}=0$ 
can be treated using the theorems of~\cite{Bringmann:2015gla}, who extend the 
the analysis of~\cite{Dabholkar:2012nd} to the negative values of the index. 
In these cases, the finite part in the decomposition~\eqref{decomppsi} identically vanishes. 
This means that the degeneracies of~$Z_{d}$, $d =1,2,3$, are simply the Fourier coefficients of 
the polar parts, which are controlled by the Laurent coefficients at the poles which have a central charge of~$36d$. 
%

In addition, we find new bound states from the polar part which have a growth controlled by:
\be
c^{bound} = 36 d \, , \qquad d \ge 1 \, . 
\ee
 In the following we present a detailed analysis of the cases $d=1, \ldots, 4$.
We then make some comments based on the numerical study of the cases~$d \le 5$.

\subsection{d=1}
For this case the curve $C$ is a genus 0 curve.  The complex moduli of the base is $d=2$, and the moduli space is itself a ${\mathbb P}^2$.  The corresponding ${\widehat C}$ which is the total space of the elliptic fibration over the ${\mathbb P}^1$,
has cohomology elements:

\begin{center}
\begin{tabular}{|c|c|c|}
\hline
$2$ & $0$ & $1$ \\
\hline
$0$ & $30$ & $0$ \\
\hline
$1$ & $0$ & $2$\\
\hline
\end{tabular}
\end{center}
which leads to the statement that the target space involves a ${\mathbb P^2}$ which is common to the left and right
movers, and in addition we have a Narain torus of $(28,4)$ dimensions fibered over it.  In addition we have  4 center of mass bosons common to left- and right-movers.  Moreover, there are the same number of right-moving
fermions as right-moving bosons, leading to $(0,4)$ supersymmetry.
So altogether we have $36$ left-moving bosons $36=4_{\mathbb P^2}+28_{T^{28}_L}+4_{CM}$ and 12-right moving bosons $12=4_{{\mathbb P}^2}+4_{T^4_R}+4_{CM}$ and 12 right-moving fermions.  The right-movers can be viewed as moving in a target which is a $T^4$ fibration over ${\mathbb P^2}$ times
$\mathbb{R}^4$.  $SU(2)_L$ index for this theory is $-1$ which comes entirely from the center of mass.
The elliptic genus can now be read off from the BPS computations of topological strings in \cite{Huang:2015sta} to be
$$Z_1= -\frac{E_4(31 E_4^3+113 E_6^2)}{48 \eta(\tau)^{36} \varphi_{-2,1}(\tau,\lambda)}$$
It would be interesting to see if one can come up with the direct definition of this $(0,4)$ sigma model and compute the
above $Z_1$ directly.
This gives the $c_L=36$ with the expected growth $d_{1,n}\sim {\rm exp}(2\pi \sqrt{n\cdot 36/6})$.

\subsection{d=2}

\begin{figure}[here!]
  \centering
	\includegraphics[width=\textwidth]{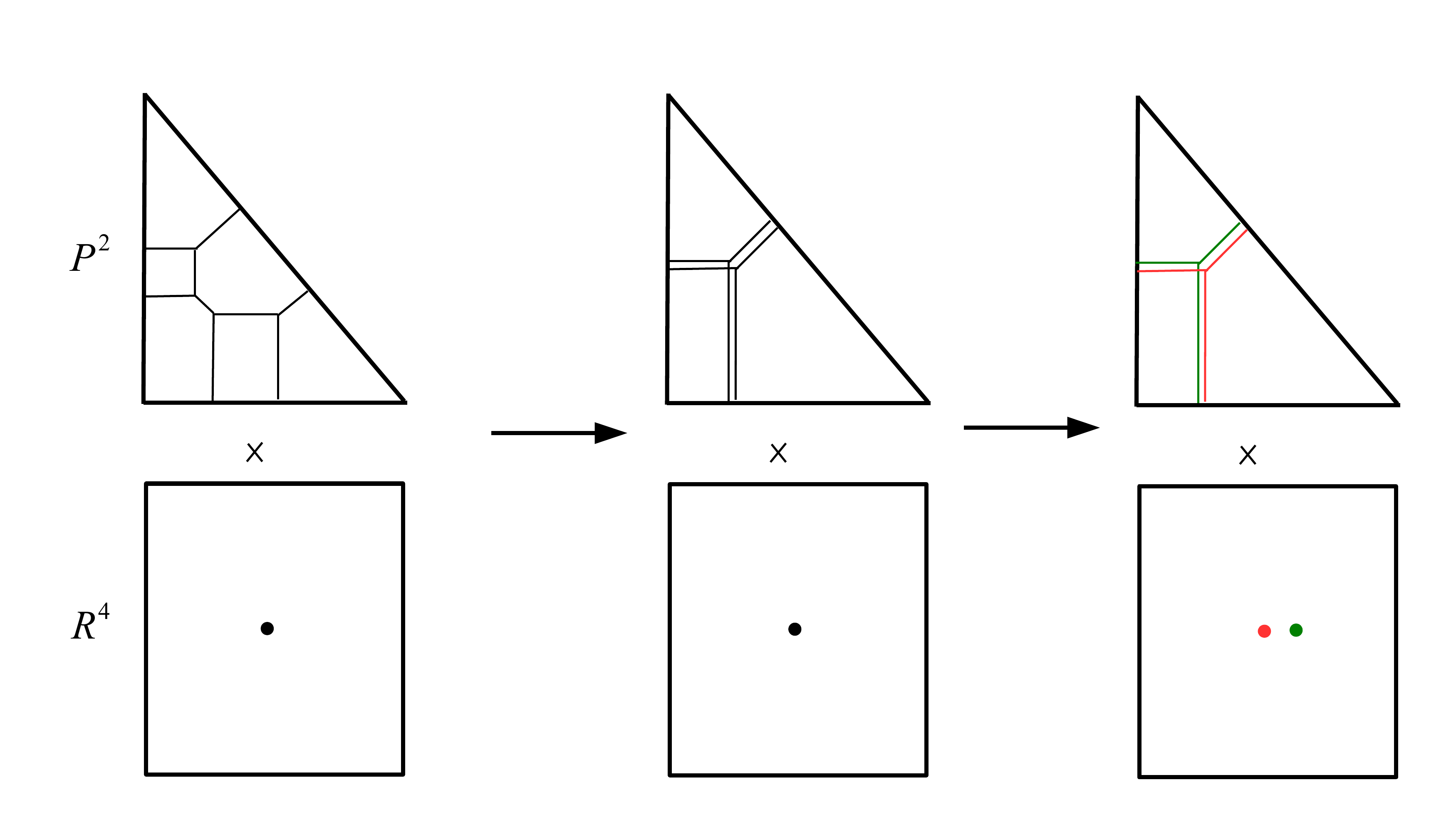}
  \caption{The toric skeleton of a degree $d=2$ hypersurface in $\mathbb{P}^2$ undergoing a phase transition. The left-most figure shows the generic case of a degree 2 curve. The central figure is a degeneration to 2 degree 1 curves, each wrapped by a D3 brane giving rise to strings which are depicted as points in $\mathbb{R}^4$. The last picture shows the Higgs branch where the two strings are separated in $\mathbb{R}^4$.}
  \label{fig:HiggsBranch}
\end{figure}

\begin{figure}[here!]
  \centering
	\includegraphics[width=0.5\textwidth]{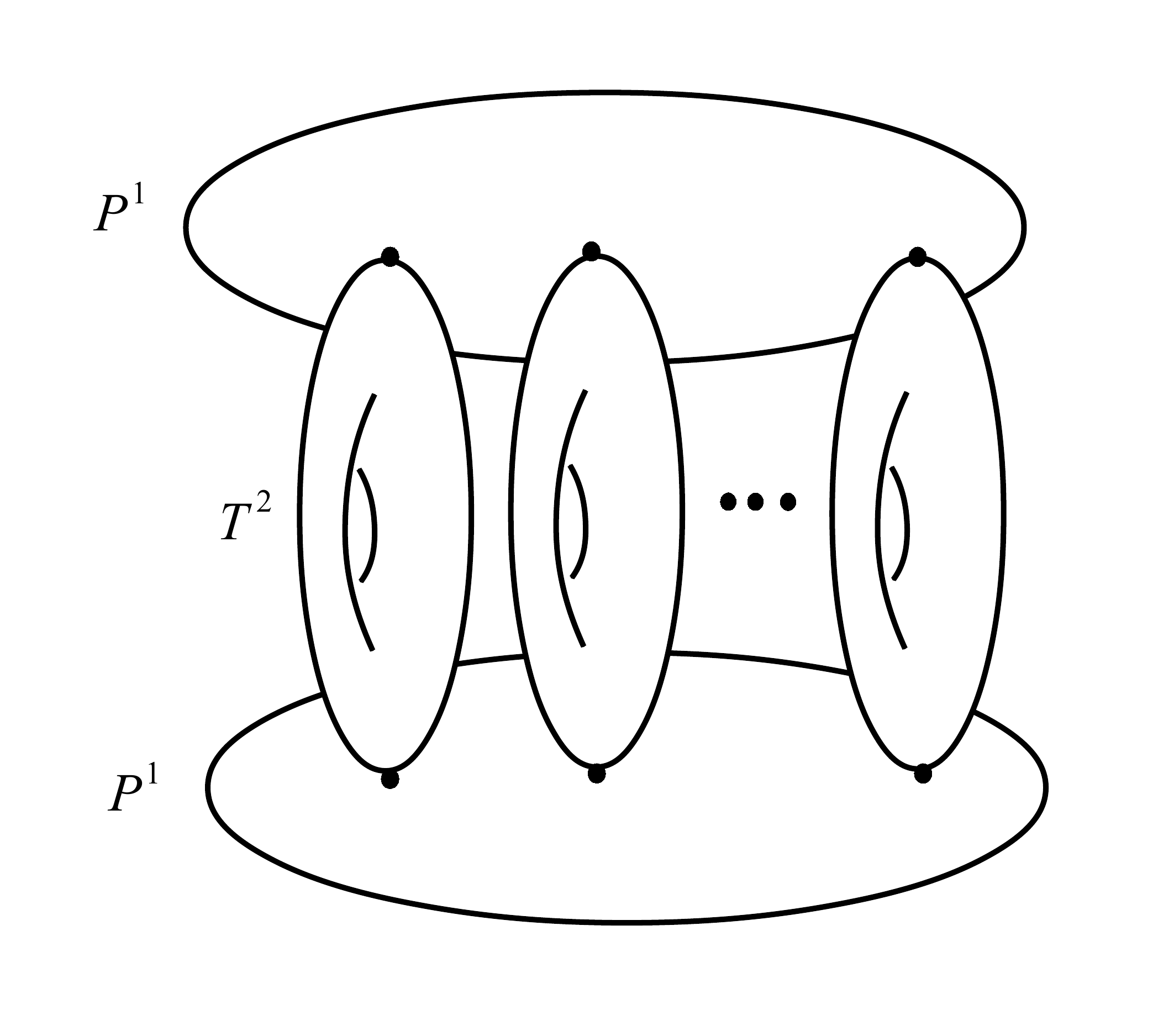}
  \caption{The 5d bound state structure as seen from the perspective of M-theory. Two $\mathbb{P}^1$'s are bound through $n$ $T^2$'s.}
  \label{fig:5dBoundState}
\end{figure}

This case is more interesting.  The generic branch involves a degree 2 curve in ${\mathbb P}^1$ which is again a ${\mathbb P}^1\subset {\mathbb P}^2$.
The moduli space for a degree 2 curve in ${\mathbb P^2}$ is 5 dimensional and is in fact a ${\mathbb P}^5$.  The cohomology of $\widehat C$ is given by
\begin{center}
\begin{tabular}{|c|c|c|}
\hline
$5$ & $0$ & $1$ \\
\hline
$0$ & $60$ & $0$ \\
\hline
$1$ & $0$ & $5$\\
\hline
\end{tabular}
\end{center}
which leads to the statement that the target space involves a ${\mathbb P^5}$ which is common to the left and right
movers, and in addition we have a Narain torus of $(58,10)$ dimensions fibered over it.  In addition we have 4 center of mass bosons common to left- and right-movers.  Furthermore, there are the same number of right-moving
fermions as right-moving bosons, leading to $(0,4)$ supersymmetry.
So altogether we have $72$ left-moving bosons $72=10_{\mathbb P^5}+58_{T^{58}_L}+4_{CM}$ and 24-right moving bosons $24=10_{{\mathbb P}^5}+10_{T^{10}_R}+4_{CM}$ and 24 right-moving fermions.  The right-movers can be viewed as moving in a target which is a $T^{10}$ fibration over ${\mathbb P^5}$ times
$\mathbb{R}^4$.  $SU(2)_L$ index for this theory is again $-1$ which comes entirely from the center of mass.  However, this is {\it not} the end of the story
for this case:  In a ${\mathbb P^2}\subset {\mathbb P}^5$ there is a singularity in this sigma model because the $D3$ branes coalesce, leading to a $U(2)$
gauge theory on the $D3$ branes, leading to a Higgs branch for this theory.  The central charge for this branch differs from
that of the Higgs branch as in the cases studied in \cite{Witten:1997yu} leading to the fact that Higgs and Coulomb branches become
disconnected in the IR.  
The partition function for the topological string for the $d=2$ case has been worked out in \cite{Huang:2015sta} leading to
$$Z_2=\frac{N_2}{\eta(\tau)^{72} \varphi_{-2,1}(\tau,\lambda)\varphi_{-2,1}(\tau,2 \lambda)},$$
where $N_2$ is given by
\begin{eqnarray}
	N_2 &  = & \frac{\varphi_{0,1}^4 E_4^2 (31 E_4^3 + 113 E_6^2)^2}{23887872} \nonumber \\
	~ & ~ & +\frac{1}{1146617856}\left(2507892 \varphi_{0,1}^3 \varphi_{-2,1} E_4^7 E_6 +9070872 \varphi_{0,1}^3 \varphi_{-2,1} E_4^4 E_6^3 \right. \nonumber \\
	~    & ~ & \left. + 2355828 \varphi_{0,1}^3 \varphi_{-2,1} E_4 E_6^5 + 36469 \varphi_{0,1}^2 \varphi_{-2,1}^2 E_4^9 + 764613 \varphi_{0,1}^2 \varphi_{-2,1}^2 E_4^6 E_6^2 \right. \nonumber \\
	~ & ~ & \left. - 823017 \varphi_{0,1}^2 \varphi_{-2,1}^2 E_4^3 E_6^4 + 21935 \varphi_{0,1}^2 \varphi_{-2,1}^2 E_6^6 - 9004644 \varphi_{0,1} \varphi_{-2,1}^3 E_4^8 E_6 \right. \nonumber \\
	~ & ~ & \left. -30250296 \varphi_{0,1} \varphi_{-2,1}^3 E_4^5 E_6^3 - 6530148 \varphi_{0,1} \varphi_{-2,1}^3 E_4^2 E_6^5 + 31 \varphi_{-2,1}^4 E_4^{10} \right. \nonumber \\
	~ & ~ & \left. 5986623 \varphi_{-2,1}^4 E_4^7 E_6^2 + 19960101 \varphi_{-2,1}^4 E_4^4 E_6^4 + 4908413 \varphi_{-2,1}^4 E_4 E_6^2 \right).
\end{eqnarray}
The growth of the BPS degeneracies is $d_{2,n}\sim {\rm exp}(2\pi \sqrt{n\cdot 72/6})$.  This expression includes
the contribution of all branches.  The most degenerate
branch is a 2-string branch and is obtained from the single string branch as is depicted in Figure \ref{fig:HiggsBranch}.
The existence of these extra BPS sates in 5d, viewed from the M-theory perspective, is described in \cite{Huang:2015sta} as coming from configurations of curves which project
to a degree 1 ${\mathbb P}^1 \subset {\mathbb P^2}$ but that are in turn connected with $n$ tori.  See Figure \ref{fig:5dBoundState}.
This in particular means that if we eliminated the tori, these BPS states would be disconnected and decay
to two BPS states.  From the viewpoint of one higher dimension, i.e. F-theory, the elliptic class translates
to the KK mode.  So this statement translates to having two strings each of which is wrapped on a circle
and are only bound because of the KK mode.  In other words the two strings together do not lead to
a bound state in 6d, but upon compactification they lead to bound states, which is reflected in the above BPS count.

\begin{figure}[here!]
  \centering
	\includegraphics[width=\textwidth]{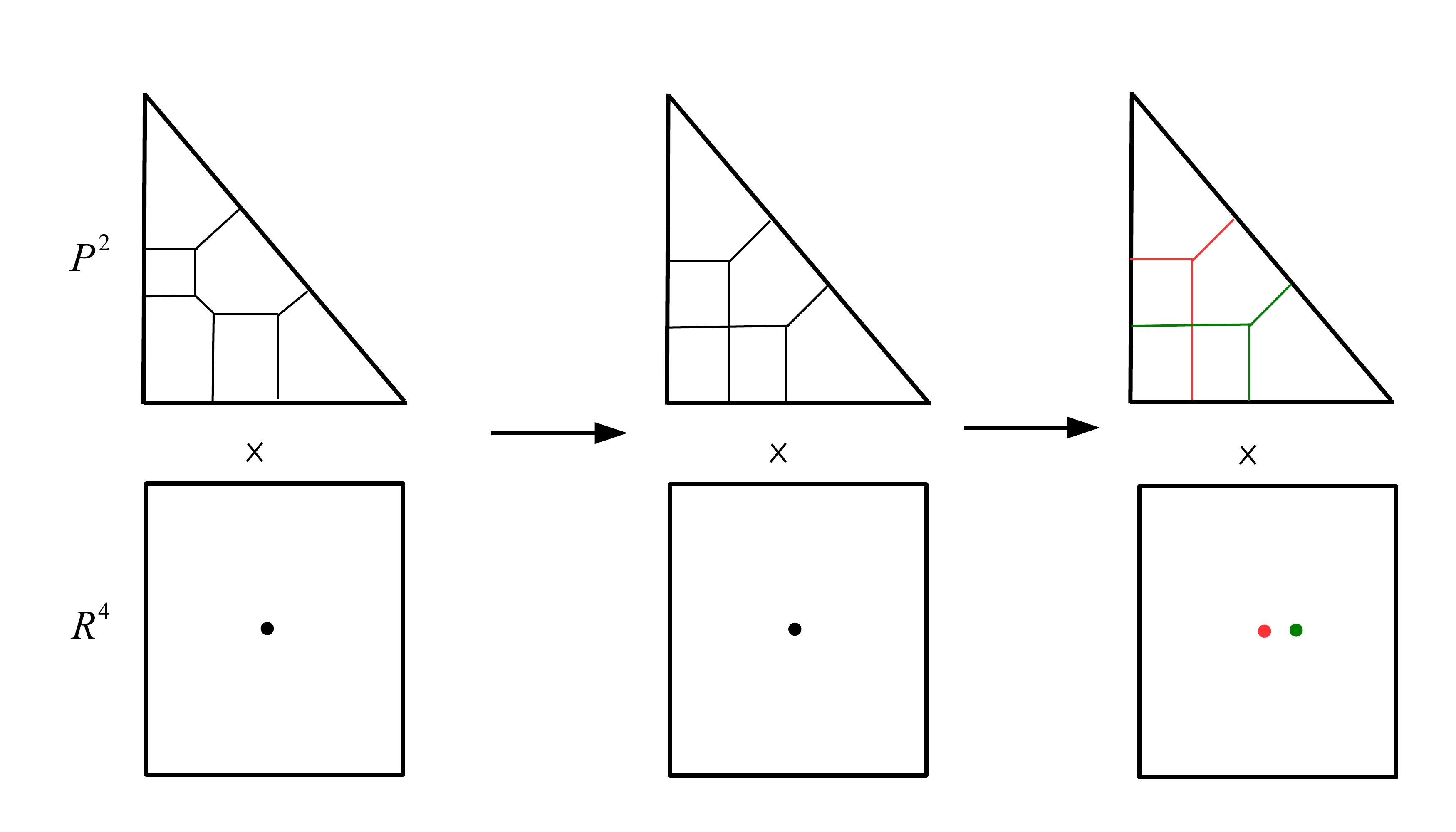}
  \caption{This branch corresponds to the degree $2$ curve splitting to two degree $1$ curves separated in $\mathbb{P}^2$ and $\mathbb{R}^4$.}
  \label{fig:branch3}
\end{figure}

There is another branch (see Figure \ref{fig:branch3}) where the degree 2 curve projects to two degree 1 curves which are again bound by the elliptic curve.
This other branch leads to a singularity in the sigma model whose locus is
$({\bf  P}^2\times {\mathbb P}^2)/{\mathbb Z}_2\subset {\mathbb P}^5$. 

Thus all told we have 3 branches, all with the same central charge $c_L=72$.

Note that the fact that the elliptic genus of the other branches do not vanish, means that unlike
the familiar case of symmetric product of $T^4$ or $K^3$ where turning on some $B$-fields can resolve
the Higgs branch singularity and make the Coulomb branch disappear altogether, the various Higgs branches
here cannot disappear by any deformation (the elliptic genus protects that).  Relatedly, this means that the
Higgs branch viewed as a sigma model, has an unremovable singularity over the locus ${\mathbb P}^2\subset {\mathbb P^5}$
and $({\mathbb P}^2\times {\mathbb P}^2)/{\mathbb Z}_2\subset {\mathbb P}^5$.

\subsection{d=3}

The generic branch here involves a degree $3$ curve in $\mathbb{P}^2$ which is a genus $1$ Riemann surface as shown in Figure \ref{fig:toric3}. The cohomology of $\widehat{C}$ in this case is given by
\begin{center}
\begin{tabular}{|c|c|c|}
\hline
$9$ & $1$ & $1$ \\
\hline
$1$ & $92$ & $1$ \\
\hline
$1$ & $1$ & $9$\\
\hline
\end{tabular}
\end{center}
We have here for the first time a situation where $h^{1,0}(\widehat{C})$ is non-zero which leads for the single string branch to left-moving fermions
realizing $SU(2)_L$ current algebra at level $k_L=g=1$. The common left/right-moving bosonic target space is $\mathbb{P}^9$ together with a Narain torus $(90,18)$ fibered over it. Again we will encounter different phases. The maximal central charge, corresponds to the $c=3d^2+27d=108$.  Here we expect to have two additional branches:  In one branch we expect two degree one curves to be on top of each other and one where all three degree one curves
wrap the same $P^1$.  The effective central charges for these two branches are expected to be $c=108$ in both cases.  So
all these branches in this case are expected to give the same growth.  The situation changes for the case of $d>3$.

\subsection{d=4}

The generic branch here involves a degree $4$ curve in $\mathbb{P}^2$ which is a genus $3$ Riemann surface. The cohomology of $\widehat{C}$ in this case is given by
\begin{center}
\begin{tabular}{|c|c|c|}
\hline
$14$ & $3$ & $1$ \\
\hline
$3$ & $126$ & $3$ \\
\hline
$1$ & $3$ & $14$\\
\hline
\end{tabular}
\end{center}
We have here for the first time a situation where $h^{1,0}(\widehat{C})$ is non-zero which leads for the single string branch to left-moving fermions
realizing $SU(2)_L$ current algebra at level $k_L=g=1$. The common left/right-moving bosonic target space is 16 complex dimensional together with a Narain torus $(124,28)$ fibered over it. Again we will encounter different phases, corresponding to decompositions $\sum_i d_i = 4$. 
The maximal central charge, corresponds to the $c=3d^2+27d=156$.  Here we expect to have additional branches corresponding to
$4=3+1=2+2=2+1+1=1+1+1+1$. The effective central charges for these four branches are expected to be $c(3)+c(1)=c(2)+c(2)= c(2)+c(1)+c(1)=c(1)+c(1)+c(1)+c(1)=144 <156$.  These additional branches give a lower growth than the single curve case.  In this case the effective central charge for these additional
branches happened to be equal.   This will not be the case for $d>4$'s and they will begin to give a diverse range of effective $c$'s which
are all smaller than the $c(d)$.

\subsection{Interpretations of the various poles \label{ObsBnd}}

In Section \ref{sec:Ftheory} we discussed the polar states of the function~$Z_{C}$ arising from the pole at~$\l=0$. 
As mentioned there,~$Z_{C}$ in fact has poles at the torsion points~$\l=\alpha\t+\beta$, 
with $\a, \b = \frac{p}{q} + \IZ$, with~$q \le k$, $0 \le p<q$. 
We now make a few comments on the interpretation of these poles, and in particular, we identify three physical sources of 
poles. We then illustrate these identifications using the explicit functions~$Z_{d}$ of this section. 

Firstly, the poles could be caused by multi-string unbound contributions. In fact we already subtracted these
contributions when we considered the free energy~$F_{C}$ in Section \ref{sec:Ftheory}. With this step, the higher order
poles at~$\l=0$ are removed as~$F_{C}$ has only second order poles there, but~$F_{C}$ still inherits the poles of~$Z_{C}$ at the 
non-zero torsion points. 

Secondly, it is known that the topological string free energy~$F$ also contains
multi-covering contributions, which can also cause some of the poles. 
In order to subtract contributions coming from multi-covering, a natural object to consider 
is the BPS generating function~$\wt F$ which is the~$k=1$ term of \eqref{defF}.  
Using a basis $\left\{C_i\right\}, i=1,\ldots,N=h_2(B)$ for the second homology of the base $B$ and the 
identity $C = \sum_{i=1}^{N} Q^i C_i$ we write
\begin{equation}
	F_C(\tau,\lambda) = F_{Q^1,\cdots,Q^{N}}(\tau,\lambda).
\end{equation}
Now we take the plethystic logarithm and arrive at the 
BPS generating function \cite{Hohenegger:2015cba,Gopakumar:1998ii,Gopakumar:1998jq}:
\begin{equation}
	\widetilde{F}_{k_1,\cdots,k_{N}} \= \sum_{d|s} \, \frac{\mu(d)}{d} \, F_{\frac{k_1}{d},\cdots,\frac{k_{N}}{d}}(d\tau,d \lambda), 
	\qquad s = \textrm{gcd}(k_1,\cdots, k_{N}) \, ,
\end{equation}
where $\mu(d)$ is the M\"obius function ($=(-1)^{n}$ if~$d$ is a product of~$n$ distinct primes, and 0 otherwise). 
As we explain below, the poles at~$\l = \frac{p}{q}$ are accounted 
for by the multi-covering. This is consistent with the relation between topological
string partition function and BPS degeneracies:  as discussed before the free energy~$F$ has a sum over 
terms with~$(\sin (\pi k \l))^{2g-2}$ for all~$k$, while the formula for~$\wt F$ has only the term with~$k=1$. 
This multi-covering formula will give poles at all $\lambda=m/k$.

Thirdly, the free energy contains multi-wound strings which can be seen by orbifold methods~\cite{Dijkgraaf:1996xw}, as
we sketched in Section \ref{sec:K3}. These are captured by 
using the Hecke-like operator~$V_{d}$ defined on Jacobi forms~\cite{Eichler:1985ja}:
\be 
 	\varphi (\tau,\lambda) | V_d  \= \sum_{AD=d \atop B \, \text{mod} \, D} \varphi \Bigl( \frac{A\t+B}{D}, A\l \Bigr) \, . 
 \ee
Indeed, the multi-wrapped contribution $F_{\frac{k_1}{d},\cdots,\frac{k_{N}}{d}}(d\tau,d \lambda)$ for $d>1$ is one 
of the terms~$(A=d,B=0,D=1)$ in the above formula for~$F_{\frac{k_1}{d},\cdots,\frac{k_{N}}{d}}(\tau,\lambda) | V_{d}$. 
In the K3 case discussed in Section \ref{sec:K3}, the partition function \eqref{defZ5d} can be written as the exponential 
of a sum of Hecke-like operators~$V_{d}$ applied to the single string index---meaning that the multi-wound strings 
are the only single-particle contributions to the index. In the F-theory compactifications, that is no longer true, and 
we have new contributions which do not come from multi-wound strings. 
We identify the poles at~$\l = \frac{p}{q} \tau + \frac{p'}{q'}$ as originating from the multi-wound strings.  

We now illustrate this using the function~$d=2$ as an example. 
The function~$Z_{2}$ has poles at~$\l=0, \frac12, \frac{\tau}{2}, \frac{\tau+1}{2}$ and their translates by the lattice~$\IZ\t+\IZ$.
At~$\l=0$ it has a fourth order and second order pole, and only second order poles at the other three points. The 
function~$\frac12 Z_{1}^{2}$ also has a fourth order pole at~$\l=0$, and the Laurent coefficients of the fourth-order 
pole of the two functions are exactly equal~\cite{Huang:2015sta}.  As a consequence,~$F_{2}=Z_{2} - \frac12 Z_{1}^{2}$ 
has only a second order pole at~$\l=0$. We note that this
does not mean that the Laurent coefficients of~$F_{2}$ at the~$\IZ \t + \IZ$ translates of~$0$ also vanish -- it would have 
if~$F_{2}$ were a true Jacobi form, but the translations~\eqref{elliptic1} of~$Z_{2}$ under~$\IZ \t + \IZ$ is governed by~$k=-1$, while those 
of~$\frac12 Z_{1}^{2}$ is governed by~$k=-2$. Nevertheless, this identification gives us a clue about the physical origin of the poles. 

Continuing in this fashion, we find that the coefficient of the second order pole of~$F_{2}$ at~$\l=\frac12$ 
exactly equals that of the function~$Z_{1}(2\t, 2\l)$ (thus implying that~$\wt F_{2}$ does not have a pole at~$\l=\frac12$).  
More precisely, the functions~$D(\t)$ and~$D'(\t)$ defined by the singular behavior~$Z_{2}(\t,\l)=D(\t)(\l-\frac12)^{-2}+O(\l-\frac12)$,
and~$Z_{1}(2\t,2\l)=D'(\t)(\l-\frac12)^{-2}+O(\l-\frac12)$ are equal. The functions~$D, D'$ are modular forms as can 
be seen from the Jacobi transformation law. 
Now, at the poles~$\l=\tau/2$, in order to get modular forms, one needs to multiply a prefactor.  
If we now define the functions~$E(\t)$ and~$E'(\t)$ by the singular 
behavior~$q^{-1/4} Z_{2}(\t,\l)=E(\t)(\l-\frac{\tau}{2})^{-2}+O(\l-\frac{\tau}{2})$,
and~$q^{-1/2} Z_{1}(\t,\l)|V_{2}=E'(\t)(\l-\frac{\tau}{2})^{-2}+O(\l-\frac{\tau}{2})$,
we find again that they are equal!

We find numerically that this pattern repeats for all the functions as far as we have checked.
The only place where this does not happen is the second order pole at~$\l=0$---for example, the Laurent coefficient of~$\wt F_{2}$ 
is not equal (even up to prefactors) to the second Laurent coefficients of any of the other functions in the game at~$\l=0$. 
This indicates that the states with growth~$36d$ second order pole at~$\l=0$ are new objects not coming from unbound or
multi-covering or multi-wrapped strings but are truly new bound states.  This is similar to what was found in 
\cite{Haghighat:2013gba} for M-strings.

\section{Macroscopics \label{macroscopics}}
In this section, we turn to the macroscopic aspects and interpretation of the results found in the previous sections. We have seen different growth formula for the finite and polar part of the elliptic genus, and wish to interpret them from the macroscopic point of view. Before going into more detail in the subsections below, let us summarize the main picture that is emerging.

The finite part in the elliptic genus, in the asymptotic regime where $n-r^2/4k \rightarrow \infty$  (see \eqref{cFinite}), is related to the large and single centered spinning black hole, which satisfies the cosmic censorship bound $M^3>J^2$, and whose entropy agrees with the Cardy formula. For the case of a single D3 brane wrapping a genus $g$ curve in $B= \mathbb{P}^2$, the entropy is controlled by the central charge 
\begin{equation}\label{cLindexP2}
c_L^{index}=36d \ ,\quad (d=1,2,3)\ , \qquad c_L^{index}=6(g-1)+36d=3d^2+27d\ ,\quad (d\geq 4)\ ,
\end{equation}
and we remind that the genus and degree of the curve are related by $g-1=\frac{1}{2}d(d-3)$. Compared to the $K3$ case, we have $c_L^{index}=6g-6$, and the spinning black hole is the BMPV black hole \cite{Strominger:1996sh,Breckenridge:1996sn,Breckenridge:1996is} which is 1/4 BPS instead 
of 1/2 BPS in F-theory. 

As we will show in the next two subsections, the leading (quadratic in $d$) term in the central charge \eqref{cLindexP2} are reproduced in two-derivative gravity, whereas the subleading linear terms follow from higher derivative corrections. Of course, there are further subleading corrections to the finite part but we have no clear macroscopic interpretation in this regime. Such configurations will have subleading entropy in any case. 

When the angular momentum approaches $J^2\rightarrow M^3$, the black hole becomes of stringy scale size with vanishing horizon and entropy. Upon further increasing the angular momentum to values $J^2>M^3$, the black hole can no longer carry the angular momentum because it would violate the cosmic censorship bound. Nevertheless, the microscopic analysis still predicts a rich phase with an entropy that is determined by the polar part of the elliptic genus. This entropy scales linearly in the charges, with central charge 
\begin{equation}
c_L=36d\ .
\end{equation}
As already discussed, the large angular momentum is generated from the zero modes in the center of mass system, and the degeneracy of these states with $J^2 > M^3$ grows exponentially. On the gravity side, the most natural thing to expect is that this phase corresponds to gravitational configurations that consist of small black holes with a stringy scale horizon.    We predict the (index) entropy of these small
black holes is then to leading order independent of the spin and given for large $d, n$ by
\begin{equation}
S=2\pi{\sqrt{6dn}}\ .
\end{equation}
For the $K3$ case, as we have seen, the $M^3 > J^2$ violating states are generated in the zero KK momentum sector. Still, they have an exponential growth $e^{4\pi{\sqrt g}}$ where $g=Q_1Q_5$ for the D1-D5 system.

We now give a few more details about the gravitational aspects. In the next two subsections, we focus on the quadratic and linear terms (in $d$) in the central charge, and show how to get them from the entropy of single centered black holes, including the effect of higher derivative terms. 
We then discuss macroscopic aspects of states with $J^2 > M^3$.

\subsection{Spinning black holes}

Macroscopically,  strings coming from wrapped D3-branes correspond to black strings in six dimensions, and after wrapping over a circle $S^1$ with $n$ units of momentum, they correspond to electrically charged and spinning five-dimensional BPS black holes. A single-centered black hole in five dimensions can have two angular momenta, coming from rotations $J_1$ and $J_2$ in two orthogonal planes in $\mathbb{R}^4$. The $U(1)_L\times U(1)_R\subset SU(2)_L\times SU(2)_R=SO(4)$ rotation generators relate to these by $J_1=J_L+J_R$ and $J_2=J_L-J_R$. Furthermore, the BPS condition requires the angular momenta of the black hole to be equal in magnitude\footnote{This is only true for spherical $S^3$ horizons, and not true e.g. for a single black ring, which has $J_1\neq J_2$ \cite{Elvang:2004rt}, or for the more recent black lens solutions of \cite{Kunduri:2014kja}. Concentrentic black rings, on the other hand, can have $J_1=J_2$ \cite{Gauntlett:2004wh}.}, see e.g. \cite{Gauntlett:1998fz}, $J\equiv J_1=J_2$  (which is the one
preserving half of the supersymmetry of the $(1,0)$ bulk) such that $J_L=2J$ and $J_R=0$. So we are led to compare the macroscopic entropy with what comes from the Cardy formula in the left moving sector of the CFT
\begin{equation}\label{S_BH}
S=2\pi{\sqrt{\frac{c_L}{6}\left(n-\frac{J^2_L}{4k_L}\right)}} = 2\pi{\sqrt{\frac{c_L}{6}\left(n-\frac{J^2}{k_L}\right)}}\ .
\end{equation}
Here we are using a convention where $J_L$ has an integer spectrum and $J$ is half-integral.
To give a concrete example, consider the case of $B=\mathbb{P}^2$, which leads to minimal supergravity in six dimensions consisting of the metric and a tensor whose field strength is selfdual. Upon compactifying to five dimensions, we get gravity coupled to a vector multiplet, and the KK charge becomes an electric charge. Then the entropy for the single string branch to leading order in the large charge limit is given by
\begin{equation}\label{Smax}
S=2\pi {\sqrt{\frac{d^2n}{2}-J^2}}\ .
\end{equation}
There is a bound on the angular momentum given by 
\begin{equation}
 J^2<\frac{nd^2}{2}\ .
 \end{equation} 
Consider as a second example, the case $B=\mathbb{P}^1\times \mathbb{P}^1$. The low-energy effective action in six dimensions is supergravity coupled to one tensor multiplet. The intersection matrix for $B=\mathbb{P}^1\times \mathbb{P}^1$ corresponds to $C\cdot C=Q^{\alpha} \eta_{\alpha\beta}Q^{\beta}=2Q_1Q_2$, and the total central charge is
\begin{equation}
c_L=6Q_1Q_2+18(Q_1+Q_2)+6\ ,
\end{equation}
where we used that $c_1(B)\cdot C=2(Q_1+Q_2)$. The level is given by $k_L=Q_1Q_2-(Q_1+Q_2)$. Hence the black hole entropy to leading order in the charges is 
\begin{equation}\label{entro-Q}
S=2\pi {\sqrt{Q_1Q_2n-J^2}}\ .
\end{equation}
The entropy formulae described above are similar to the ones obtained from the macroscopic BMPV black holes that have been studied in \cite{Strominger:1996sh,Breckenridge:1996sn,Breckenridge:1996is}. To leading order in the charges, the entropy scales like $S\propto N^{3/2}$ as we rescale all charges uniformly with a factor $N$. In an appropriate normalization for the mass of the black hole, the entropy can be written as $S=2\pi {\sqrt {M^3-J^2}}$, and hence the presence of a horizon requires
$M^3 > J^2$. Beyond the cosmic censorship bound, our microscopic analysis suggest that there is still entropy that scales like $S\propto N$.  This is somewhat similar to the analysis done in \cite{moulting}. We note, however, that the linear scaling that we find beyond the 
cosmic censorship bound exists for arbitrary large $J$. 
We explain their macroscopic interpretation in section 6.4. 

So far, we have only reproduced the entropy to leading, quadratic order in the charges, which is similar to the macroscopic spinning BMPV black hole. 
In the next subsection, we derive the leading and subleading (linear in the charges) contributions to the central charges and levels from a supergravity analysis on $AdS_3 \times S^3$ backgrounds, the near-horizon geometry of a black string in six dimensions. We include effects coming from higher derivative terms in six dimensions and match the gravity prediction with the microscopic prediction for the single-centered spinning black hole including these subleading corrections. The center of mass contributions to the central charges and levels cannot be derived from this analysis, as they are not captured by near-horizon degrees of freedom living on the bulk of $AdS_3$. Instead, using the terminology of \cite{Dabholkar:2010rm}, they should correspond to exterior degrees of freedom that live outside the near-horizon geometry.

\subsection{$AdS_3\times S^3$ and central charges}

A generic F-theory compactification on an elliptic CY 3-fold $X$ produces (1,0) six-dimensional supergravity with gauge groups and matter \cite{Vafa:1996xn,Morrison:1996na,Morrison:1996pp}.   For simplicity let us assume the gauge group is abelian and denote by $n_T,n_V,n_H$  the number of tensors,
vectors and hypermultiplets respectively.   In six dimensions we have
\begin{equation}
n_T=h^{1,1}(B)-1\ ,\qquad n_V=h^{1,1}(X)-h^{1,1}(B)-1\ ,\qquad n_H=h^{2,1}(X)+1\ .
\end{equation}
Besides the gravity multiplet, only the tensor multiplets will be relevant for our discussion.
We collectively denote all tensors (one from the gravity multiplet that is selfdual, and $n_T$ are anti-selfdual) by $B^\alpha$ and their field strengths $H^\alpha=dB^\alpha; \alpha=1,\cdots, n_T+1$. They descend from the RR four-form field strength in type IIB expanded in a basis of (1,1)-forms on the base $B$.
%
%
We can write down an action in six dimensions if we relax the duality constraints on the tensors, and impose them by hand in the equations of motion. Most relevant for our discussion is not only the two-derivative action, but also a gravitational ``Chern-Simons term" in six dimensions needed for gravitational anomaly cancellation. The relevant terms, following \cite{Ferrara:1996wv,Bonetti:2011mw}, read
$$S=\int_{M_6}\,\left[ \frac{1}{2}R*1-\frac{1}{4}g_{\alpha\beta}G^\alpha\wedge *G^\beta-\frac{1}{2}g_{\alpha\beta}dj^\alpha\wedge *dj^\beta-\frac{1}{4}\eta_{\alpha\beta}K^\alpha B^\beta \wedge \tr {\cal R}\wedge {\cal R}+\cdots \right] \ .$$
Here, $g_{\alpha\beta}$ is the metric on the tensor branch parametrized by the scalars $j^\alpha$, and $\eta_{\alpha\beta}$ is the intersection matrix on the base $B$ which is $SO(1,n_T)$ invariant. The scalars satisfy $\eta_{\alpha\beta}j^\alpha j^\beta=1$ and the metric is $g_{\alpha\beta}=2j_\alpha j_\beta-\eta_{\alpha\beta}$, where indices are lowered by $\eta_{\alpha\beta}$. The selfduality constraints on the gauge invariant tensors are $g_{\alpha\beta}*G^\beta=\eta_{\alpha\beta}G^\beta$. These three-form tensors are $G^\alpha=H^\alpha+\frac{1}{2}K^\alpha \omega^{CS}_{grav}+\cdots\ ,$
where $\omega^{CS}_{grav}$ is the gravitational Chern-Simons three-form build out of the spin connection one-form, and the dots refer to terms involving the Chern-Simons forms of the gauge fields such that the $G^\alpha$ are  invariant under both local Lorentz and gauge transformations.  These terms are not important for our purpose, since we do not switch on any charges in the vector multiplet sector. The gravitational Chern-Simons term contains the curvature two-form ${\cal R}$, which satisfies $d\omega^{CS}_{grav}=\tr {\cal R}\wedge {\cal R}$. 

Finally, the coefficients $K_\alpha=\eta_{\alpha\beta}K^\beta$ appearing in the Chern-Simons term were found to be \cite{Sadov:1996zm,Bonetti:2011mw}
$$K_\alpha = -\int_{C_\alpha}\,c_1(B)|_{C_\alpha}=\frac{1}{12}
\int_X\,\omega_\alpha \wedge c_2(X)\ ,$$
with second Chern class $c_2(X)$ and $\omega_\alpha$ the dual of the pull back $\pi^*(C_\alpha)$ for a basis $C_\alpha$ in $H^{1,1}(B)$. This relation can be determined from the matching with the M-theory Chern-Simons terms using M-theory/F-theory duality. There are other Chern-Simons like terms in six dimensions of the form $B\wedge F \wedge F$ mixing the tensors with the gauge sector, but we did not write them explicitly since we also suppressed the $F\wedge *F$ terms in the action.

We now formulate the theory on $AdS_3\times S^3$, the near horizon geometry of a black string, and determine the levels and central charges of the dual CFT following the ideas of \cite{Kraus:2005vz,Kraus:2005zm,Hansen:2006wu,Dabholkar:2010rm} which we adapt here to our F-theory setup. Since we are focusing on the near horizon geometry, we will only reproduce the microscopic formulas for the central charges and levels coming from the single string conformal field theory, and exclude the contributions from the center of mass. The latter would be captured by exterior degrees of freedom outside the horizon, as in \cite{Dabholkar:2010rm}. We first focus on the two-derivative theory. The $AdS_3\times S^3$ geometry is supported by the fluxes
$$4\pi^2Q^\alpha\equiv \int_{S^3} H^\alpha\ .$$
The charges associated with the dual field strength are related to $Q^\alpha$ by the selfduality constraint. After reducing on $S^3$, one produces $SO(4)=SU(2)_L\times SU(2)_R$ gauge fields with three-dimensional Chern-Simons terms in $AdS_3$. These are not gauge invariant, but transform into total derivatives with anomaly coefficients that determine the levels of the dual CFT \cite{Hansen:2006wu,Dabholkar:2010rm}. It was found that the levels scale quadratically in the charges which, translated to our conventions, are
$$k_L=k_R=\frac{1}{2}Q^2\ ,\qquad Q^2\equiv Q^\alpha \eta_{\alpha\beta}Q^\beta\ .$$
For example, when the base is $B=\mathbb{P}^1\times \mathbb{P}^1$, one has $n_T=1$ and $n_V=0$, hence two charges with intersection matrix corresponding to $Q^\alpha \eta_{\alpha\beta}Q^\beta=2Q_1Q_2$, and so $k_L=k_R=Q_1Q_2$ which agrees with \cite{Dabholkar:2010rm} for integer charges. Hence we checked the overall normalization coefficient in the formula for the levels. The result confirms the leading order result for the microscopic formulas for the levels by identifying 
$C\cdot C =Q^2\ .$
At two-derivative level, the left and right central charges are still equal and since there is supersymmetry in the right moving sector we have 
$$c_L=c_R=6k_R=3Q^2=3C\cdot C\ ,$$ 
as expected from the microscopics to leading order. 

Now we switch on the higher derivative Chern-Simons term proportional to $B\wedge \tr {\cal R}\wedge {\cal R}$. Because of the gravitational anomaly, all charges and levels will get corrections and, following Appendix A of \cite{Dabholkar:2010rm}, the following relations hold:
$$c_L-c_R=12 \Delta k_R=-12 \Delta k_L=12\beta_\alpha Q^\alpha\ = 6c_1(B)\cdot C,$$
where $\beta_\alpha =-\frac{1}{2}K_\alpha= \frac{1}{2}c_1(B)_\alpha$ appears as the coefficient multiplying the gravitational Chern-Simons term in \cite{Dabholkar:2010rm}. These relations are indeed satisfied by our microscopic index-analysis.  Hence we get (using
the fact that by $(0,4)$ supersymmetry $c_R=6k_R$) to first subleading order in $C$
$$c_L=3C\cdot C + 9 c_1(B)\cdot C\ ,$$
$$k_L=\frac{1}{2}C\cdot C-\frac{1}{2}c_1(B)\cdot C\ ,$$
$$c_R=6k_R=3C\cdot C+ 3 c_1(B)\cdot C\ .$$
Altogether, the supergravity analysis shows that type IIB on
\begin{equation}
(AdS_3\times S^3)_{Q^\alpha} \times B\ ,
\end{equation}
or more precisely F-theory on $(AdS_3\times S^3)_{Q^\alpha} \times X$, with $Q^\alpha$ units of flux through $S^3$ from the three-forms $H^\alpha$, is dual to a $(0,4)$ conformal field theory with $SU(2)_L\times SU(2)_R$ symmetries with levels and central charges given above.  Here the coupling constant of type IIB varies over $B$ and undergoes $SL(2,{\bf Z})$ monodromy.
 That such conformal field theories had to exist was already anticipated in \cite{Vafa:1997gr,deBoer:1998ip}.  However, here we are predicting that these conformal theories must have, in addition to the center of mass, additional continuum spectrum, which
signals the existence of other CFT branches which are disconnected, similar to the ideas studied in \cite{Seiberg:1999xz} in the
context of $B=K3,T^4$.  

\subsection{Exterior center of mass modes}

After having reproduced the quadratic and linear terms in the central charge from the macroscopic viewpoint, we now focus on the constant terms (independent of the charges) in the total central charge. Microscopically, for single centers, they are related to the center of mass modes in the CFT. Macroscopically, these modes are generated by acting with the broken supercharges on the black hole solution, and form some type of hair that lives exterior to the horizon. Because of these modes, there is a difference between the absolute degeneracy and the index-degeneracy \cite{Banerjee:2009uk,Jatkar:2009yd, Dabholkar:2010rm}. Microscopically, we have seen that in F-theory, this difference is a factor of +6, whereas it is a factor of +12 for $K3$ compactifications in type IIB. 

We now give an argument how to understand this, following Section 4 of~\cite{Dabholkar:2010rm}. Macroscopically, the BPS string is a solitonic BPS black string in six dimensions and we can quantize it semiclassically. We introduce a macroscopic index $Tr[(-)^{J_R} J_R^2\, e^{2\pi i p \tau +2\pi i J_L \lambda}]$, where $J_L$ and $J_R$ are as before the Cartan generators of the rotation group in the perpendicular dimensions $\mathbb{R}^4_\perp$ with integer eigenvalues.  The index only receives contributions from the left-moving modes on the worldsheet of the $(0,4)$ black string. For left-moving fields with nonzero $J_R$ charge, this leads to an index-degeneracy $c_L^{eff}$ that can be different from the central charge $c_L$. For $SU(2)_R$ doublets, $c_L^{eff}$ can be computed by realizing that bosons contribute to the index like the inverse of the contribution to the partition function from a fermion, and vice-versa  \cite{Dabholkar:2010rm}.
For a left-moving $(2_L,2_R)$ bosonic center of mass system, one has $c_L=4$ and $c_L^{eff}=-2$, so the difference is $+6$. For left-moving $(1_L,2_R)$ fermionic center of mass degrees of freedom , one has $c_L=1$ and $c_L^{eff}=-2$, so the difference is $+3$. In F-theory, there are no such left-moving fermions, so the difference remains $+6$, but for $K3$, we have a pair of them, giving an additional factor of $+6$ in the difference between absolute and index-degeneracy. 

\subsection{Comments on states with $J^2 > M^3$}

\begin{figure}[here!]
  \centering
  \subfloat[Entropy of a black hole with $J_{tot}=J_{spin}$]{\label{fig:entropyA}\includegraphics[width=0.5\textwidth]{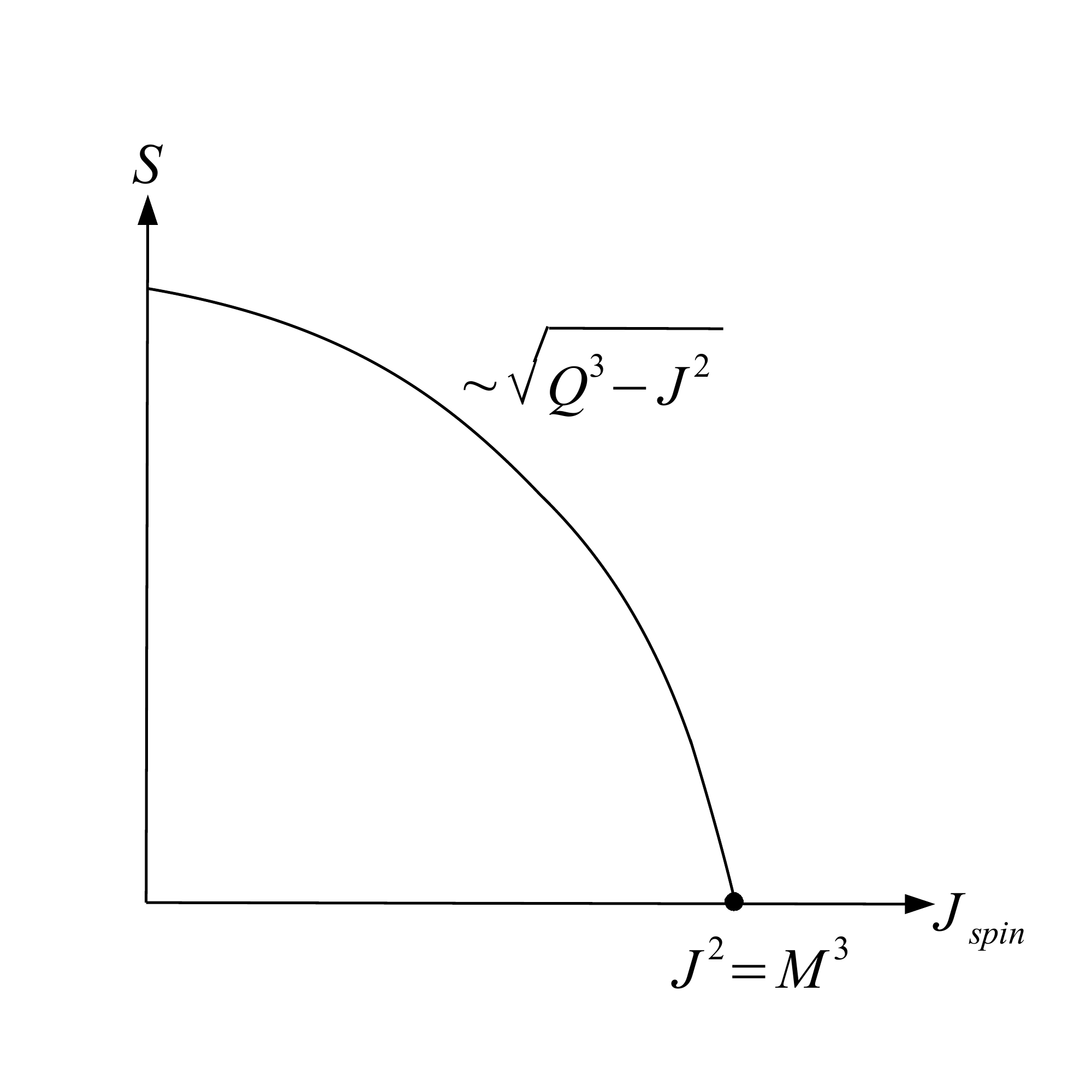}}     
  \subfloat[Entropy of a black hole with $J_{tot}=J_{orb}$]{\label{fig:entropyB}\includegraphics[width=0.5\textwidth]{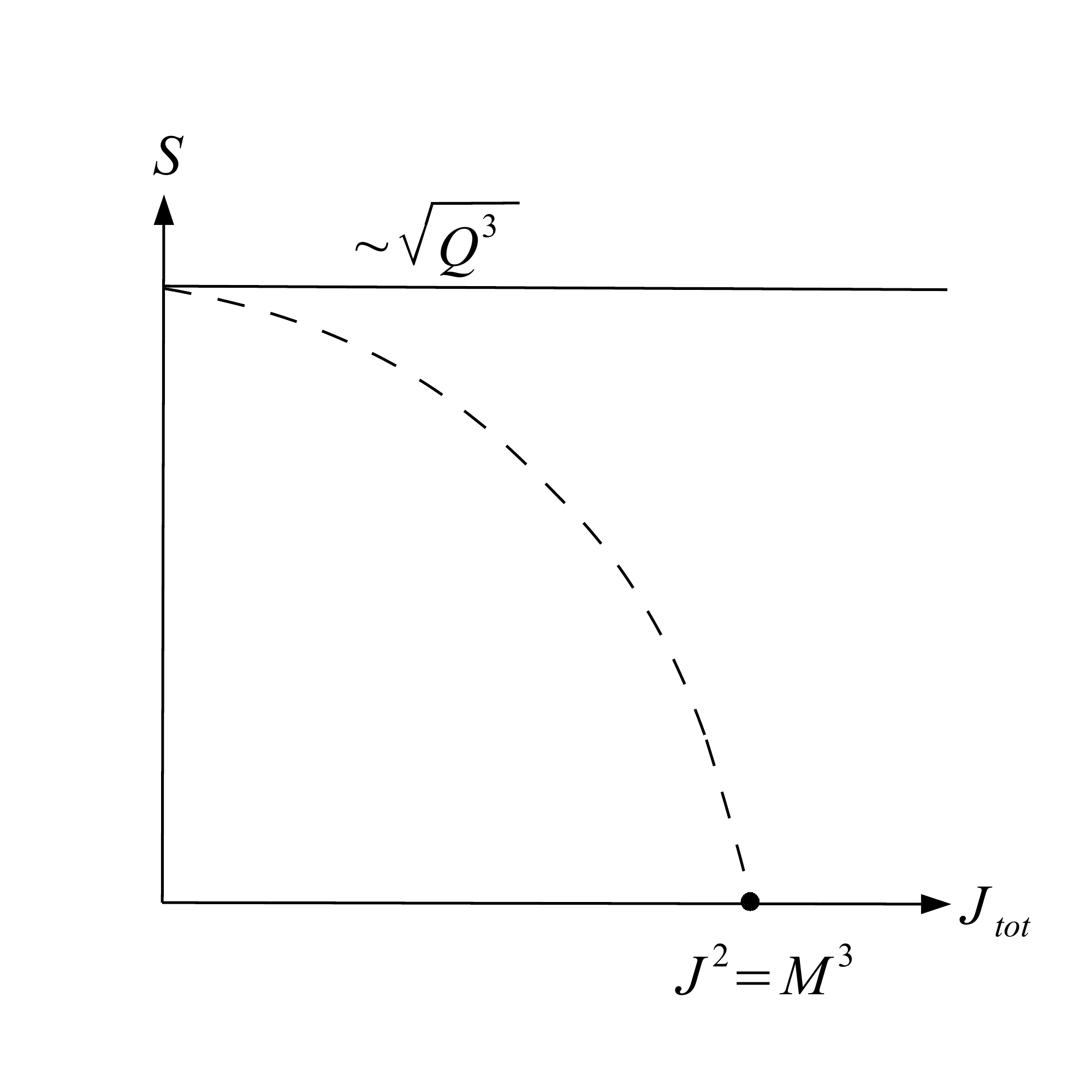}}
  \caption{Entropy of a black hole with $J_{tot} = J_{spin} + J_{orb}$. In part (a) we see the entropy of the large spinning black hole as a function of internal angular momentum. In part (b) we see the expected entropy of the black hole in flat space as a function of total angular momentum as measured from infinity.}
  \label{fig:entropy}
\end{figure}

In this section we comment on the macroscopic interpretation of states which violate the bound $M^3 > J^2$.
Consider a spinning black hole.  It clearly has macroscopic
angular momentum as it is spinning.  In fact this leads to the simplest explanation of CCB, i.e. the horizon of a macroscopic black hole cannot
move faster than speed of light.  How could this be violated?  
The answer is due to orbital angular momentum. As already stated previously the bound $M^3 > J^2$ is allowed to be violated if particles are allowed to have arbitrarily high orbital angular momentum. This is what is happening in flat space as wave-functions are non-normalizable and therefore give rise to arbitrarily high angular momenta. However, this raises a puzzle: If there are always states in the ensemble which can have arbitrarily large orbital angular momentum, then the entropy as a function of angular momentum measured at infinity will be always constant. This is because the entropy will pick the largest possible value for given angular momentum which is precisely the large black hole without internal spin. This phenomenon is depicted in Figure \ref{fig:entropy}.
The resolution of the puzzle comes from the fact that we are putting our ensemble in a Taub-NUT background. That is, we take our six-dimensional spacetime to be 
\begin{equation}
	TN \times \mathbb{R} \times S^1,
\end{equation}
where $TN$ denotes the Taub-NUT space which can be considered as an $S^1$ fibration over $\mathbb{R}^3$, where the circle shrinks to zero at the origin and attains a finite radius $R$ at infinity. The metric can be written with $\chi \in S^1$, and $\vec x \in \mathbb{R}^3$, as
\begin{equation}
	ds_{TN}^2 = R^2 \left[\frac{1}{V}(d\chi + \vec{A} \cdot d\vec{x})^2 + V d \vec{x}^2\right],
\end{equation}
with 
\begin{equation}
	V = 1 + \frac{1}{|\vec{x}|}, \quad \vec{\nabla} \times \vec{A} = \vec{\nabla} V.
\end{equation}	
For an earlier analysis, but unrelated to the work in the present paper, of 5d spinning black holes in the presence of Taub-NUT geometry see \cite{Bena:2006qm}.

\begin{figure}[here!]
  \centering
  \subfloat[Taub-NUT geometry]{\label{fig:Taub-NUT}\includegraphics[width=0.5\textwidth]{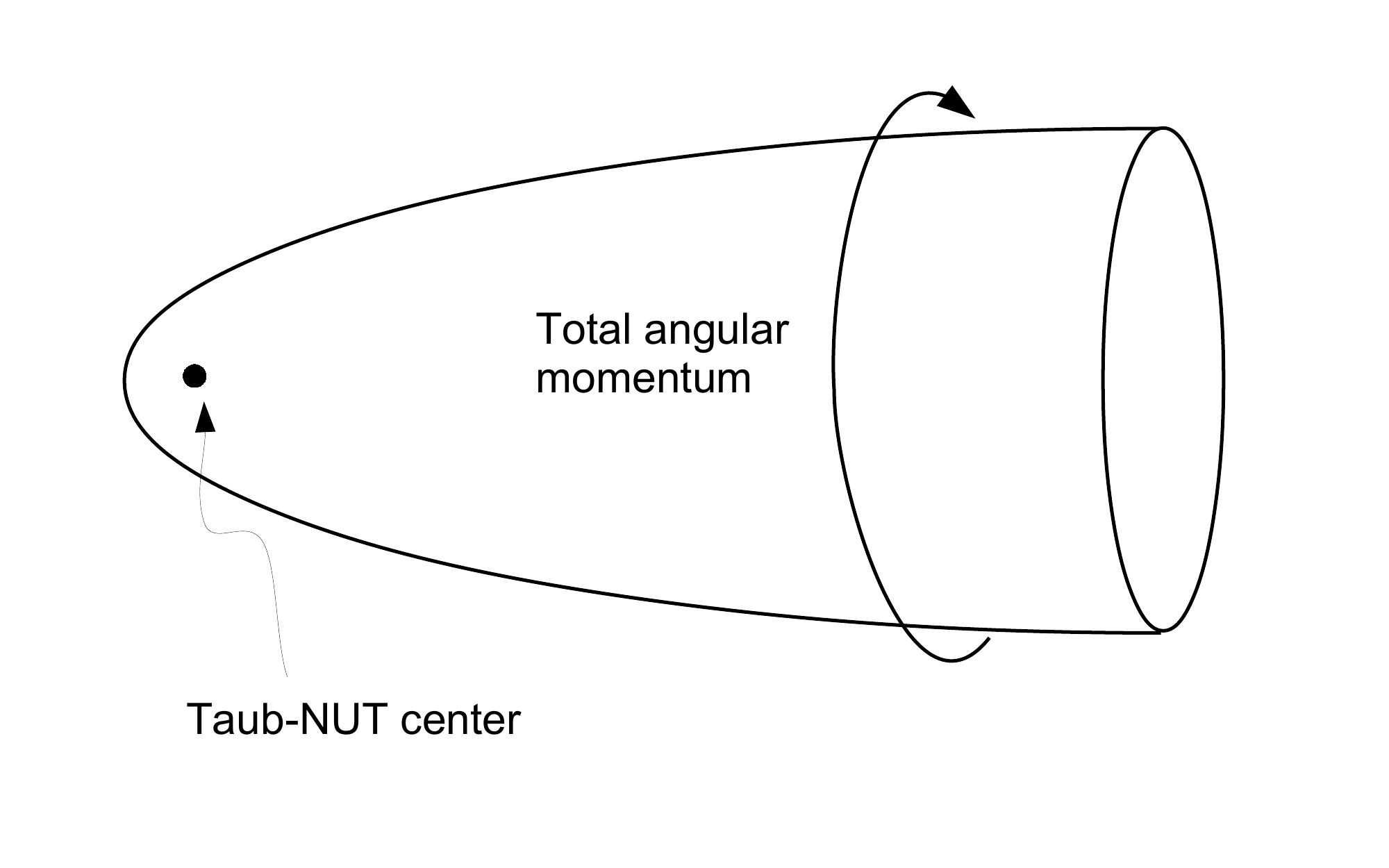}}     
  \subfloat[Entropy which contributes to the index]{\label{fig:entropyB}\includegraphics[width=0.5\textwidth]{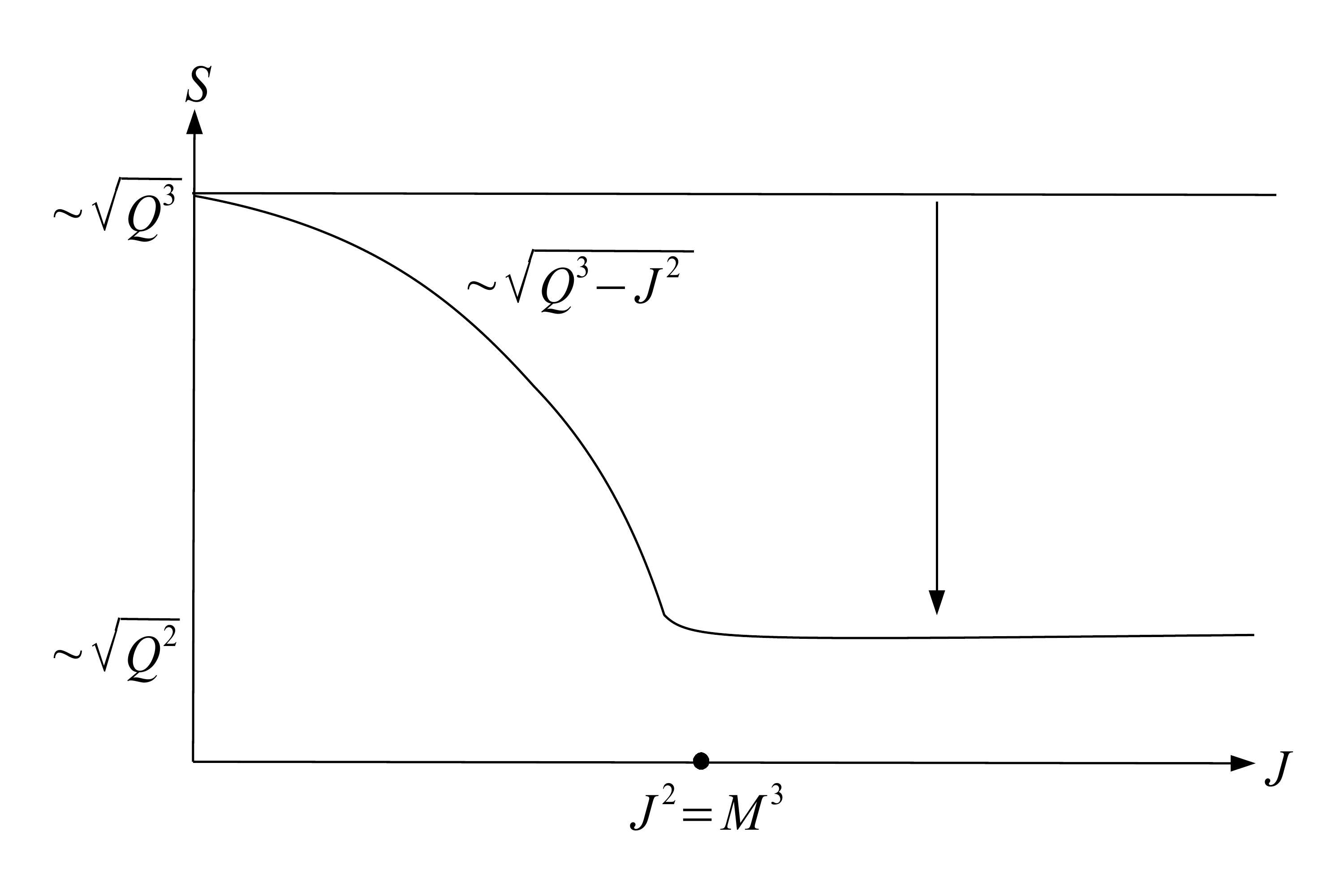}}
  \caption{In part (a) we see the Taub-NUT geometry with total angular momentum identified with momentum along the Taub-NUT circle. In part (b) we see the entropy of the black hole in $\mathbb{R} \times TN$ as given by the index.}
  \label{fig:entropytn}
\end{figure}

As described in section \ref{sec:INTRO} the Taub-NUT geometry leads to an effective decrease of entropy for large values of angular momentum as measured from infinity. In the Taub-NUT background the orbital angular momentum becomes momentum along the Taub-NUT circle and leads to a regularization of the corresponding wave functions. Microscopically, this leads to a cancellation of states in the index as illustrated for an example in equation (\ref{eq:microexa}). Macroscopically, this cancellation in the index leads to the entropy profile shown in Figure \ref{fig:entropytn}.
In other words, the particle states that contribute to the macroscopic black holes are the ones contributing to the index $N^Q_g$ for $g>0$
discussed in section 4.  On the other hand the $g=0$ BPS states, counted by $N^Q_0$ are the ones which lead to states with $J^2 > M^3$ but their growth is not big enough to count as macroscopic black holes.  This gives a physical explanation to the puzzle raised in \cite{Vafa:1997gr} as to why the genus 0 curves have considerably lower entropy than the macroscopic black hole states.

In fact we can prove this is the case for the simplest example of these states, namely the ones
that arise from Type IIB on $K3\times S^1$.   For this case, 
the  states with $J^2 > M^3$ we have been considering are
dual to Dabholkar-Harvey states for heterotic strings on $T^5$  \cite{Dabholkar:1989jt}.  To see this recall that type IIB on $K3 \times S^1$
is dual to type IIA on $K3\times S^1$ and is dual to heterotic strings on $T^5$
and the CCB violating D3 branes which are wrapped around genus $g$ curves times $S^1$ with no KK momentum,
are dual to D2 branes wrapped around genus $g$ curves in $K3$ and these are
dual to Dabholkar-Harvey states  \cite{Vafa:1995zh} which are perturbative string states of
heterotic strings on $T^5$.  
In particular for a class of F-theory models which are dual to heterotic strings (for example when the base is $\mathbb{P}^1\times \mathbb{P}^1$
and the heterotic dual is $K3\times S^1$), a subset of the CCB violating states (in the $\mathbb{P}^1\times \mathbb{P}^1$ example, the
D3 branes wrapped around only one of the $\mathbb{P}^1$'s) can again be mapped to perturbative DH states
of the heterotic string.

\section{Concluding Thoughts}

In this paper we have elucidated the properties of 6d strings which arise from F-theory on elliptic
CY 3-folds and have seen how they lead to BPS spinning black holes in 5d.  They arise from D3 branes wrapping
the curves in the base of the elliptic CY and in addition wrap a circle when we go down in dimension from
6 to 5.  The usual spinning BPS black hole in 5d arise from curves which are not degenerate.  However
we also found that degenerate curves lead to multi-string configurations which, even though they
do not form bound states in 6d, KK momenta binds them when they wrap a circle.  Even though the entropy
they lead to is generically lower than that of a single black hole they are expected to lead to interesting
new bound states. Moreover we found that a subclass of BPS states (those that degenerate  to genus 0 curve
configurations in the M-theory setup) lead to $J^2>M^3$ states.

  By duality between F-theory and M-theory, this also implies that coincident
configurations of M5 branes, which have extra light modes, give significant modification to the entropy in certain phases of the
4d black holes.  For example the 5d BPS states can also be viewed as 4d black holes which have lower
growth than the macroscopic 4d black hole.  It is natural to expect that in the context
of \cite{Maldacena:1997de} where M5 branes wrap an ample divisor in a CY 3-fold, the coincident configurations lead
to new branches which can dominate the single string configuration in certain phases just as the CCB violating states
do in the 5d case in the phase with large angular momentum.  This could be relevant for the multi-centered black hole
configurations in 4 dimensions \cite{Denef:2007vg}. 

Clearly our work indicates an intricate phase structure for black holes in 5 dimensions based on microscopic counts and we expect
this to carry over to 4 dimensions as well.  It would be interesting to also study this intricate phase structure from the macroscopic side for
4 and 5 dimensional black holes.

\section*{Acknowledgements}

We would like to thank Iosif Bena, Clay Cordova, Atish Dabholkar, Gary Horowitz, Sheldon Katz, Albrecht Klemm, Guglielmo Lockhart, Samir Mathur, Martin Ro\v cek, Maria Rodriguez, Ashoke Sen, Phil Szepietowski, Andrew Strominger, Nick Warner, Martin Westerholt-Raum, and Don Zagier for valuable discussions. S.V. would like to thank the Harvard Center for the fundamental Laws of Nature for hospitality during the course of this project.  We would also like to thank SCGP for hospitality during the 2015 summer workshop.

This research is supported in part by NSF grant PHY-1067976, by EPSRC First Grant UK EP/M018903/1 and by the Netherlands Organisation for Scientific Research (NWO) under the VICI grant 680-47-603.

\appendix

\section{Aspects of modular forms and Jacobi forms}

We begin this appendix with a very brief discussion of the asymptotic growth of Fourier coefficients of 
modular forms~$f(\t)$ in one variable~$\t \in \IH$. We then turn to Jacobi forms that are functions of 
another \emph{elliptic variable}~$\lambda \in \IC$ in addition to~$\t$. 
The classical theory of Jacobi forms~\cite{Eichler:1985ja} considers functions that are holomorphic in~$\l$. 
We briefly describe some properties of such holomorphic Jacobi form and, in particular, their relation to 
modular forms of one variable. 
This relation leads to a formula for the growth of their Fourier coefficients.

In the main text of this paper we encounter Jacobi forms that are meromorphic in the elliptic variable~$\l$.
In this appendix we describe a few relevant results in the theory 
of meromorphic Jacobi forms following~\cite{Dabholkar:2012nd}. We then describe the growth of their  
Fourier coefficients that follow from these results, which we use in the main text. 

Throughout this appendix, we set $q =  e^{2 \pi i \tau}$, $y =  e^{2 \pi i \l}$. 

\subsection{Modular forms \label{appmodular}}

Modular forms~$f(\t)$ are holomorphic functions of one variable~$\t \in \IH$ 
and often appear as the holomorphic partition function~$f(\t)={\rm Tr}_{\CH} \, q^{L_{0}}$ 
of a two-dimensional CFT with Hilbert space~$\CH$ on a torus with modular parameter~$\t$. 
The degeneracies of states~$a(n)$ with~$L_{0}=n$ are summarized  in the Fourier expansion:
\be\label{deffCFT}
f(\t) \= \sum_{n= - n_{0}}^\infty a(n)\,q^n \, , \qquad a(n_{0}) \neq 0 \, , 
\ee
Here the number $-n_{0}$ is the lowest energy level of the CFT, and we have~$n_{0}> 0$. 
The asymptotic growth as~$n \to \infty$ of the Fourier coefficients
\be \label{defan}
a(n) \= \int_{C} e^{-2\pi i n \t} f(\t) \, d\t \, 
\ee
can be derived using the familiar modular transformation law
  \be\label{modtransform0} f(\frac{a\t + b }{c\t + d}) = (c\t + d)^w\,f(\t) \qquad \forall \;\left(
       \begin{array}{cc} a & b \\ c & d \\  \end{array} \right) \in SL(2,\Z)\, . \ee

For the leading estimate, we use the saddle point method and deform the contour~$C$ in~\eqref{defan}. 
As~$n\to \infty$, the integral is dominated by small values of~$\t$, or~$q \to 1$. The modular 
transformation~$\tau \to -1/\tau$ relates this to~$\t \to i \infty$, or $q \to 0$, for which~$f(\t)$ is dominated 
by its most polar term~$q^{-n_{0}}$. We thus obtain the leading behavior:
\be \label{angrowth}
a(n) \sim \exp(4 \pi \sqrt{n_{0}\, n}) \,  \qquad n \to \infty \, .
\ee
For a generic CFT with central charge~$c$, the vacuum energy is~$-n_{0}=-c/24$, and 
the estimate~\eqref{angrowth} is simply the familiar  Cardy formula~$a(n) \sim \exp(2 \pi \sqrt{c \, L_{0}/6})$.

From the mathematical perspective, the exponential growth~\eqref{angrowth} of the Fourier coefficients
is controlled by the exponential growth~$q^{-n_{0}}$ of the function~$f(\t)$ as~$\t = i \infty$. 
This is reflected in the nomenclature\footnote{Modular forms with~$n_{0} = 0$, or $n_{0} < 0 $ are called 
holomorphic modular forms, and cusp forms, respectively, and in these cases the asymptotic growth of~$a(n)$ is 
indeed polynomial in~$n$.}\emph{weakly holomorphic modular forms} for modular forms obeying~$n_{0} >0$.

\subsection{Jacobi forms \label{appJacobi}}

The classical theory of Jacobi forms~\cite{Eichler:1985ja} treats holomorphic 
functions $\varphi(\tau,\lambda)$ from $\mathbb{H} \times \mathbb{C}$ to $\mathbb{C}$ 
and appear, for example, as the elliptic genus of~$\CN=(2,2)$ SCFTs 
$\varphi(\t,\l) = {\rm Tr}_{\CH} \, (-1)^{F} \, q^{L_{0}} \, y^{J_{0}} $ 
where~$L_{0}$ and $J_{0}$ are the left-moving Hamiltonian and~$U(1)_{R}$ charge of the SCFT. 
The degeneracies~$c(n,r)$ of eigenstates with~$L_{0}=n, J_{0}=r$ are summarized in the Fourier expansion:
\be \label{Jac}
\varphi(\t,\l)  \= \sum_{n,r} c(n,r) \, q^n \, y^r \, .
\ee 
For usual SCFTs with compact target space like the~$\CN=2$ minimal models, the elliptic genus is 
indeed holomorphic in~$\l$.

Jacobi forms transform under the modular group as
  \be\label{modtransform}  \varphi\Bigl(\frac{a\t+b}{c\t+d},\frac{\l}{c\t+d}\Bigr) \= 
   (c\t+d)^w\,e^{\frac{2\pi ikc \l^2}{c\t+d}}\,\varphi(\t,\l)  \qquad \forall \quad
   \Bigl(\begin{array}{cc} a&b\\ c&d \end{array} \Bigr) \in SL(2; \Z) \, , \ee
and under the translations of $\l$ by $\mathbb{Z} \tau + \mathbb{Z}$ as
  \be\label{elliptic}  \varphi(\t, \l+a\tau+b)\= e^{-2\pi i k(a^2 \t + 2 a \l)} \, \varphi(\t, \l)
  \qquad \forall \quad a,\,b \in \Z \, . \ee
The numbers $w$, and $k$, which we take to be integers are called the \emph{weight}, and respectively the \emph{index},
of the Jacobi form~$\varphi$. 
The definition of a Jacobi form, in addition to the above transformation laws, include a growth condition that 
can be stated in terms of its Fourier coefficients. The condition often appearing in physics 
specifies\footnote{The condition $c(n,r) = 0$ unless~$4k n \geq r^2$ specifies a class of functions that 
are traditionally called holomorphic Jacobi forms, and similarly, Jacobi cusp forms are defined by the 
condition $c(n,r) = 0$ unless~$4k n > r^2$. This notion of holomorphicity refers to the~$\t$ variable, and 
we shall mostly not be concerned with such functions.} 
a \emph{weak Jacobi form} that obeys~$c(n,r) = 0$ unless $n \geq 0$ in the Fourier expansion~\eqref{Jac}.

The periodicity property~\eqref{elliptic} implies that the Fourier coefficients have the form
\be
c(n, r) \= C_{r}(4 n m - r^2) \ ,
  \qquad \mbox{where} \; C_{r}(\D) \; \mbox{depends only on} \; r \text{mod}\,{2m} \, .
\ee
The quantity~$\D=4mn-r^{2}$ is called the \emph{discriminant} and is a U-duality invariant in terms of the charges 
in the relevant physical set up~\cite{Dabholkar:2012nd}. Note that this implies that~$c(n,r)=0$ unless~$n-r^{2}/4k \ge -\mu^{2}/4k$,
where~$\mu \equiv r \, \text{mod}(2m)$. In the~$\CN=2$ SCFT, the shift of~$-\mu^{2}/4k$ reflects the factorization of the boson 
representing the~$U(1)_{R}$ current algebra. 

The periodicity property~\eqref{elliptic} also implies the theta expansion 
 \be\label{jacobi-theta} \varphi(\t,\l) \= \sum_{\ell\inn \Z/2k\Z} h_\ell(\t) \, \vth_{k,\ell}(\t, \l)\,, \ee
where $\vth_{k,\ell}(\t,\l)$ denotes the standard index $k$ theta function 
$ \vth_{k,\ell}(\t, \l) = \sum_{r\,\equiv\,\ell\,(\rm{mod}\,2k)} q^{r^2/4m} \, y^r$. 
The theta coefficients have the expansion
\be\label{defhltau}  h_{\ell}(\t) \= \sum_{\Delta \ge - \Delta^{\ell}_{0}} C_{\ell}(\Delta) \,  q^{\Delta/4k} 
\, \qquad (\ell \inn \Z/2k \Z)\, ,  \ee
and are modular forms\footnote{These are weakly holomorphic, holomorphic or cuspidal if $\varphi$ 
is a weak Jacobi form, a Jacobi form or a Jacobi cusp form, respectively. The Fourier coefficient~$c(n,r)$ 
of a weak Jacobi form therefore has exponential growth as~$\Delta=4mn-r^{2}$ becomes large.}
 of weight $w-\frac12$. 
More precisely, the vector $h := ( h_1, \ldots, h_{2m})$ transforms like a modular form of weight $w-\frac 12$ 
under $SL(2,\Z)$ with the modular transformation law now including a linear transformation that is independent of~$\t$.
This linear transformation is the S-matrix that relates the different characters of the~$\CN=2$ SCFT.

The derivation for the asymptotic growth of the Fourier coefficients ($\ell \inn \Z/2k \Z$) is similar to the modular 
form case, but now we also have to take into account the CFT S-matrix that relates~$h_{\ell}(\-1/\t)$ and~$h_{\ell'}(\t)$,
different values of~$\ell'$.  The leading growth is controlled by the most polar term~$q^{-\D^{\ell'}_{0}/4k}$ 
which appears in the expansion of~$h_{\ell'}$ for some~$\ell'$. 
We have (see e.g.~\cite{Dijkgraaf:2000fq} for a derivation\footnote{As mentioned in ~\cite{Dijkgraaf:2000fq},
one can go much beyond the leading order estimates that concern us here, and in fact one can get
exact convergent expansions for the Fourier coefficients of modular and Jacobi forms using the so-called
\emph{circle method} and the \emph{Hardy-Ramanujan-Rademacher expansion}.}):
\be \label{cnrgrowth}
C_{\ell}(\D) \sim S^{-1}_{\ell \ell'} \, \exp \biggl(4 \pi  \sqrt{\bigl(\frac{\D^{\ell'}_{0}}{4k} \bigr)\,
\bigl(\frac{\D}{4k} \bigr)} \biggr) \, , 
\qquad \D \to \infty \, . 
\ee

To find the most polar term for a weak Jacobi form, we note that the condition~$n\ge 0$ in the 
Fourier expansion~\eqref{Jac}, combined with the behavior~$\vth_{k,\ell} = O(q^{\ell^{2}/4k})$, implies that 
its theta coefficients~$h_{\ell}$ have~$\D^{\ell}_{0} = -\ell^{2}$ in the expansion~\eqref{defhltau}.  
Choosing~$\ell$ in the coset representative $-k+1, \cdots k$, we see that, for a generic 
weak Jacobi form, the growth is controlled by~$\D^{k}_{0}/4k=k/4$, with the most polar term being~$q^{-k/4}$. 
In the interpretation as the elliptic genus of a SCFT, the terms with~$n=0$ in~\eqref{Jac} are the 
Ramond-Ramond ground states, with~R-charge~$\ell$. 
Indeed the maximum R-charge is~$\pm k$ (this corresponds to the spectral flow of the NS vacuum which is 
present in the spectrum for usual compact SCFTs), so that the most polar term is~$q^{-k/4}$, and the 
growth in this case is 
\be \label{cnrgrowth2}
C_{\ell}(\D) \sim \exp \bigl(\pi  \sqrt{\D} \bigr)  \; \; \Longleftrightarrow \; \; c(n,r) \sim \exp \bigl(\pi  \sqrt{4kn-r^{2}} \bigr) \, , 
\qquad \D = 4kn-r^{2} \to \infty \, . 
\ee
For a generic SCFT where the level of the~$U(1)$ R-current is~$k$, the central charge is~$c=6k$,
and the above growth estimate~\eqref{cnrgrowth2} is, again, simply the Cardy formula.

We mention one other type of Jacobi form that is relevant in physics, and will be useful for us. 
Often the partition function is given by the product of a weak Jacobi form with~$1/\eta(\t)^{b}$
(this could arise from~$b$ bosonic oscillator modes for a center of mass degree of freedom.) Such a function
falls under the category of \emph{weakly-holomorphic Jacobi forms}---this is a much larger space than the 
space of weak Jacobi forms, and correspondingly their structure is less rigid. For practical purposes, e.g.~to
analyze the growth of the Fourier coefficients, one can simply treat the~$\eta$-functions separately, and 
the weak Jacobi forms separately. In this case, formula~\eqref{cnrgrowth2} is modified because the 
most polar state now has energy~$-\frac{k}{4}-\frac{b}{24}$.

\subsection{Meromorphic Jacobi forms \label{meroJac}}

Our main aim here is to study the structure and growth of Fourier coefficients of 
Jacobi forms~$\varphi(\t,\l)$ that have poles in~$\l$. In particular, we are interested in meromorphic Jacobi forms defined in the main text which have the form:
\be \label{Zdbapp}
Z_{d}(\t,\l) \= \frac{N_{d}(\t,\l)}{\eta(\t)^{36 d} \, \prod_{j=1}^{d} \varphi_{-2,1}(\t,j\l)} \; . 
\ee
Here the numerator~$N_{d}(\t,\l)$ is a holomorphic Jacobi form of weight~$w=16d$ and 
index~$k_\text{num}=d(d-1)(d+4)/3$. The function~$\varphi_{-2,1}$ of weight~$-2$, index~1 appearing in the denominator is
\be\label{phiminus2} \varphi_{-2,1}(\t, \l) \, := \, \frac{\vth_1(\t, \l)^{2}}{\eta(\t)^{6}} \, . \ee 
The denominator thus has~$w=16d$, $k_\text{den}=\frac16 \, d(d+1)(d+\frac12)$, so that~$Z_{d}$ has~$w=0$, 
$k_\text{mero}=\frac12 \, d(d-3)$.
Our analysis below follows~\cite{Dabholkar:2012nd},
and is more an illustration of the points relevant to this paper than a complete 
rigorous treatment.

The numerator $N_{d}$ in~\eqref{Zdbapp} is a weak Jacobi form holomorphic in~$\l$, and from the
discussion of the previous subsection, the leading exponential growth of its Fourier coefficients is 
determined by its most polar term~$q^{-\D_{0}}$. 
Using the product formula
\be
\vartheta_1(\tau,\l) \= -i q^{1/8} \, \zeta^{1/2} \, \prod_{n=1}^\infty (1-q^n) (1-\zeta q^n) (1-\zeta^{-1} q^{n-1}) \, , 
\ee
the denominator can be written as:
$$ \eta(\t)^{32 d} \, \prod_{n=1}^\infty \, \prod_{j=1}^{d} \, (1-y^{j} q^n)^{2} \, (1-y^{-j} q^n)^{2} 
$$
The structure of the denominator is that of a bunch of bosonic oscillators carrying different charges (some zero).  
Thinking purely of the growth of states (without considering signs), one may be tempted to conclude that the 
Fourier coefficients of the ratio~$Z_{d}$ grows faster than the numerator~$N_{d}$ because the bosons that 
the denominator represents will only add to the growth. 
This is, however, not quite correct because we are really computing an index.
The numerator generically has both positive and negative signs in its Fourier expansion. Asymptotically, 
as~$n \to \infty$, each power of~$y$ typically has a fixed sign. The expansion of the denominator 
gives a growth of bosonic states (all positive signs) weighted by different powers of~$y$. 
These states, when combined with the negative sign states in the numerator, can thus reduce the growth of states 
in the ratio~\cite{Bringmann:2012zr}. 

\vspace{0.3cm}
\noindent {\bf Decomposition of meromorphic Jacobi forms}

\noindent One thus needs to consider the Fourier expansion of the full meromorphic Jacobi 
form~$Z_{d}$, which is a non-trivial problem. The usual method of using the ellipticity property to first write 
a theta-expansion as in~\eqref{jacobi-theta} and then using the modular properties of the 
theta-coefficients~$h_{\ell}(\t)$ does not work---the right-hand side of~\eqref{jacobi-theta} is holomorphic in~$\l$, 
while~$Z_{d}$ has poles in~$\l$. 
Further, the theta-expansion~\eqref{jacobi-theta} is equivalent to the statement that the Fourier coefficients only 
depend on~$4mn-r^{2}$, and~$r\, \text{mod} \, 2m$. A meromorphic Jacobi form near a pole will behave locally
like~$(1-y)^{-\a}$, where $\a>0$ is the order of the pole. This implies an unbounded Fourier expansion of the 
type~$\sum_{n=0}^{\infty} a(\alpha) \, y^{\alpha}$, with~$a(\alpha)$ having polynomial growth in~$\alpha$. 

The main decomposition theorems (Theorems 8.1--8.3) of~\cite{Dabholkar:2012nd} present a solution of this problem. 
With the conditions as specified there, the theorems say that a meromorphic Jacobi form~$\psi$ of weight~$w$
and index~$k>0$ can be decomposed uniquely into a finite piece~$\psi^{F}_{k}$ and a polar piece~$\psi^{P}_{k}$
\be \label{decomp} 
\psi_{k}(\tau,\l) \; = \;  \psi_{k}^{\rm F}(\tau,\l) \, + \, \psi^{\rm P}_{k}(\tau,\l) \, ,
\ee
where the definitions of the two pieces on the right hand side is a priori unambiguously given in terms of the left-hand side. 
All three parts of the equation~\eqref{decomp} have the same elliptic transformation property~\eqref{elliptic} governed 
by the index~$k$. The polar part has the same poles and residues as the meromorphic Jacobi form, and can be written 
as sums of products of the negative Laurent coefficients of~$\psi$ with explicit functions of~$\t$ and $\l$ 
(called \emph{Appell-Lerch sums}). 
Having separated out the polar pieces, we find that 
the finite piece is holomorphic in~$\l$, and therefore has a theta expansion with theta-coefficients~$h^{F}(\t)$. 
The main statement of the theorem is that the finite part~$\psi^{F}_{k}(\t,\l)$ is a \emph{mixed mock Jacobi form}, or  
equivalently, its coefficients~$h^{F}(\t)$ are vector-valued \emph{mixed mock modular forms}.  

We refer to~\cite{Dabholkar:2012nd} for a discussion of the notion of mock modular forms and mock Jacobi forms detail.
Here we only state the bare minimum facts. A mock modular form is holomorphic 
function~$f(\t)$ which is almost modular, but not quite---its lack of modularity is governed by another anti-holomorphic modular 
form~$\overline{g}$ which is called the \emph{shadow} of~$f$. The function~$\widehat{f} = f + g^{*}$, called the \emph{completion}
of~$f$ transforms as a holomorphic modular form, but it is not holomorphic. The function~$g^{*}$, and therefore the 
function~$\widehat{f}$ obeys a holomorphic anomaly equation:
    \be\label{ddtbarh}   (4\pi\t_2)^w\,\,\frac{\partial \widehat f(\t)}{\partial \overline{\tau}} \= -2\pi i\;\overline{g(\tau)}\;.  \ee
The additional adjective \emph{mixed} means that the mock modular form can be multiplied by a true modular form. In this 
case, the shadow is of the form~$h(\t) \, \overline{g(\t)}$, where~$h$ is a holomorphic modular form.
Mock Jacobi forms are holomorphic functions of~$\l$ obeying the elliptic property and therefore have a theta-expansion. 
Their theta coefficients are now vector-valued \emph{mock} modular forms~$h_{\ell}$ with a corresponding vector of
shadows~$\overline{g}_{\ell}$, $\ell \in \IZ/2k\IZ$. 
For the meromorphic Jacobi forms discussed in the text, the shadows~$g_{\ell}$ are simply the theta functions~$\vartheta_{k,\ell}(\t,0)$,
or suitable derivatives thereof. When the meromorphic Jacobi form of index ~$k$ has a second order pole with Laurent coefficient~$D(\t)$, 
the shadow function is~$D(\t) \, \sum_{\ell \in \IZ/2k\IZ} \overline{\vartheta_{k,\ell}(\t,0)} \, \vartheta_{k,\ell}(\t,z)$. This relates to the 
functions discussed in \S\ref{sec:BHCen}.

\vspace{0.3cm}
\noindent {\bf Fourier coefficients}

\noindent With this decomposition in hand, we can analyze the growth of the Fourier coefficients of the 
meromorphic Jacobi form. 
We have three main conclusions:
\begin{enumerate}
\item The finite part~$\psi^{F}_{k}$ has a theta-expansion of the form~\eqref{jacobi-theta}. In particular, 
the coefficients of the Fourier expansion~$\psi_{k}^{F}(\t,\l) = \sum_{n,r} \, c_{k}^{F}(n,r) \, q^{n} \, y^{r}$ 
vanish for large negative values of~$\D=4kn-r^{2}$. The Fourier coefficients of the polar parts are non-zero
for unbounded negative values of~$\D=4kn-r^{2}$. 
\item The Fourier coefficients of the finite part~$\psi^{F}_{k}$ grow much faster 
than those of the polar part~$\psi_{k}^{P}$, in any expansion region. 
\item The Fourier 
coefficients~$c^{F}_{k}(n,r)$ of the finite part~$\psi^{F}_{k}$ have the same growth as those of a true holomorphic 
Jacobi form of the same index:
\be \label{cfest}
c_{k}^{F}(n,r) \sim \exp\Bigl(2 \pi \sqrt{\frac{(6k+b)}{6} \bigl(n- r^{2}/4k \bigr)} \Bigr) \, , \quad n \to \infty \, , \; r \; \text{fixed,}
\ee
where the most polar term in~$\psi^{F}_{k}$ is~$O(q^{-k/4-b/24})$, as at the end of~Section \ref{appJacobi}.
\end{enumerate}

We now give some details of the arguments that support these conclusions. 
The first conclusion essentially follows from the statement of the decomposition theorem above: as discussed above,
the unbounded behavior in negative discriminant~$\D$ comes from the existence of poles in~$y$, 
and these are completely separated out  by the polar term~$\psi_{k}^{P}$. The finite part is holomorphic 
in~$\l$, and therefore, has a bounded, periodic, Fourier expansion as for a true Jacobi form. 
In order to explain the second and third conclusions about the growth of the Fourier coefficients, we shall analyze each piece in~\eqref{decomp} in turn, illustrating our statements with the reference case of the K3 theory which we discuss in Section \ref{sec:K3}. 
There,~$\psi_{k}$ has weight~$-12$, index~$k$, with only one double pole at~$\l = 0+ \IZ\t+\IZ$, with a constant Laurent coefficient.

We first discuss the finite part. 
The asymptotic growth of the Fourier coefficients can be determined by the use of the modular transformation
applied to the most polar term in~$q$ as for true modular and Jacobi forms (\S\ref{appmodular}, \S\ref{appJacobi}). 
The main point is that the holomorphicity in~$y$ guarantees that there \emph{is} a most polar term in 
the~$q$-expansion. 
Thus, although the modular transformation properties of the mock modular and Jacobi forms differs from that of true 
modular and Jacobi forms, it only does so at sub-leading order, and leading exponential 
estimate\footnote{This was, in fact, already pointed out by Ramanujan.} remains the 
same\footnote{In fact one can do much better. Since mock modular/Jacobi forms are only defined modulo 
true modular/Jacobi forms, one can try to subtract true Jacobi forms from~$\psi^{F}_{k}$ to define a canonical 
mock Jacobi form of slow growth for each~$k$. In the~K3 situation, it was shown in~\cite{Dabholkar:2012nd}
that this can be systematically done for every~$k$ so that the leading growth of the black hole is really controlled 
by a true Jacobi form of index~$k$, and the ``mock piece'' is really small 
(roughly $e^{\frac{1}{k}\sqrt{4kn-r^{2}}}$ instead of~$e^{\sqrt{4kn-r^{2}}}$).} 
as in \eqref{cfest}. 
This statement is illustrated by the explicit sub-leading estimates that have been worked 
out for the functions arising in the K3 theory~\cite{Bringmann:2012zr}.

We now discuss the polar part. 
The polar part is completely determined by the poles of~$\psi_{m}$ and its Laurent coefficients 
at the poles. Suppose there is a pole at~$\l=\l_{P}=\a \t + \b$, $\a,\b \in \mathbb{Q}$, of order~$a \in \IN$, 
the meromorphic Jacobi form with singular part~$D_{P}(\t)\, (\l-\l_{P})^{-a}$ near the pole,
where the Laurent coefficient~$q^{m\a(\a\t+\b)} \, D_{P}(\t)$ is\footnote{More precisely, it is a finite 
combination of quasi modular forms that are determined by the details of the function~$\psi_{k}$.}, 
a modular form of weight~$w-a$. In such a situation, the polar part is a product of~$D_{P}(\t)$ 
and explicit functions called Appell-Lerch sums that are essentially index-$k$ averages over the 
lattice~$\IZ\t+\IZ$ of the basic pole~$(\l-\l_{P})^{-a}$. The full polar part is a sum of such products  
over all the poles of~$\psi_{k}$.

In the~K3 situation, there was one pole of order~$2$ at~$\l=0$,
and the relevant Appell-Lerch sum is presented in~\eqref{simfun}, \eqref{FexpAL}. 
As explained there, its Fourier coefficients grow extremely slowly. 
In our F-theory situation, the functions~$Z_{d}$ have higher poles of even orders from~$\a=2$ to~$\a=2 d$. 
(See~\cite{BringFol, Olivetto} for a more detailed analysis of the higher pole cases.)
In each case, the simple Appell-Lerch sums~$\CA_{2,k}(\t,\l)$ are replaced by more complicated polar functions, 
but the conclusion of slow (polynomial) growth is valid in each of the cases. So, for our question of interest,
one can ignore the Appell-Lerch sums, and the only exponential growth comes from the Laurent coefficients.

\end{document}